\documentclass[useAMS,usenatbib,usegraphicx,usegraphicx]{mn2e}

\usepackage{multirow}
\usepackage{amssymb}
\usepackage[normalem]{ulem}
\usepackage{xcolor}

\long\def\Ignore#1{\relax}

\newcommand{\degrees} {^\circ}

\newcommand{\cut}[1]{}

\title[Spatial and kinematic segregation in star cluster merger
  remnants]{Spatial and kinematic segregation in star cluster merger
  remnants} \author[D. R. Cole et al.]{David
  R. Cole$^{1,2}$\thanks{E-mail: {\tt david.cole@physics.ox.ac.uk}},
  Victor P. Debattista$^{2}$\thanks{E-mail: {\tt
      vpdebattista@gmail.com}}, Anna-Lisa Varri$^{3}$\thanks{E-mail:
    {\tt annalisa.varri@gmail.com}}, Markus
  Hartmann$^{4}$\thanks{E-mail: {\tt hartmann@ari.uni-heidelberg.de}}
  \and and Anil C. Seth$^{5}$\thanks{E-mail: {\tt aseth@astro.utah.edu}}
  \\ $^{1}$Rudolf Peierls Centre for Theoretical Physics, Keble Road, Oxford, OX1 3NP, United Kingdom 
  \\ $^{2}$Jeremiah Horrocks Institute, University of Central
  Lancashire, Preston, PR1 2HE, United Kingdom \\ 
$^{3}$School of Mathematics and Maxwell Institute for Mathematical Sciences, University of Edinburgh, Edinburgh, EH9 3JZ, United Kingdom \\ 
  $^{4}$Astronomisches Rechen-Institut, Zentrum f\"ur Astronomie der Universit\"at Heidelberg (ZAH), M\"onchhofstr. 12-14, 69120 Heidelberg, Germany \\
  $^{5}$Department of Physics and Astronomy, University of Utah, Salt Lake City, UT 84112, USA}

\begin{document}

\date{Accepted xxx Received xxx ; in original form \today}

\maketitle

\label{firstpage}

%===========================================================================
% Abstract
%===========================================================================

\begin{abstract}
  Globular clusters which exhibit chemical and dynamical complexity
  have been suggested to be the stripped nuclei of dwarf galaxies
  (e.g., M54, $\omega$ Cen).  We use $N$-body simulations of nuclear
  star clusters forming via the mergers of star clusters to explore
  the persistence of substructure in the phase space.  We find that
  the observed level of differentiation is difficult to reconcile with
  the observed if nuclear clusters form wholly out of the mergers of
  star clusters.  Only the star clusters that merged most recently
  retain sufficiently distinct kinematics to be distinguishable from
  the rest of the nuclear cluster though the critical factor is the
  number of merger events not the elapsed time.  In situ star
  formation must therefore be included to explain the observed
  properties of nuclear star clusters, in good agreement with previous
  results.
  
\end{abstract}


%==========================================================================
% Introduction
%==========================================================================

\section{Introduction}  
\label{sec:intro}

High resolution \textit{Hubble Space Telescope} (\textit{HST}) observations
have shown that many low to intermediate mass galaxies across the
Hubble sequence contain a dense star cluster at their centre, a
nuclear star cluster (NSC) \citep{Carollo1997, Boker2002, Cote2006,
  Turner2012}.

NSCs in late-type galaxies are found to have complex star formation
histories with mean luminosity-weighted ages ranging from 10 Myr to 10
Gyr \citep{Rossa2006}. Observations frequently show that the star
formation is bursty, recurring on a timescale of the order of 100 Myr
with the most recent episodes in the last 100 Myr \citep{Walcher2005,
  Walcher2006}. One example is the NSC in M33 which had periods of
star formation 40 Myr and 1 Gyr ago \citep{Long2002}.
\citet{Georgiev2014} studied 228 late-type galaxies and found that
recent star formation is common and their stellar populations had a
range of ages. \citet{Carson2015} found increasing roundness at longer
wavelengths in {\it HST} WFC images of the 10 brightest and nearest
NSCs. They inferred that the NSCs contained discs with younger stellar
populations. Colour-colour diagrams for most of these NSCs also show
evidence for two populations, a younger one of the order of a few
hundred Myr old and an older one more than a Gyr
old. \citet{Pfuhl2011} studied the Milky Way's NSC and found that
$\sim 80\%$ of its stars are more than 5 Gyr old but there was a deep
minimum in star formation 1 to 2 Gyr ago followed by an increase in
star formation in the last few hundred Myr. NSCs in late-type galaxies
are often made of an older spheroidal component with a younger, bluer
disc embedded in it, with the disc approximately aligned with the
plane of the main galactic disc \citep{Seth2006, Seth2008}. The NSC in
NGC~4244 has such a structure and the stars in the disc are less than
100 Myr old. Integral field spectroscopy indicates that the disc is
rotating in the same sense as the main galactic disc and is misaligned
by only $\sim15\degrees$. The NSC in the elliptical galaxy FCC~277 also has
the spheroid$+$disc structure with stars younger than those in the main
galaxy \citep{Lyubenova2013}.

Two principal formation mechanisms have been proposed to explain the
formation of NSCs: the merging of GCs, and in situ star formation. In
the GC merger scenario the GCs' orbits decay due to dynamical friction
and then they merge at the centre of galaxies \citep{Tremaine1975,
  CapuzzoDolcetta1993, Miocchi2006, CapuzzoDolcetta2008a,
  CapuzzoDolcetta2008b, Antonini2012, Antonini2013, Gnedin2014}.  In
situ star formation could occur due to a variety of mechanisms but
would require a process whereby gas is driven to the nuclear regions
of galaxies \citep{Milosavljevic2004, Bekki2007}. These include the
action of re-ionisation epoch radiation fields \citep{Cen2001} and
compressive tidal fields \citep{Emsellem2008}. \citet{Georgiev2014}
found that the half-light radius, $r_{eff}$, of their sample of NSCs
increases with wavelength and argue that this could be explained if
NSCs form from gas which falls to the centre and forms stars, meaning
that younger populations will be more centrally concentrated than
older ones. The most direct evidence for the need of in situ star
formation comes from modelling the kinematic data for the NSC in
NGC~4244.  Simulations by \citet[][see also
  \citet{deLorenzi2013}]{Hartmann2011} find that though the globular
cluster (GC) merger scenario can reproduce many of the density and
kinematic properties of NSCs, mergers give rise to a central peak in
$v_{\rm rms} = \sqrt{\sigma_{\rm los}^2 + v_{\rm los}^2}$, which is not
observed in the data.  Based on this, they conclude that less than
$50\%$ of the mass of the NSC could have been assembled from the
mergers of GCs, with the majority due to in situ star formation.

\begin{table}
  \begin{center}
    \begin{tabular}{lcccc} 
      \hline
%      \hline#
      \multicolumn{1}{l}{Cluster} &
      \multicolumn{1}{c}{Mass} &
      \multicolumn{1}{c}{Absolute} &
      \multicolumn{1}{c}{Half} &
      \multicolumn{1}{c}{[Fe/H]} \\
      \multicolumn{1}{l}{} &
      \multicolumn{1}{c}{$\times10^6$ M$_\odot$} &
      \multicolumn{1}{c}{visual mag.} &
      \multicolumn{1}{c}{mass radius} &
      \multicolumn{1}{c}{} \\
      \multicolumn{1}{l}{} &
      \multicolumn{1}{c}{} &
      \multicolumn{1}{c}{} &
      \multicolumn{1}{c}{pc} &
      \multicolumn{1}{c}{} \\[-0.5ex]
      \hline
\hline
$\omega$ Cen & 2-5$^1$ & -10.24 & 6.20 & -1.62 \\
47 Tuc & 0.7-1.45$^{2,3}$ & -9.37 & 3.49 & -0.76 \\
NGC~1851 & 0.561$^3$ & -8.35 & 1.85 & -1.26 \\
M54 & 1.45$^3$ & -9.96 & 3.76 & -1.59 \\
M22 & 0.536$^3$ & -8.45 & 3.03 & -1.64 \\
Terzan~5 & $\sim2^4$& -7.86 & 1.93 & -0.28\\
\hline
    \end{tabular}
  \end{center}
  \caption[]{\label{tab:props} Properties of $\omega$ Cen, 47 Tuc,
    NGC~1851, M54, M22 and Terzan~5 from
    \citet{Harris1997}. $^1$\citet{Meylan1995,vandeVen2006,DSouza2013},
    $^2$\citet{Marks2010}, $^3$\citet{Gnedin1997},
    $^4$\citet{Lanzoni2010}. Although 47 Tuc does not show any
    evidence of enhancement or spread in its iron abundance (see
    \cite{Marino2016}, we include it in this list in light of the
    estimated total mass, its complex light elements abundance
    patterns \citep[e.g., see][]{Cordero2014,Kucinskas2014}, and its
    rich internal dynamics \citep[e.g.,
      see][]{Richer2013,Bianchini2013}.}
\label{tab:gcs}
\end{table}

Turning our attention to globular clusters, the interpretative
paradigm for their formation and dynamical evolution is even more
puzzling. At one time the Milky Way’s GCs were thought to consist of a
single stellar population, but the availability, over the past decade
or so, of high quality and homogeneous photometric and spectroscopic
datasets has revealed a much more complex picture of the star
formation history of this class of stellar systems. In particular,
there is now clear evidence that most Galactic GCs exhibit light
elements abundance patterns and colour-magnitude diagram morphology
indicative of the existence of multiple stellar populations \cite[see
  e.g.,][]{Gratton2012,Piotto2015}.  A number of possible scenarios
have been proposed to provide an interpretation of such an ubiquitous
and puzzling phenomenon, often invoking the presence of two (or more)
generations of stellar populations, with several different possible
sources for the gas out of which second population stars form. These
sources include rapidly rotating massive stars, massive binary stars,
and intermediate-mass asymptotic giant branch (AGB) stars \citep[see
  e.g.,][]{Ventura2001,Prantzos2006,deMink2009,DErcole2008,DErcole2010,DErcole2012}.
Alternative scenarios further elaborate on the role of the ejecta from
massive interacting binaries, in the context of the formation of
circumstellar disks of young, low mass stars \citep{Bastian2013}, as
a possible origin for the observed abundance anomalies.

One crucial insight into this problem may arise from the investigation
of the structural \citep[see e.g.,][]{Vesperini2013} and kinematical
properties \citep[see
  e.g.,][]{MastrobuonoBattisti2013,HenaultBrunet2015} of multiple
stellar populations. \cite{Lardo2011} has studied nine Galactic GCs
with Sloan Digital Sky Survey (SDSS) data, and found that there is a
statistically significant spread in $u - g$ colour, corresponding to
variations in the abundances of light elements, with the redder stars
being more centrally concentrated than the bluer ones. They concluded
that there are distinct populations which have different radial
distributions. From the kinematic perspective, \cite{Richer2013}
analysed the proper motions of main sequence stars in 47 Tuc by
dividing them into four colour bands, assuming that the colour bands
represent stars with different chemical composition. They found that
the main sequence stars in 47 Tuc have anisotropic proper motions, and
that such a feature is correlated with their colours. They also found
that the bluest stars are also the most centrally concentrated,
confirming that, also in the case of 47 Tuc, different stellar
populations can be distinguished by their spatial distribution. More
recently, \cite{Bellini2015} have studied the kinematic properties of
multiple populations in NGC 2808 on the basis on high-precision Hubble
Space Telescope proper-motion measurements, and they found that the
helium-enriched populations are more radially anisotropic. All aspects
of the formation, chemistry, and dynamical evolution of GCs are
currently intensely debated \citep[see
  e.g.,][]{Renzini2015,Bastian2015,DAntona2016}, and only the synergy
between state-of-the-art photometric \cite[especially the {\it HST} UV
  Legacy Survey of Galactic GCs, presented by][]{Piotto2015},
spectroscopic \citep[see ][]{ Carretta2015,Lardo2015}, and proper
motion \citep[from {\it HST} and {\it Gaia}, e.g.,
  see][respectively]{Watkins2015,Pancino2013} information will allow
us to address many of these open questions.

One additional (and older) puzzle is the existence of globular
clusters with significant variations in their heavy elements
abundances. In this respect, evolutionary scenarios that include one
or more merger events have been often formulated as a possible
formation channel of these "multimetallic clusters"
\citep{vandenBergh1996,Catelan1997,Lee1999,Carretta2010c,Carretta2011,Bekki2012,AmaroSeoane2013}. The
scenarios in this class have often been considered rather unlikely in
the Galactic environment, but not unrealistic in other settings, such
as in interacting galaxies (e.g., the Antennae) or in the core of a
dwarf galaxy (e.g., Sagittarius). In particular, it has been envisaged
that GCs may come close and merge due to galaxy interactions or where
GCs have fallen to the centre of the host system due to dynamical
friction. \cite{AmaroSeoane2013} investigated this process using
$N$-body simulations and found that the radial distribution of
different populations are similar to those in multimetallic GCs. In
particular, they found that the distribution of stellar populations in
their dynamical models had some resemblance to the observed
distribution in $\omega$ Cen. However \cite{Catelan1997} found that a
merger of two GCs would produce a red giant branch with bi-modal
colours and no such bimodality had been seen in Galactic GCs, which
led to the conclusion that they are unlikely to be formed by
mergers. \citet{Ferraro2009} has found that Terzan~5 shows bimodality
in the red clump and red giant branch.

Interestingly, Galactic globular clusters that are characterised by
anomalous metallicity distributions tend to be also particularly
massive. These two aspects, coupled with additional signatures of
dynamical complexity, have often been interpreted as possible
indications that these stellar systems may be remnants of dwarf
galaxies, which have been tidally stripped by the potential of the
Milky Way \citep[e.g.][]{vandenBergh1996,Bekki2006}. In this context,
it has also been speculated that these objects are actually able to
retain fast supernovae ejecta (hence the spread in heavy elements), as
they were much more massive at their birth, further supporting the
possibility of identifying them as nuclei of disrupted dwarf
galaxies. Notable cases, as characterized by a very wide or even
multi-modal metallicity distribution, include $\omega$ Cen
\citep{Lee1999,Bekki2006,Carretta2010b}, M54
\citep{Carretta2010c,Sarajedini1995,Siegel2007}, and
Terzan~5\citep{Ferraro2009,Origlia2011,Massari2014}.

Significant intrinsic iron spreads have been measured also in M22
\citep{DaCosta2009,Marino2009,Marino2011}, M2 \citep{Yong2014}, NGC
1851 \citep{Carretta2010c,Carretta2011,Bekki2012}, and NGC 5286
\citep[][see their Table 10]{Marino2015}. In this context, it should
also be emphasized that the analysis of GCs identified as having an
intrinsic Fe spread deserves particular care, especially with respect
to non-local thermodynamical equilibrium effects driven by
over-ionization mechanisms in the atmosphere of AGB stars, which may
lead to spurious metallicity assessments
\citep[e.g. see][]{Lapenna2014,Mucciarelli2015}.

Indeed, the half mass radii and masses of many clusters listed above
are consistent with those of NSCs \citep[see Table 1, and also Fig. 3
  of ][]{Walcher2005}. In particular, $\omega$ Cen has been often
identified as a very peculiar star cluster, not just in consideration
of its chemical complexity, but also for a number of interesting
kinematical and dynamical features \citep[such as the strong internal
  rotation][]{Sollima2009,Bianchini2013}, especially the possibility
of the presence of a central, dynamically decoupled substructure, as
revealed by the Schwarzschild model proposed by
\cite{vandeVen2006}. M54 is thought to be the NSC of the Sagittarius
dSph and, as such, to be in the process of being stripped by the
Galactic potential \citep[but][find that M54 may be 2 kpc in the
  foreground of the centre of the Sagittarius dSph, although this
  would require an unusual alignment]{Siegel2011}. There is also
evidence for the presence of an intermediate mass black hole in both
M54 and $\omega$ Cen, which also favours them being stripped NSCs
\citep{Ibata2009,Wrobel2011,Noyola2008,Miocchi2010}, although,
especially in the case of $\omega$ Cen, this issue is still highly
debated \citep{Anderson2010,vanderMarel2010}.

Inspired by the new recognition of chemical and dynamical complexity
which seems to characterise these stellar systems, possibly at the
interface between globular clusters and nuclear star clusters, we wish
to perform an investigation of a number of structural, kinematical,
and phase space properties of the products of numerical experiments of
globular clusters mergers, as a possible formation scenario of
NSCs. In particular, we wish to assess the persistence of any
structural and kinematical distinction between the different
components, associated with the original globular clusters, within the
stellar system resulting from the merger process. An analysis, devoted
to the exploration of the spatial and age differences among different
mass components, has recently been presented by \cite{Perets2014}; in
the present investigation we wish to devote our attention in
particular to the kinematical and dynamical properties.

To test whether different components can be distinguished spatially
and kinematically in the case of GC merging, we have studied two
simulations of this process. In Simulation 1, a number of GCs merge to
form a larger object, while in Simulation 2 we start with a
pre-existing star cluster, and then add several GCs, one at a time, to
merge to the central object. We then study the spatial distributions
and the kinematics of the stars originating in different GCs, and
investigate how well mixed they are. The article is organised as
follows: Section 2 describes the simulation methods, Section 3
describes the resultant star cluster for simulation 1 and the remnant
star cluster for simulation 2, and finally Section 4 presents a
discussion of our conclusions.

%%%%%%%%%%%%%%%%%%%%%%%%%%%%%%%%%%%%%%%%%%%%%%%%%%%%%%%%%%%%%%%%%%%%%%

\section{The simulations}
\label{sec:simulation}

\begin{figure}
\begin{tabular}{ll}
\includegraphics[width=0.5\hsize,angle=0]{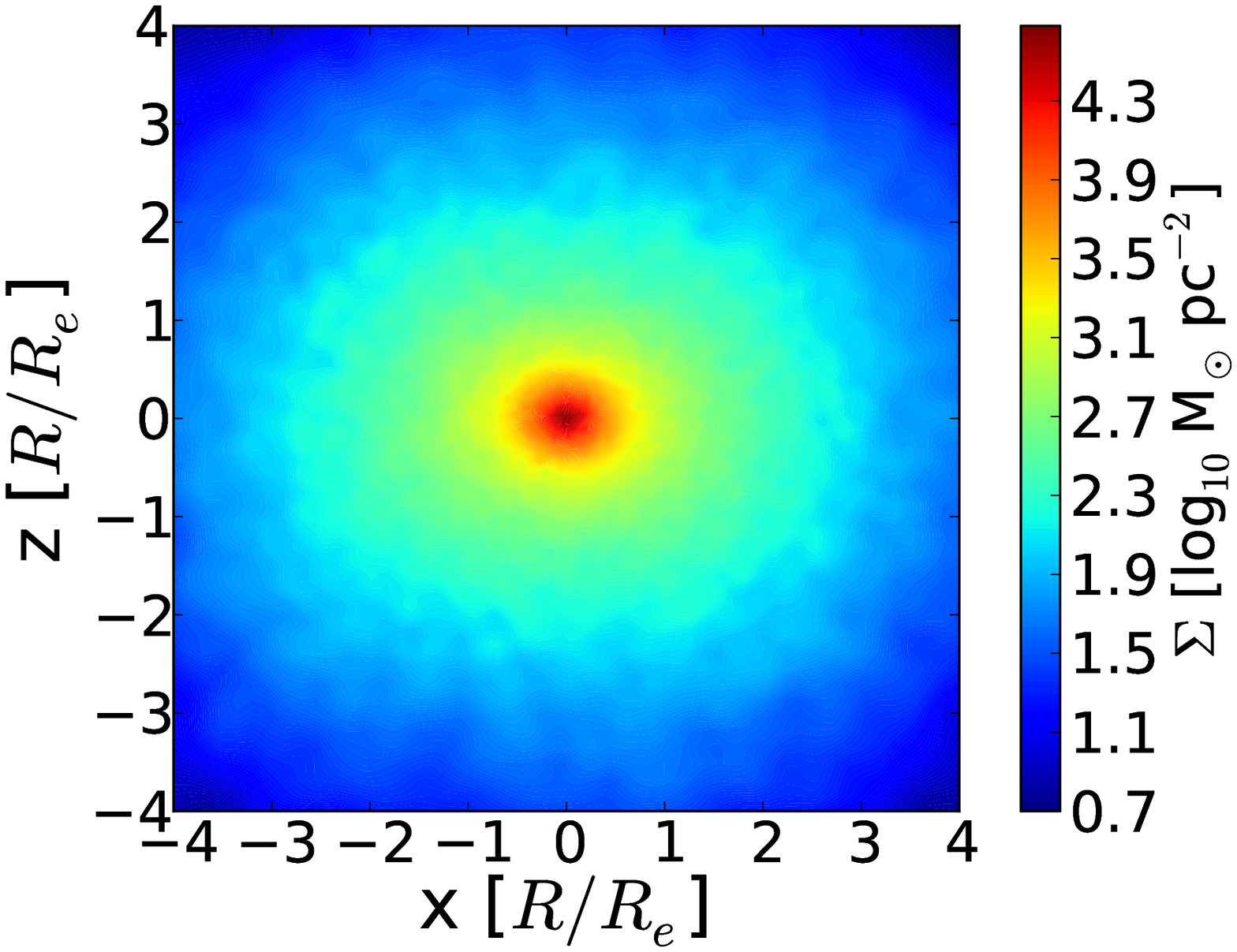} \\
\includegraphics[width=0.5\hsize,angle=0]{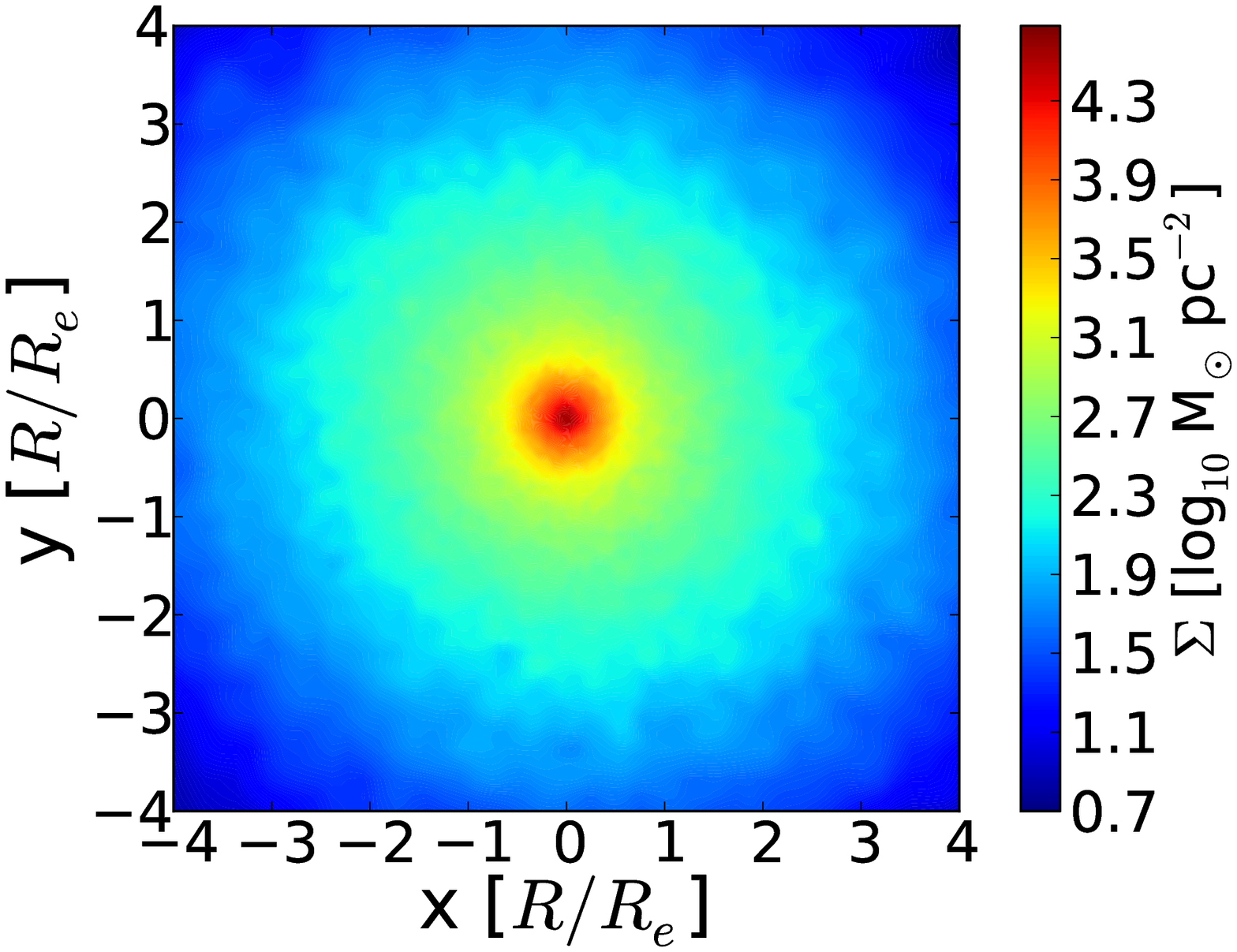} 
\includegraphics[width=0.5\hsize,angle=0]{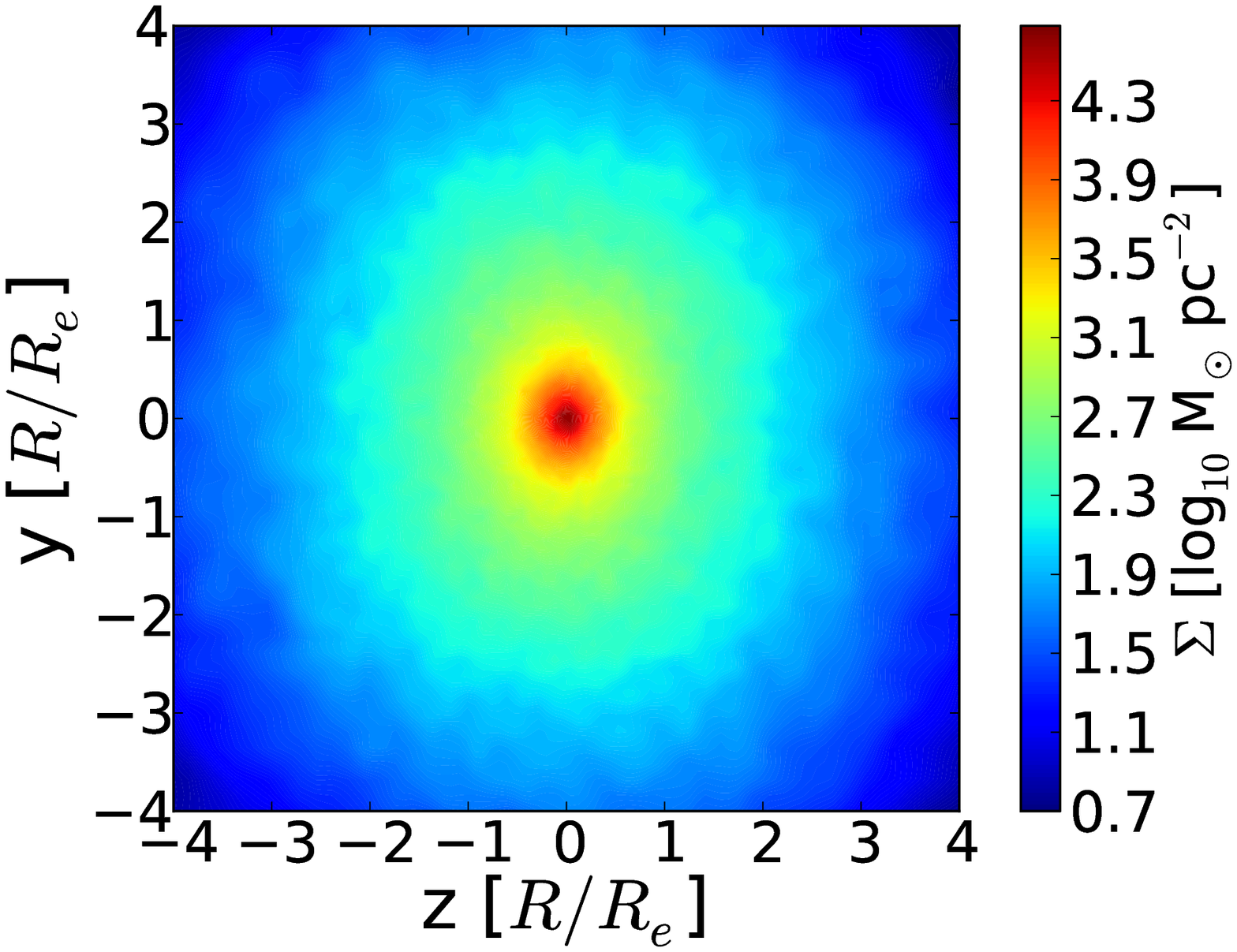} \\
\end{tabular}
\caption{Surface density map of the final merger remnant from
  simulation 1 showing three orthogonal projections. The system has
  been oriented so that the resulting angular momentum is about the
  z-axis. }
\label{fig:sdens}
\end{figure}

Our $N$-body simulations were run using the efficient parallel tree
code PKDGRAV \citep{Stadel2001} suitable for studying collisionless
dynamics.  Simulation S1 has not been described previously and we
provide a description here. Simulation S2 was described by
\citet{Hartmann2011} and we only provide a brief description of it
here.  Both simulations S1 and S2 evolve within a bulge model. The
bulge model has a \citet{Hernquist1990} profile:
\begin{equation}
\rho(r)=\frac{aM_b}{2{\pi}r(r+a)^3},
\end{equation}
where the mass, $M_b=5\times10^9M_\odot$, scale radius $a=1.7$ kpc and
is truncated at $r>15a$ \citep{Sellwood2009}. The bulge is made up of
$3.5\times10^6$ particles with masses ranging from 40 $M_\odot$ at the
centre to $3.9\times10^5M_\odot$ further out giving increased mass
resolution inside of 160 pc \citep{Sellwood2008}. The bulge has no
strong instabilities meaning that the distribution of particles
remains unchanged on timescales up to a Gyr, more than adequate to model
multiple accretions of GCs. 

In simulation S1 there was no initial structure at the centre of the
bulge and star clusters were placed on central orbits which decayed
due to dynamical friction, falling to the centre, where six of
initially ten SCs merged to form a nuclear remnant. The orbits of the
ten SCs were found by selecting particles of the bulge with high
angular momentum within a radius of 100 pc. We placed the SCs at the
same position and with the same velocities as these bulge
particles. Afterwards we rotated the SC system by 180$^\circ$ around
the centre. The initial distances from the centre of the SCs range
from 50 to 100 pc, with velocities in the range of 18 to 121 km
s$^{-1}$. The model SCs have a mass of $4\times10^4M_\odot$,
comparable to young star clusters found in the Milky Way and the Local
Group \citep{Figer1999,Figer2002,Mackey2003,McLaughlin2005}. The SCs
are composed of particles of equal mass of 1 M$_\odot$ and softening
of 0.04 pc. The SC model is a isotropic distribution function of a
lowered polytrope with index $n=2$:
\begin{equation}
f(x,v)\propto[-2E(x,v)]^{1/2}-[-2E_{max}]^{1/2}
\end{equation}
An iterative process is used to produce equilibrium models
\citep{Debattista2000}.

Simulation S2 is the same as run A1 of \citet{Hartmann2011} which was
also studied by \citet{Portaluri2013}.  The SC models have particles
of equal mass (15 M$_\odot$) and equal softening ($\epsilon$ = 0.13
pc). The concentration c = 0.16 is defined as c = log$(R_{eff}/R_c)$,
where $R_{eff}$ = 1.11 pc is the half-mass radius (effective radius)
and $R_c$ is the core radius, where the surface density drops to half
of the central. This model is also comparable to massive young star
clusters in the Milky Way and the Local Group
\citep{Figer1999,Figer2002,Mackey2003,McLaughlin2005}. We create a NSC
for the SCs to accrete onto by letting a massive star cluster of
similar profile and mass 2 $\times10^6$ M$_\odot$, c = 0.12 and
$R_{eff}$ = 2.18 pc fall to the centre of the bulge. This star cluster
was allowed to settle to the centre from a circular orbit at 127 pc,
which takes 65 Myr, before we started the accretion of 27 GCs,
starting them on circular orbits at a distance of 32 pc from the
centre. In total, the mass accreted corresponds to $\sim$8.1 times
the NSCs initial mass. Each accretion is allowed to finish before a
new GC is inserted. A single accretion on average requires $\sim$20
Myr and the 27 GCs are accreted in 810 Myr.

%%%%%%%%%%%%%%%%%%%%%%%%%%%%%%%%%%%%%%%%%%%%%%%%%%%%%%%%%%%%%%%%%%%%%%

\begin{figure}
\includegraphics[width=\hsize,angle=0]{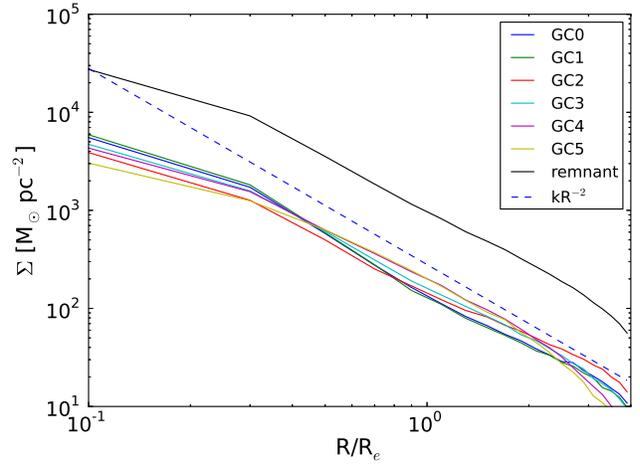} 
\caption{Cylindrical density profiles for the final merger remnant for
  model 1 showing the overall density and the densities contributed by
  stars originating in each progenitor GC. For comparison the dashed
  (blue) line shows a power law with an exponent of -2.}
\label{fig:profiles}
\end{figure}

\begin{figure*}
\centering
\begin{tabular}{c}
\includegraphics[width=0.33\hsize,angle=0]{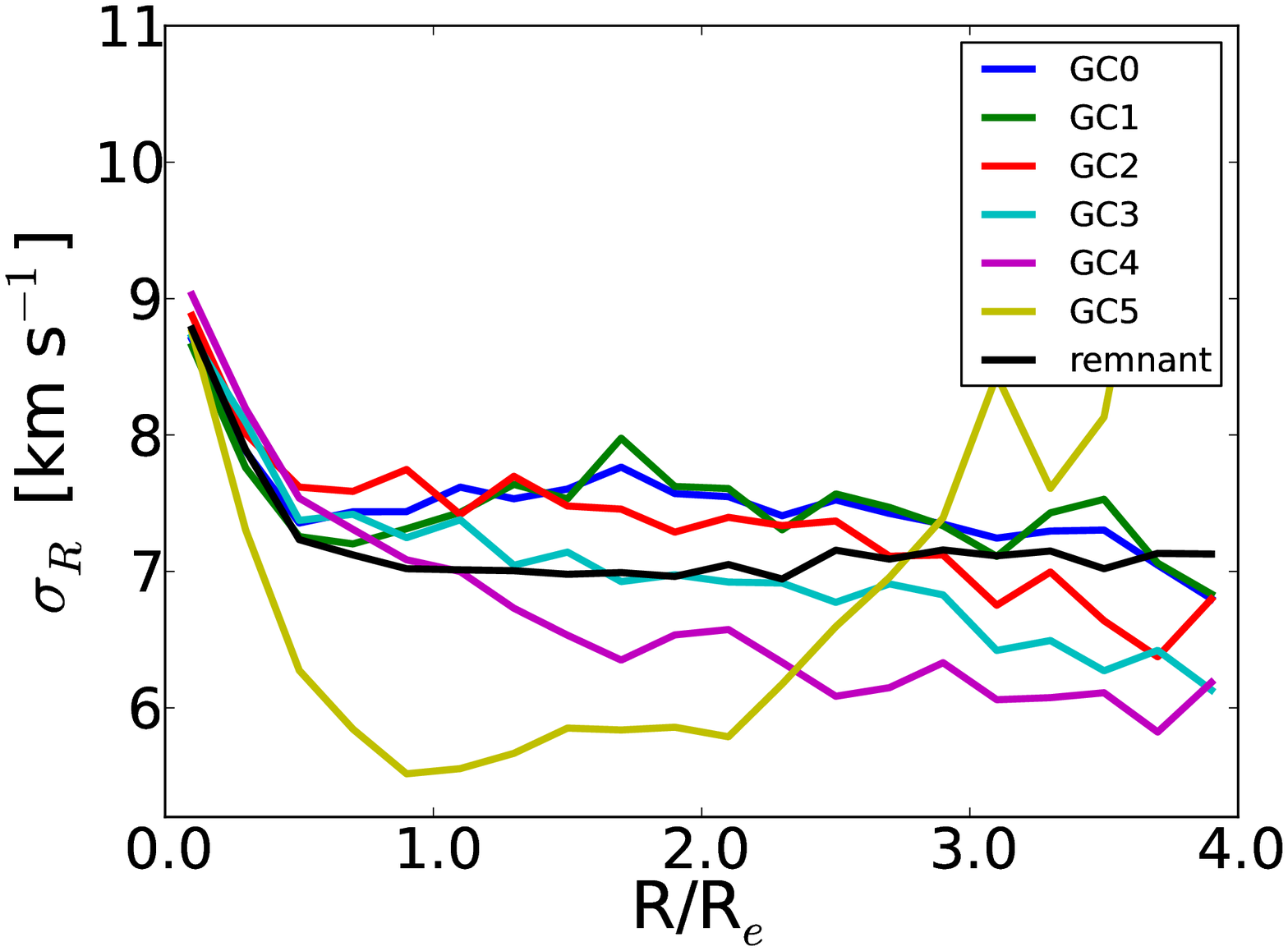}
\includegraphics[width=0.33\hsize,angle=0]{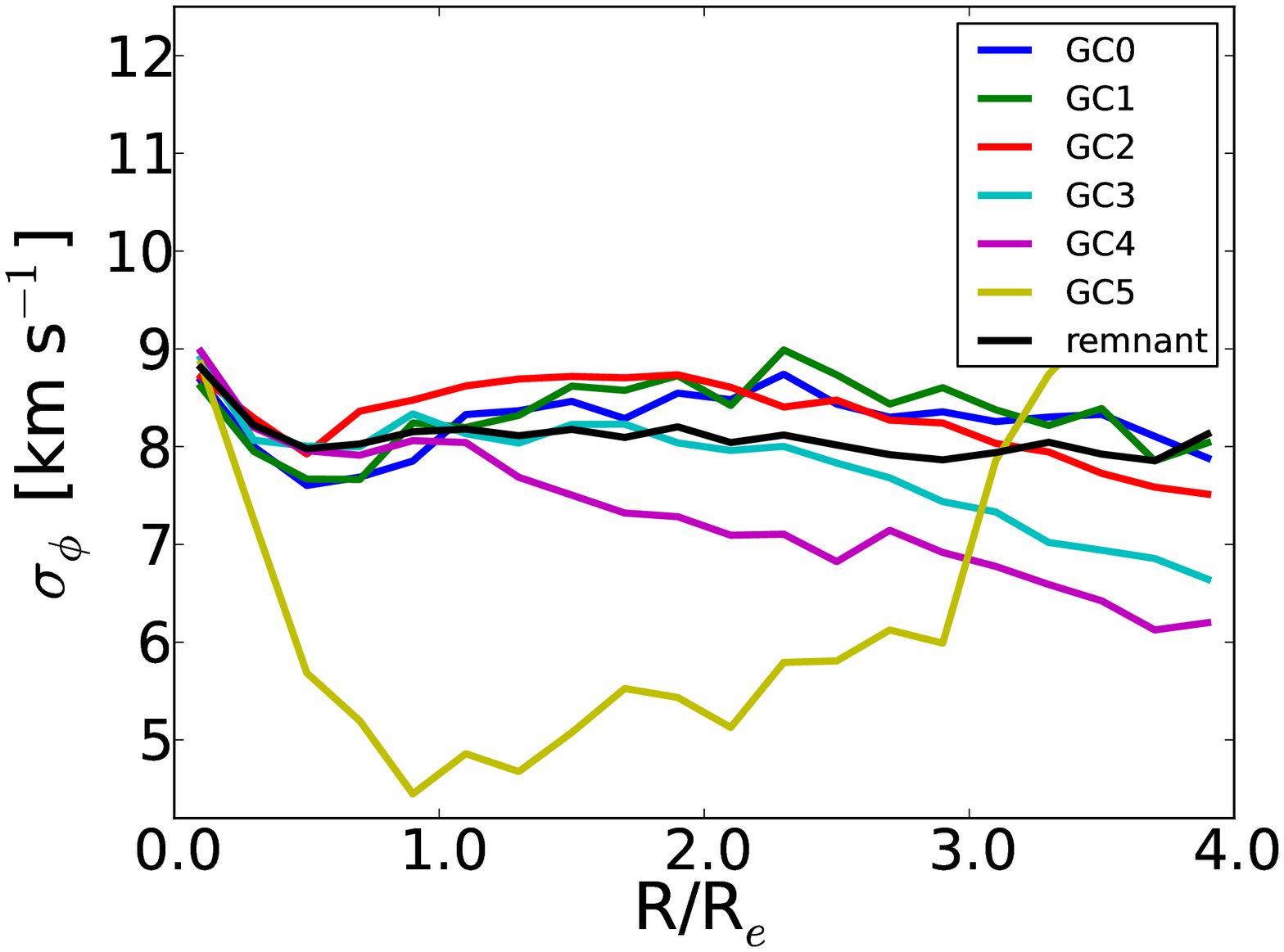}
\includegraphics[width=0.33\hsize,angle=0]{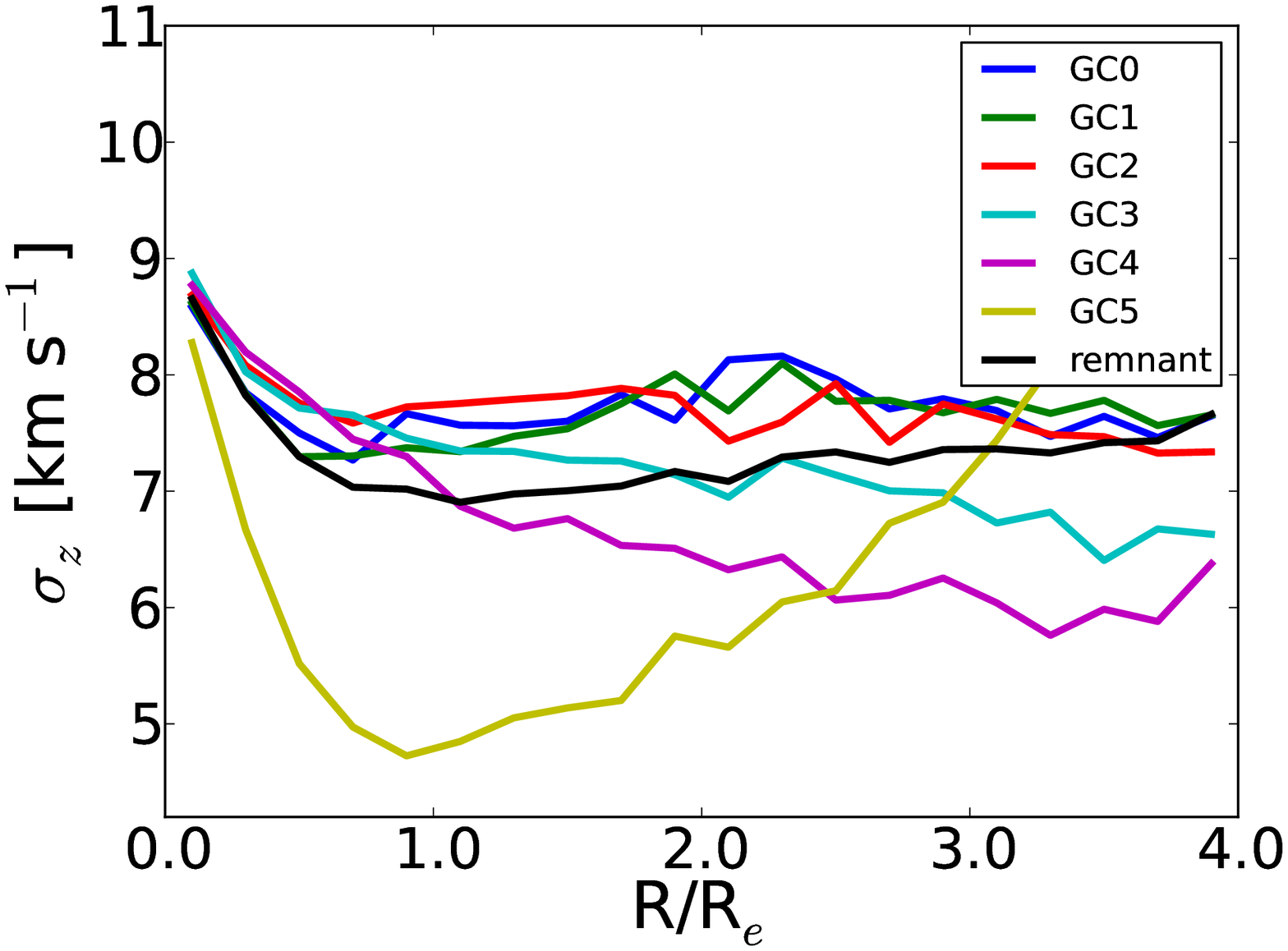}
\end{tabular}
\caption{Velocity dispersions for the final merger remnant in model 1
  at 1.56 Gyr and for the the stars originating in each progenitor GC.
  Left: $\sigma_R$, Middle: $\sigma_\phi$, Right: $\sigma_z$. }
\label{fig:disp}
\end{figure*}

\section{Results}
\label{sec:results}

\subsection{Properties of the the merger remnant in simulation 1}
\label{sec:sim1}

\subsubsection{Final properties}
\label{sec:final}

In this section we examine the properties of the merger remnant
formed in simulation 1 in terms of the stars which originated in each
progenitor GC. The aim will be to see if we can distinguish the stars
which originated in individual GCs as populations with distinct
distributions and kinematics within the merger remnant.

Table \ref{tab:evo} shows the mass and the half mass radius $R_e$ at
five times during the simulation. We measure the surface density
profile at each time and find the total mass and the radius enclosing
half this mass. It should be noted that there have been recent mergers
at $t=200$, 400 and 1560 Myr but not at 600 and 800 Myr which explains
why there is very little change in $R_e$ between the latter
times. Figure \ref{fig:sdens} shows a map of the merger remnant's
surface density within 4 $R_e$ after 1.56 Gyr. It has a mass of $\sim$
2.4$\times$10$^5$ M$_\odot$ and a half mass radius $R_e\sim$ 3.2 pc
with a mildly oblate shape.

Figure \ref{fig:profiles} shows the density profiles and Figure
\ref{fig:disp} shows the velocity dispersion profiles in cylindrical
coordinates for the merger remnant at the end of the simulation. We
separate the stars by their progenitor GC and labelled the GCs in the
order in which they merged, from GC0 to GC5. The contribution of each
GC is distinguished as a different coloured line. It is apparent that
it is difficult to identify stars visually from different progenitor
GCs on the basis of the density profile and velocity dispersion,
though GC5 looks distinguishable on the basis of its velocity
dispersion. The final density distribution is very similar for the
stars from all GCs. Even stars from the most recently merged GC5 have
a very similar density profile to the stars originating in the other
GCs. All velocity dispersions are similar for all the groups of stars
originating in each GC except for GC5. GC5 shows the most significant
deviations from the behaviour of all the other components; in
particular, the radial profiles of the velocity dispersion tensor are
systematically lower in the central to intermediate regions
($R<2R_e$), while they become comparable to the values of the other
components in the outer parts ($R>2R_e$). The differences in
dispersion between GCs 0, 1, 2 and 3 are $\sim$1 km s$^{-1}$ which
would be hard to detect. GC4 has a maximum difference of $\sim$2 km
s$^{-1}$, which may be detectable, and the maximum difference of GC5's
velocity dispersion from the mean is for $\sigma_\phi$ and is
$\sim$3.5 km s$^{-1}$ at $R\sim R_e$.

\subsubsection{Merger remnant evolution in simulation 1}
\label{sec:evo}

\begin{table}
  \begin{center}
    \begin{tabular}{lccc} 
      \hline
%      \hline#
      \multicolumn{1}{l}{Time} &
      \multicolumn{1}{c}{Mass} &
      \multicolumn{1}{c}{Half mass radius} & 
      \multicolumn{1}{c}{$N_{GC}$} \\
      \multicolumn{1}{l}{Myr} &
      \multicolumn{1}{c}{$\times10^5$ M$_\odot$} &
      \multicolumn{1}{c}{pc} &
      \multicolumn{1}{c}{} \\ 
\hline
\hline
200 & 1.22 & 1.25 & 3 \\
400 & 2.02 & 2.2 & 5 \\
600 & 2.03 & 2.25 & 5 \\
800 & 2.03 & 2.25 & 5 \\
1560 & 2.4 & 3.2 & 6 \\
\hline
    \end{tabular}
  \end{center}
  \caption[]{\label{tab:evo} Properties of the merger remnant in
    simulation 1 at several times. $N_{GC}$ is the number of globular
    clusters that have merged to that point.}
\end{table}

\begin{figure}
\centering
\begin{tabular}{c}
\includegraphics[width=\hsize,angle=0]{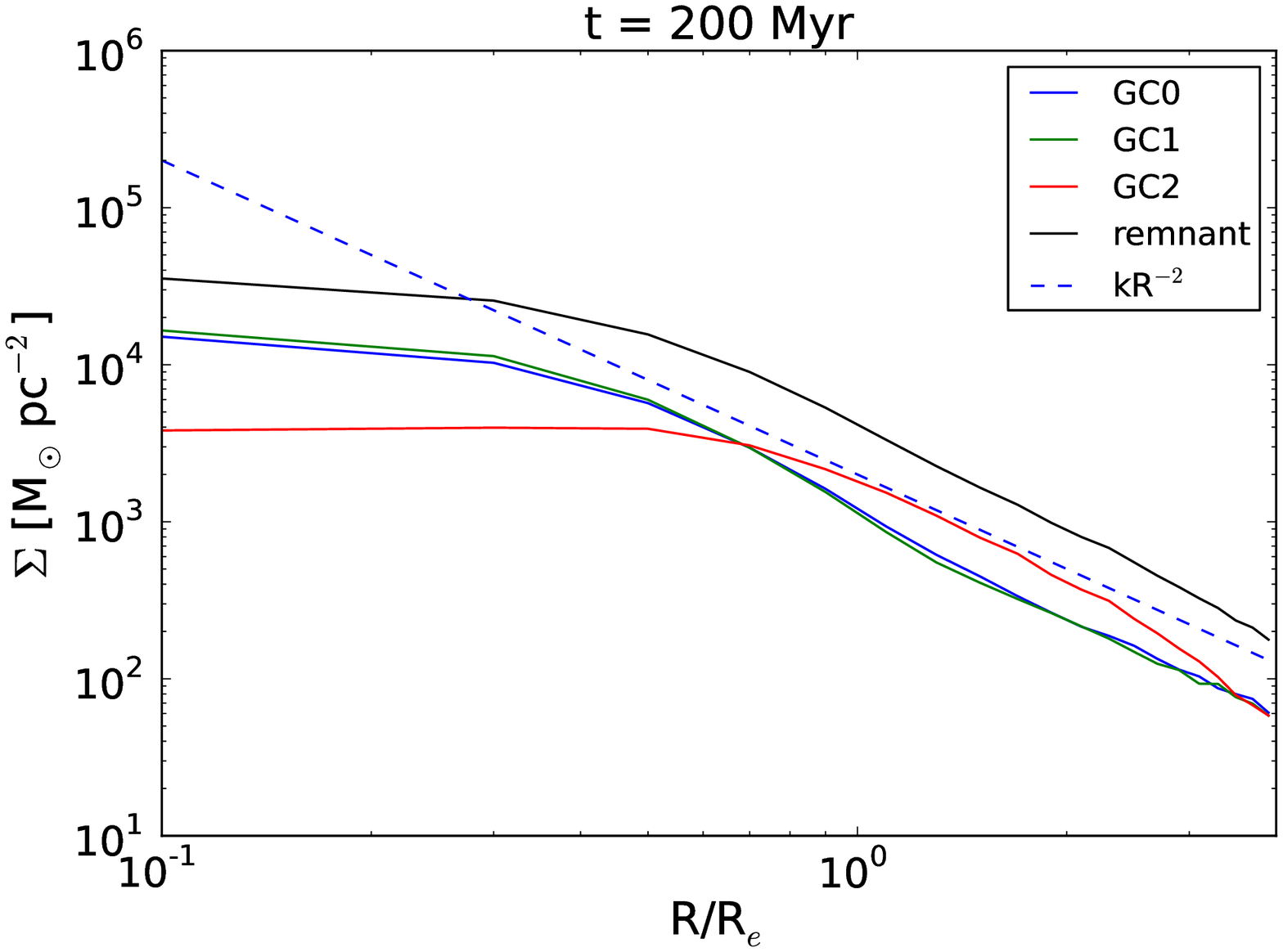}  \\
\includegraphics[width=\hsize,angle=0]{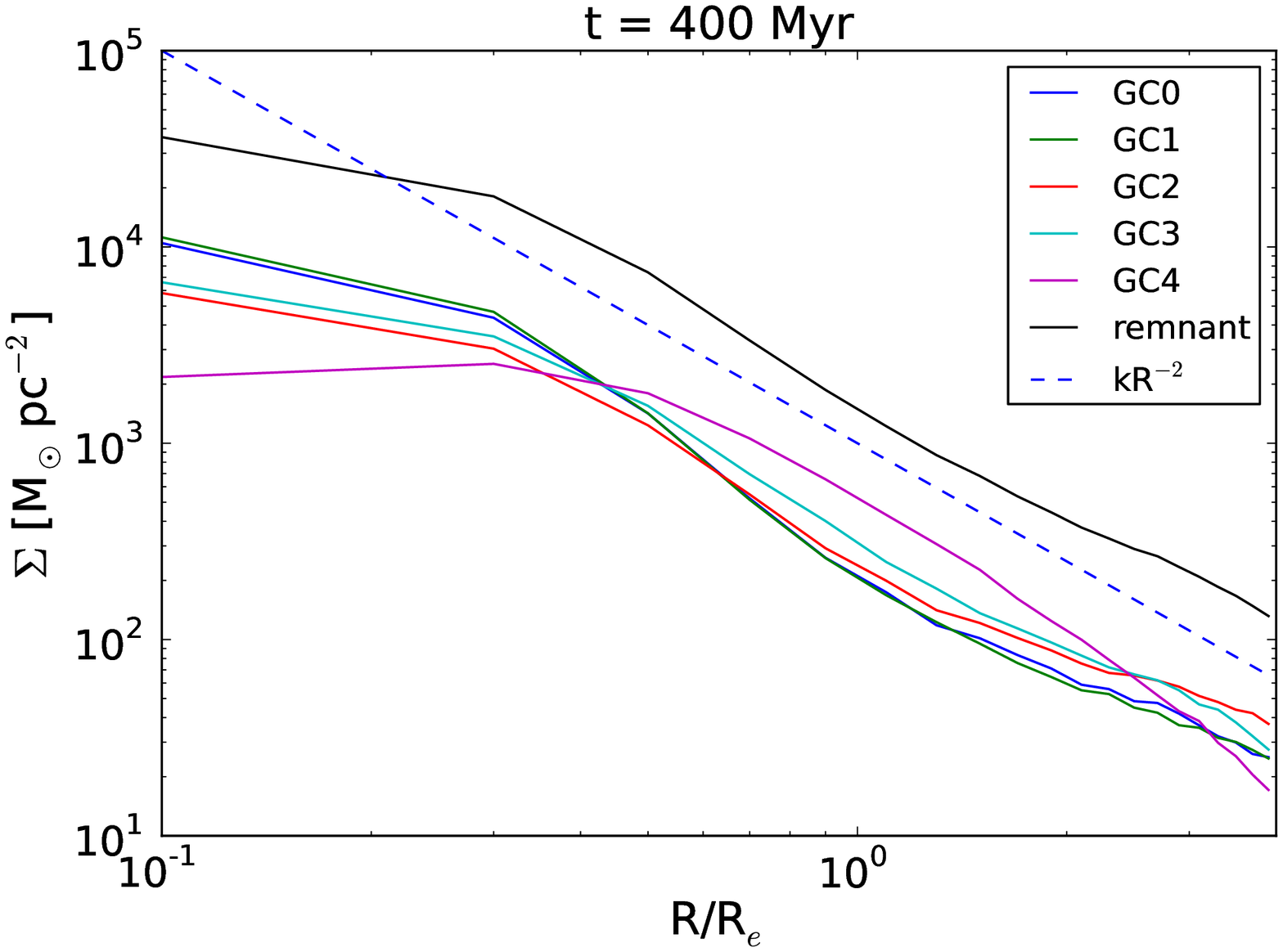} \\
\end{tabular}
\caption{ The density profile of the stars from individual GCs soon
  after a merger has occurred in model 1. Top: After 200 Myr the first
  three GCs have merged with GC2 being the most recent. Bottom: After
  400 Myr GC3 and GC4 have now merged with GC4 being the most
  recent. For comparison the dashed (blue) lines show a power law
  relation with an exponent of -2. }
\label{fig:densevo}
\end{figure}

We now consider whether it is possible to distinguish different
progenitor populations at earlier times in the simulation. Figure
\ref{fig:densevo} shows the merger remnant density profile at 200 Myr
and at 400 Myr, times at which GCs have recently merged into the
remnant. When GC0, GC1 and GC2 have merged the merger remnant has a
mass of 1.22$\times$10$^5$ M$_\odot$ and a half mass-radius of $\sim$
1.25 pc. After GC3 and GC4 have also merged the remnant's mass has
increased to 2.02$\times$10$^5$ M$_\odot$ and the half-mass radius to
$\sim$ 2.2 pc. There is a noticeably larger difference in the density
profiles of the most recently merged GCs, having a flatter central
density. By the time 5 GCs have merged the density profile for GC2,
which was quite different from the rest when just 3 GCs had merged,
has become close to the typical density profile of the other GCs.

\begin{figure*}
\centering
\begin{tabular}{c}
\includegraphics[width=0.33\hsize,angle=0]{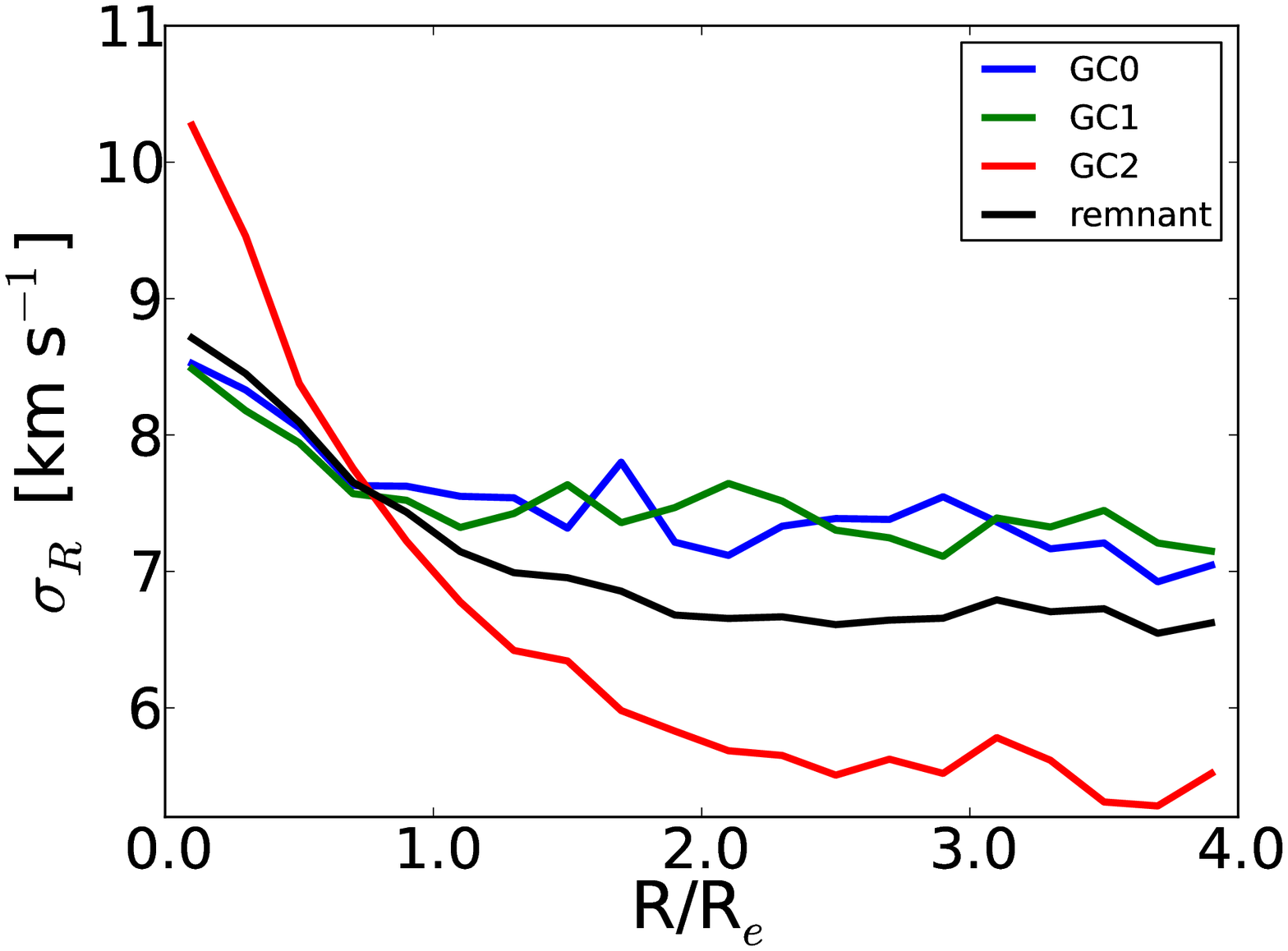}
\includegraphics[width=0.33\hsize,angle=0]{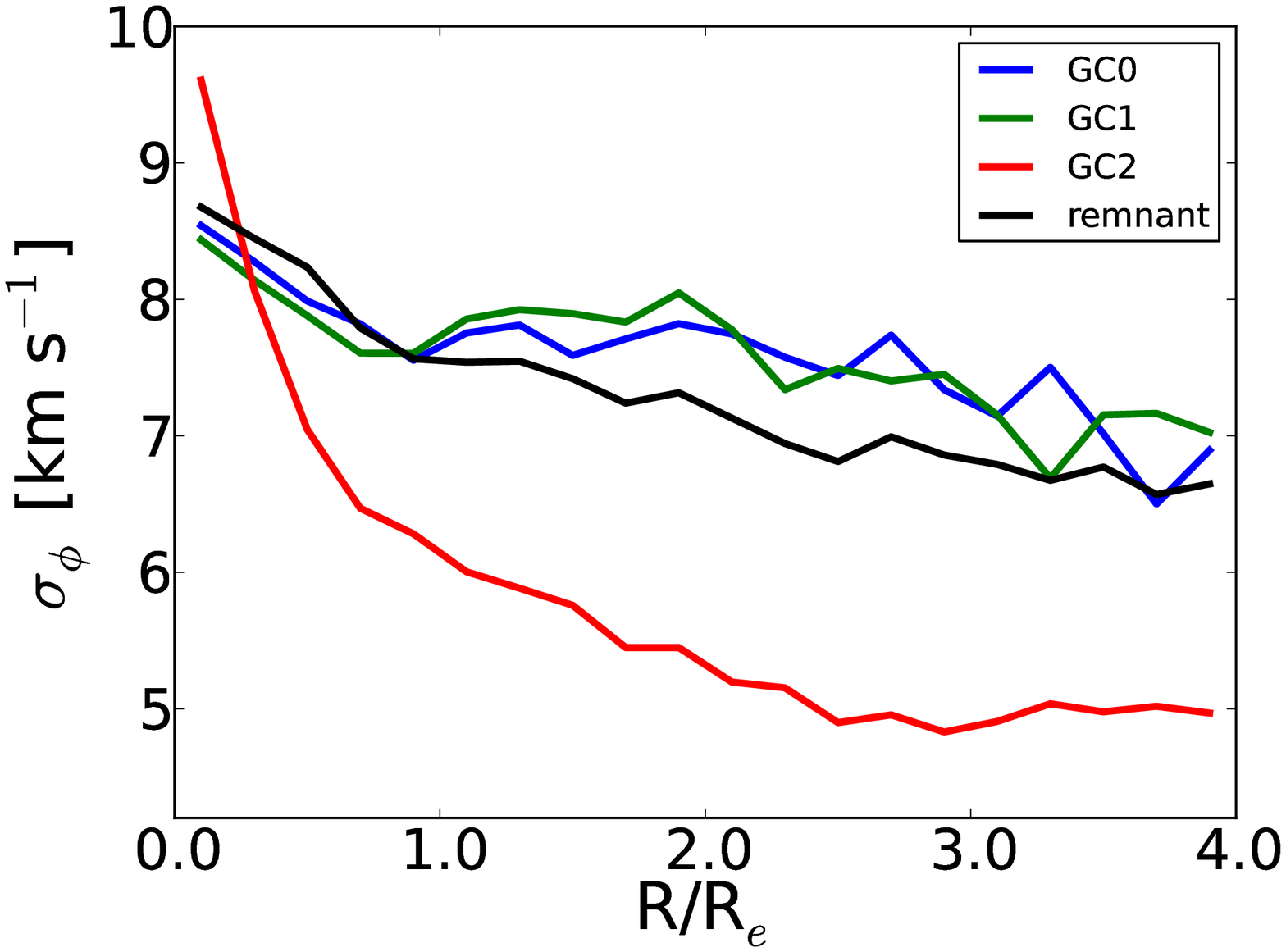}
\includegraphics[width=0.33\hsize,angle=0]{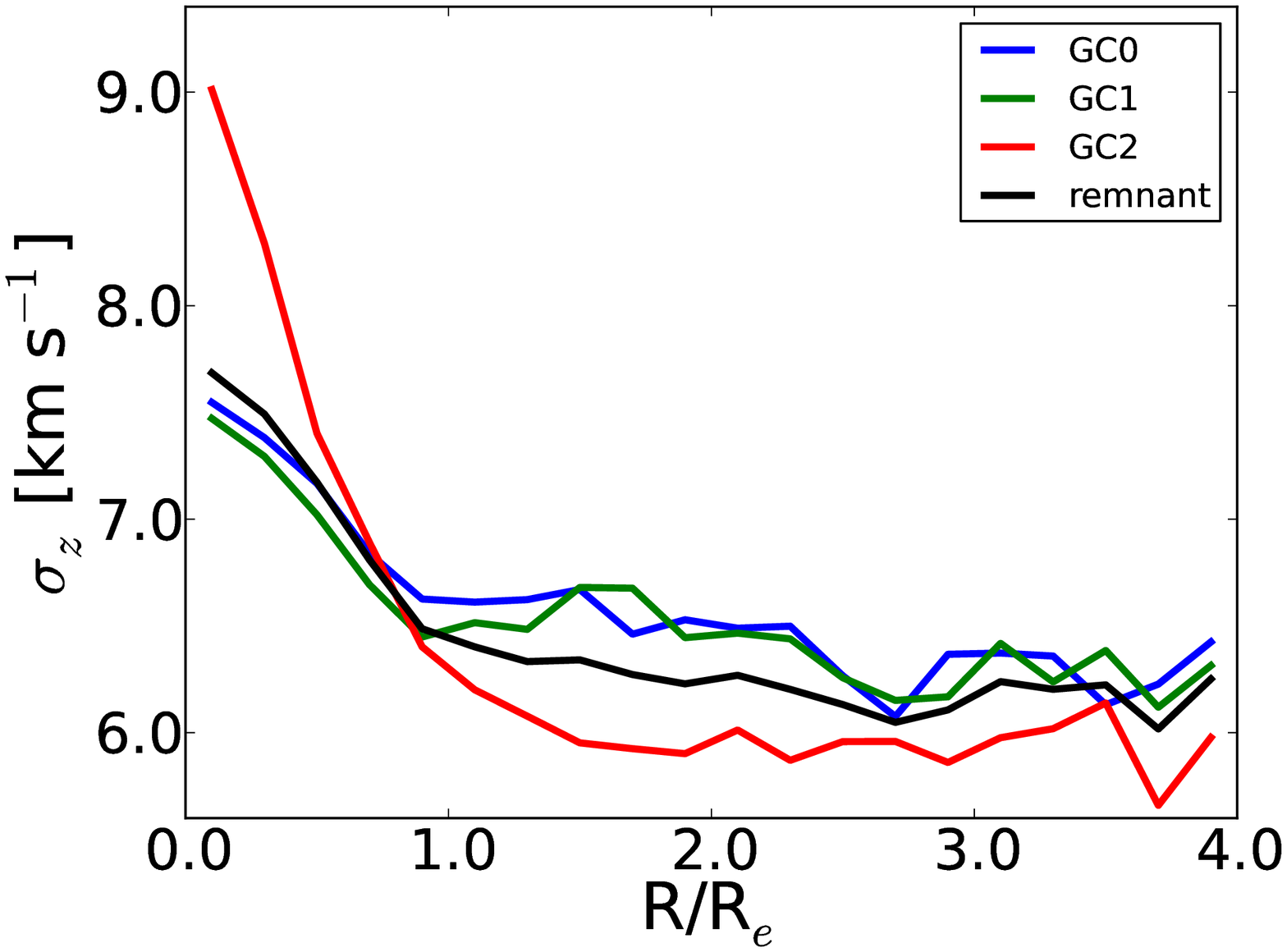} \\
\includegraphics[width=0.33\hsize,angle=0]{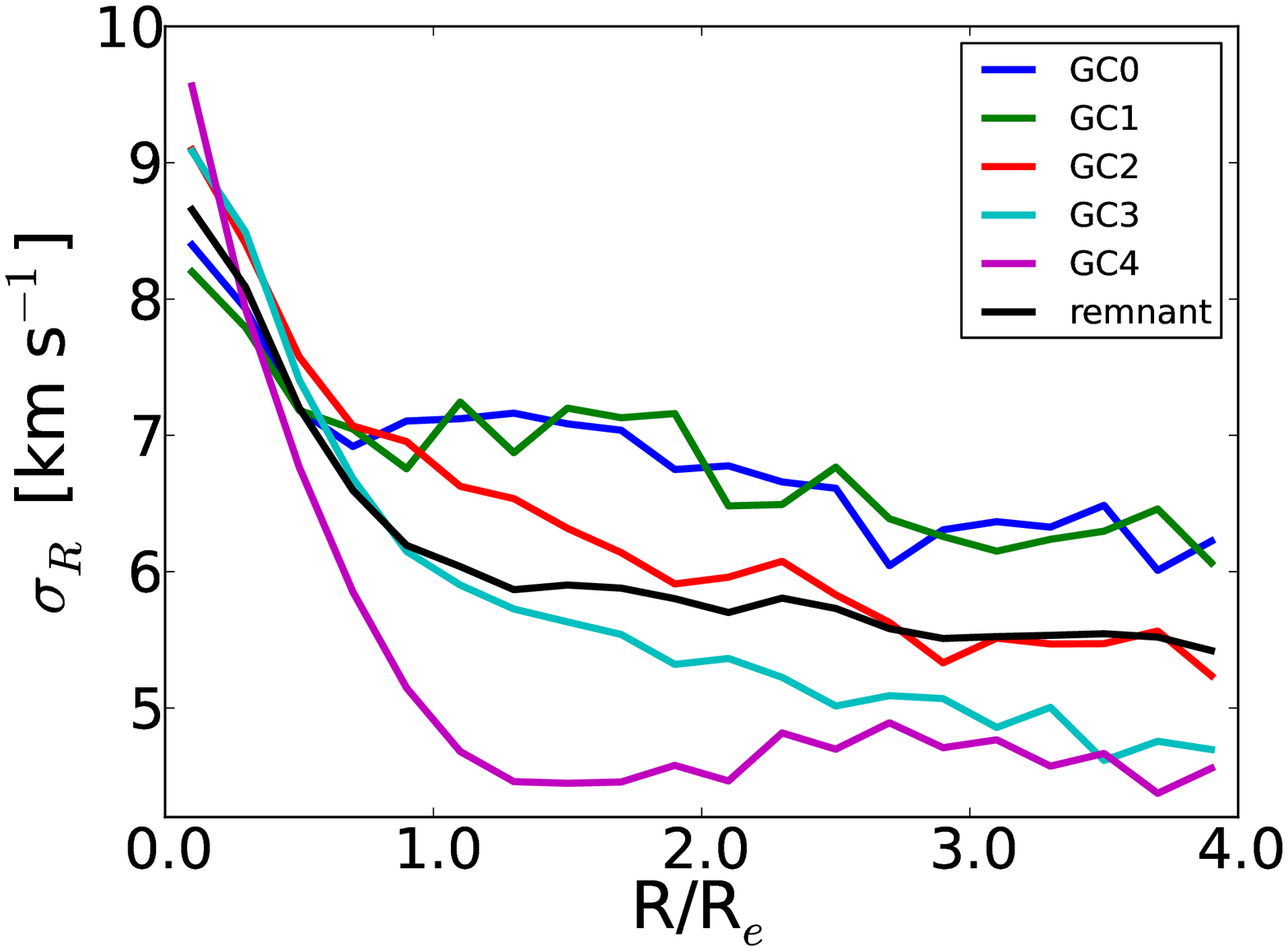} 
\includegraphics[width=0.33\hsize,angle=0]{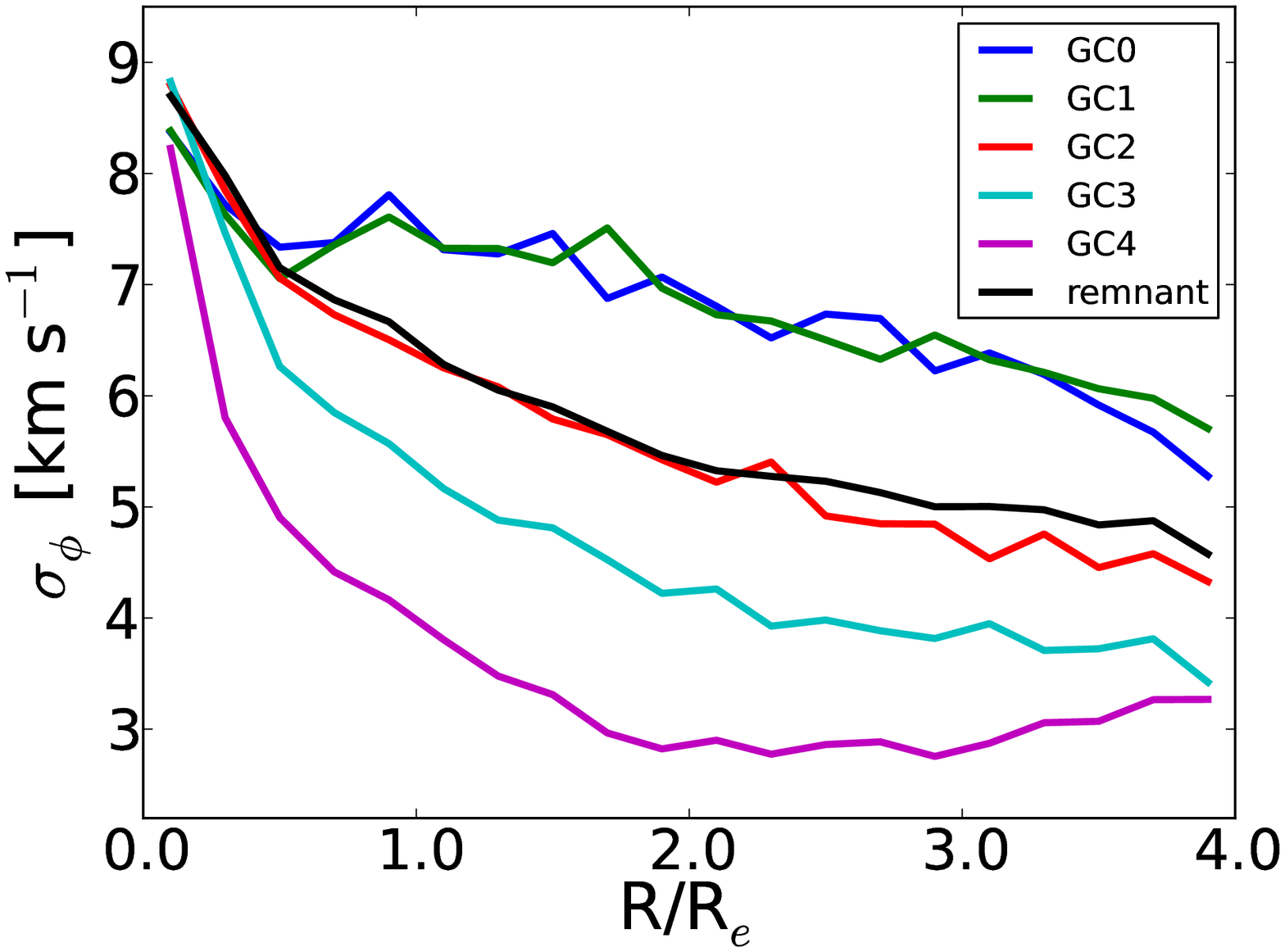} 
\includegraphics[width=0.33\hsize,angle=0]{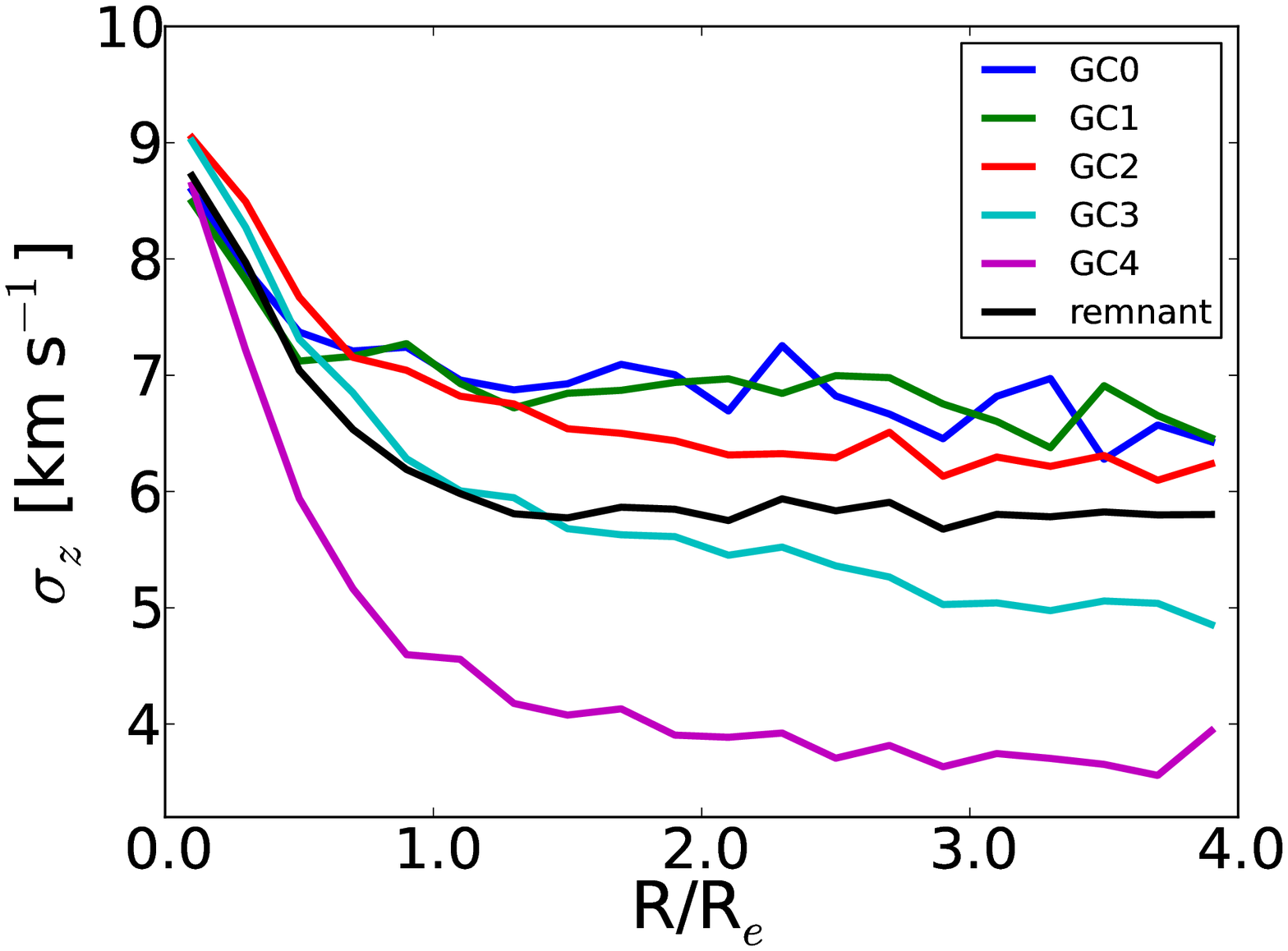} 
\end{tabular}
\caption{ The velocity dispersion profile of the stars originating
  from individual GCs after 3 GCs have merged (top) and after 5 GCs
  have merged (bottom) in model 1.  $\sigma_R$ left, $\sigma_\phi$
  middle and $\sigma_z$ right. }
\label{fig:sigevo200}
\end{figure*}

Figure \ref{fig:sigevo200} shows the merger remnant velocity
dispersion profiles within $4R_e$\footnote{We choose this radius
  because \citet{Kucinskas2014} include stars out to 4$R_e$ in their
  work on 47 Tuc.} after 3 GCs have merged and after 5 GCs have
merged.  These plots generally show the velocity dispersion profiles
are similar in shape and magnitude for all stars originating in each
GC. After 3 GCs have merged the most recently merged GC2 shows the
greatest deviation from the mean at $\lesssim$ 3 km s$^{-1}$. After
the merger of two additional GCs, it is again the most recently merged
GC4 which shows lower velocity dispersions with a maximum difference
of $\sim$ 2 km s$^{-1}$ from the mean. GC3 and GC4 are quite
distinguishable here, indicating that mixing is less complete.

\begin{figure}
\centering
\begin{tabular}{c}
\includegraphics[width=1.0\hsize,angle=0]{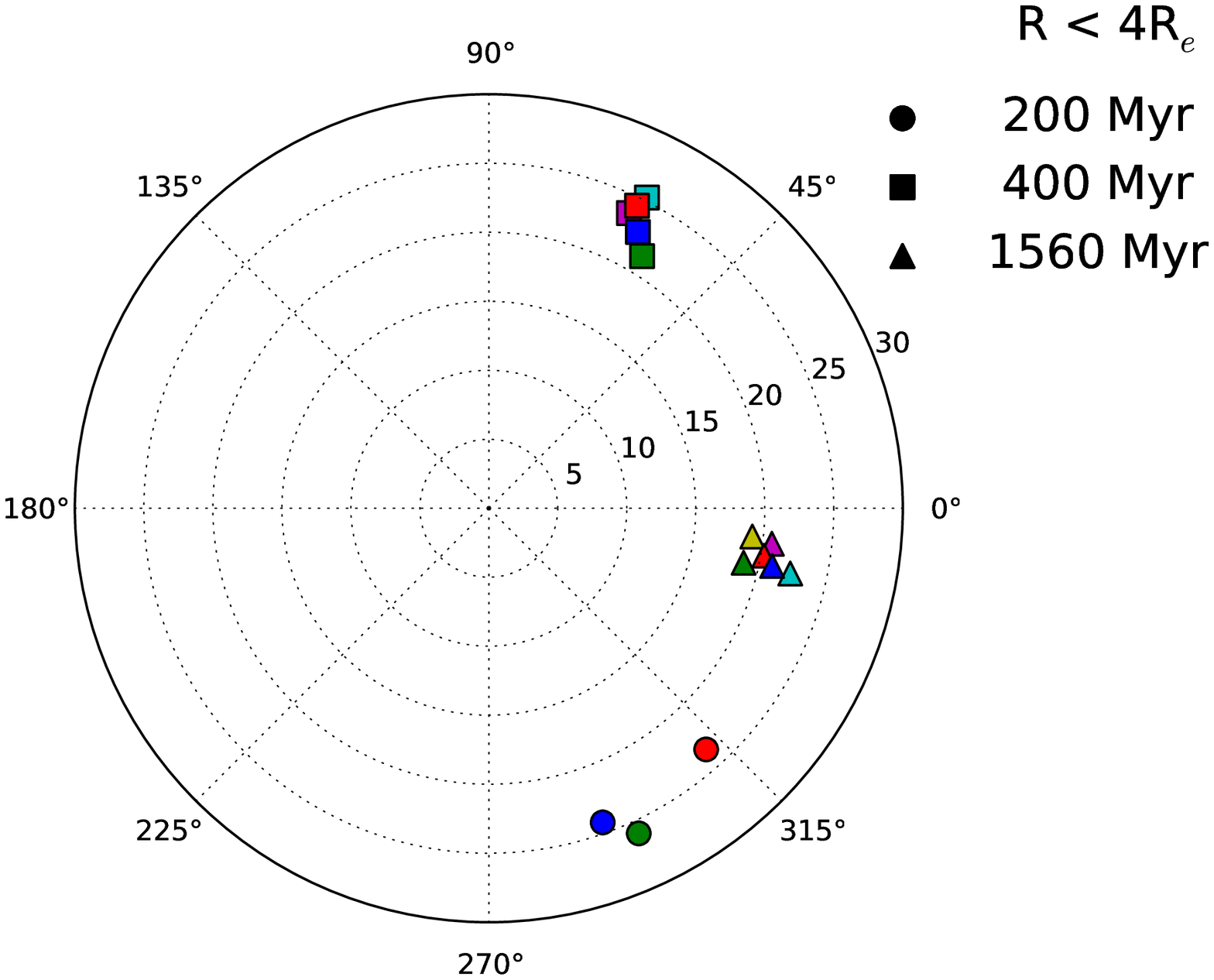} 
\end{tabular}
\caption{Briggs figure for the stars from different GCs in model
  1. The direction of the angular momentum vector for each GC is
  indicated by the position of the symbol on the plot. Circles are
  values after 3 GCs have merged, squares after 5 GCs have merged and
  triangles after 6 GCs have merged. The plot shows the values for
  stars inside of 4$R_e$. We use the same colour code as in other
  figures: GC0 blue, GC1 green, GC2 red, GC3 cyan, GC4 magenta, GC5
  yellow. }
\label{fig:angmom}
\end{figure}

Figure \ref{fig:angmom} shows the evolution of the angular momentum of
the merger remnant. We use Briggs figures \citep{Briggs1990} which are
2-D polar coordinate representations of vector directions where the
two spherical angle coordinates relative to a fixed reference frame,
$\theta$ and $\phi$, are plotted as the radial and angle coordinates,
respectively, on a 2-D polar plot. The plot shows a Briggs figure for
the stars from different GCs. It can be seen that the angular momentum
vectors are well aligned to better than 10$^{\circ}$ for stars inside
of 4$R_e$. The one exception is GC2 after 3 GCs have merged which is
misaligned by $\sim$20$^\circ$ from GC0 and GC1 probably due to its
recent merger.

\begin{figure*}
\centering
\begin{tabular}{c}
\includegraphics[width=0.33\hsize,angle=0]{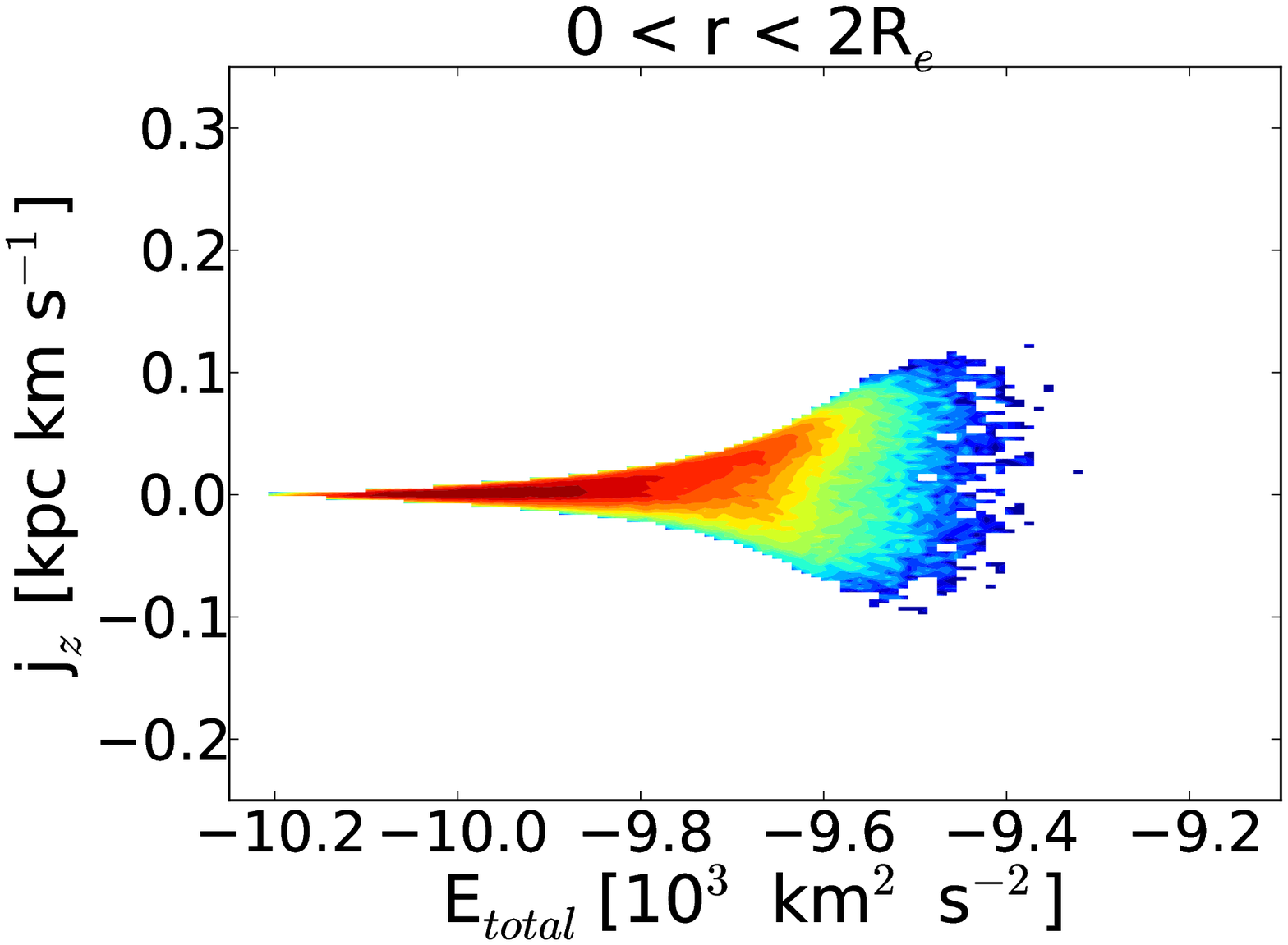} 
\includegraphics[width=0.33\hsize,angle=0]{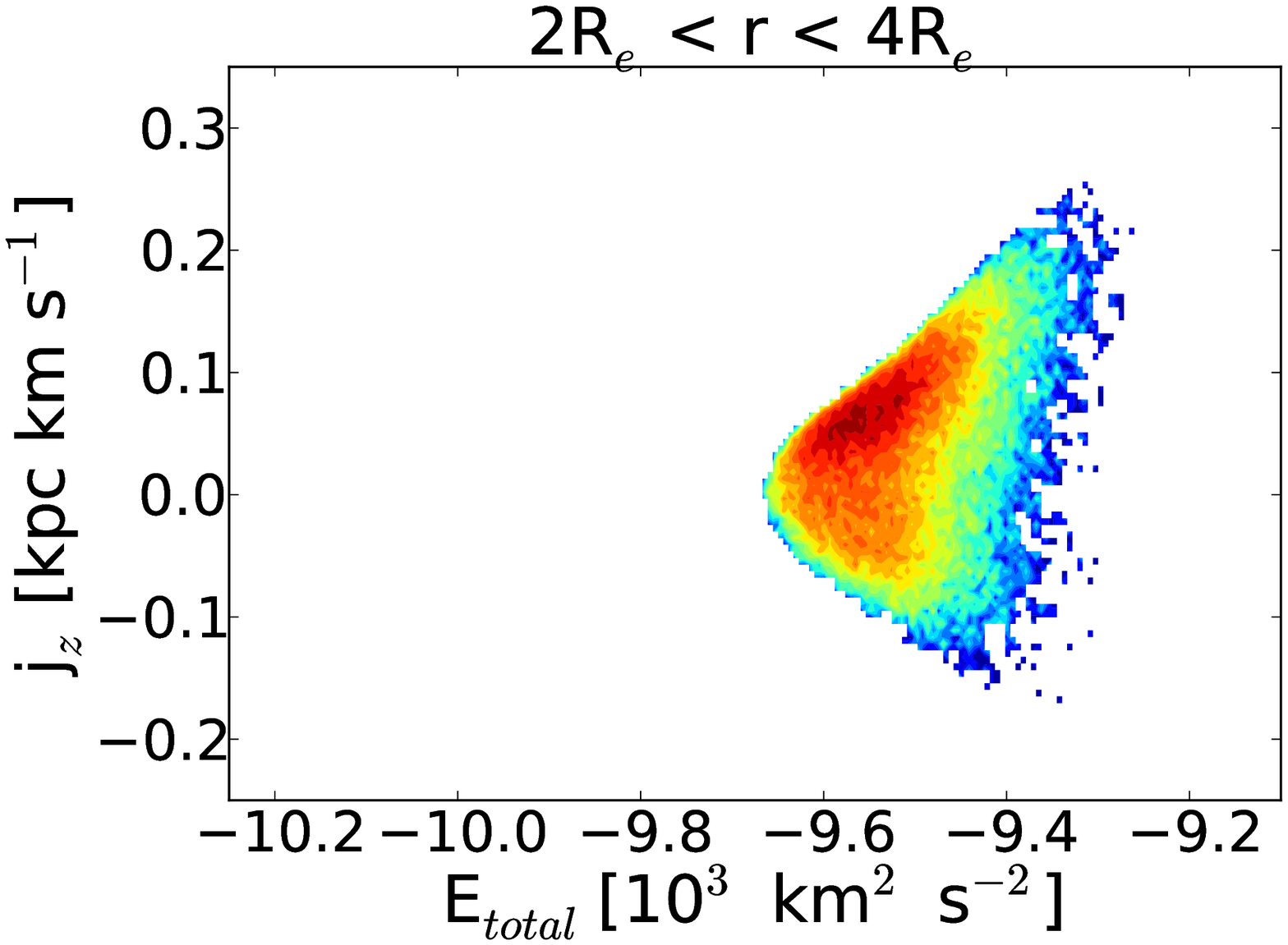}
\includegraphics[width=0.33\hsize,angle=0]{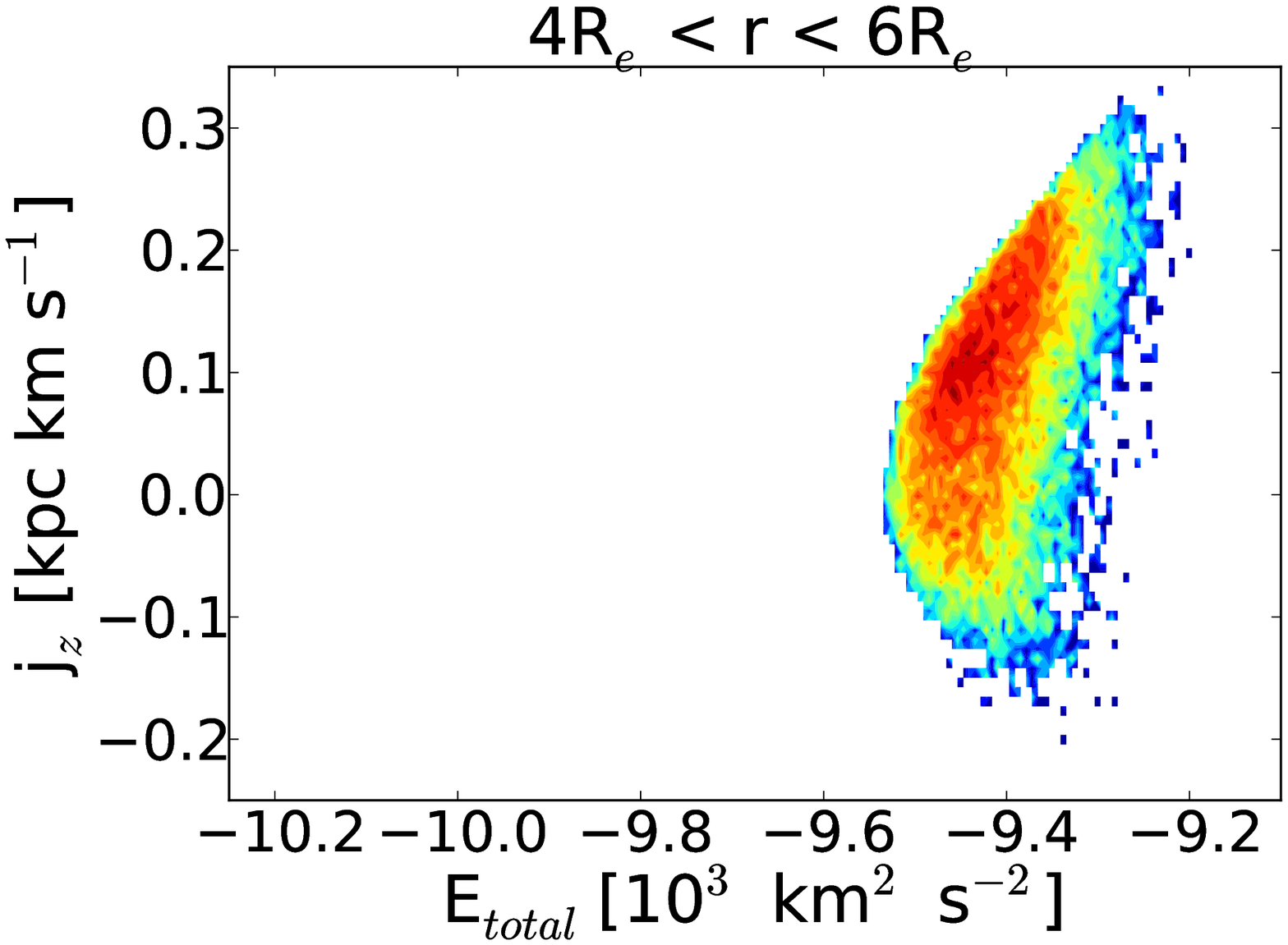} \\
\includegraphics[width=0.33\hsize,angle=0]{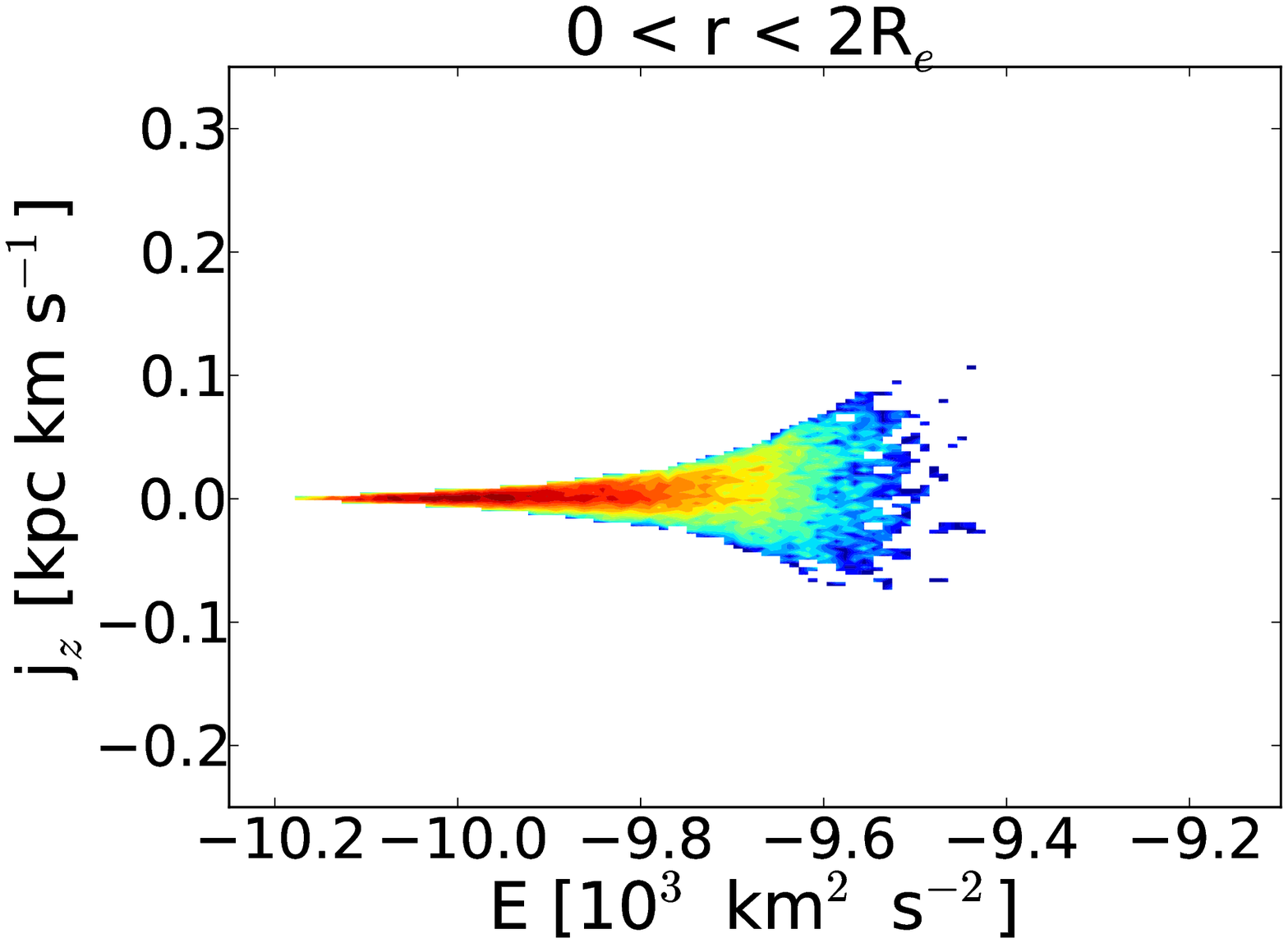} 
\includegraphics[width=0.33\hsize,angle=0]{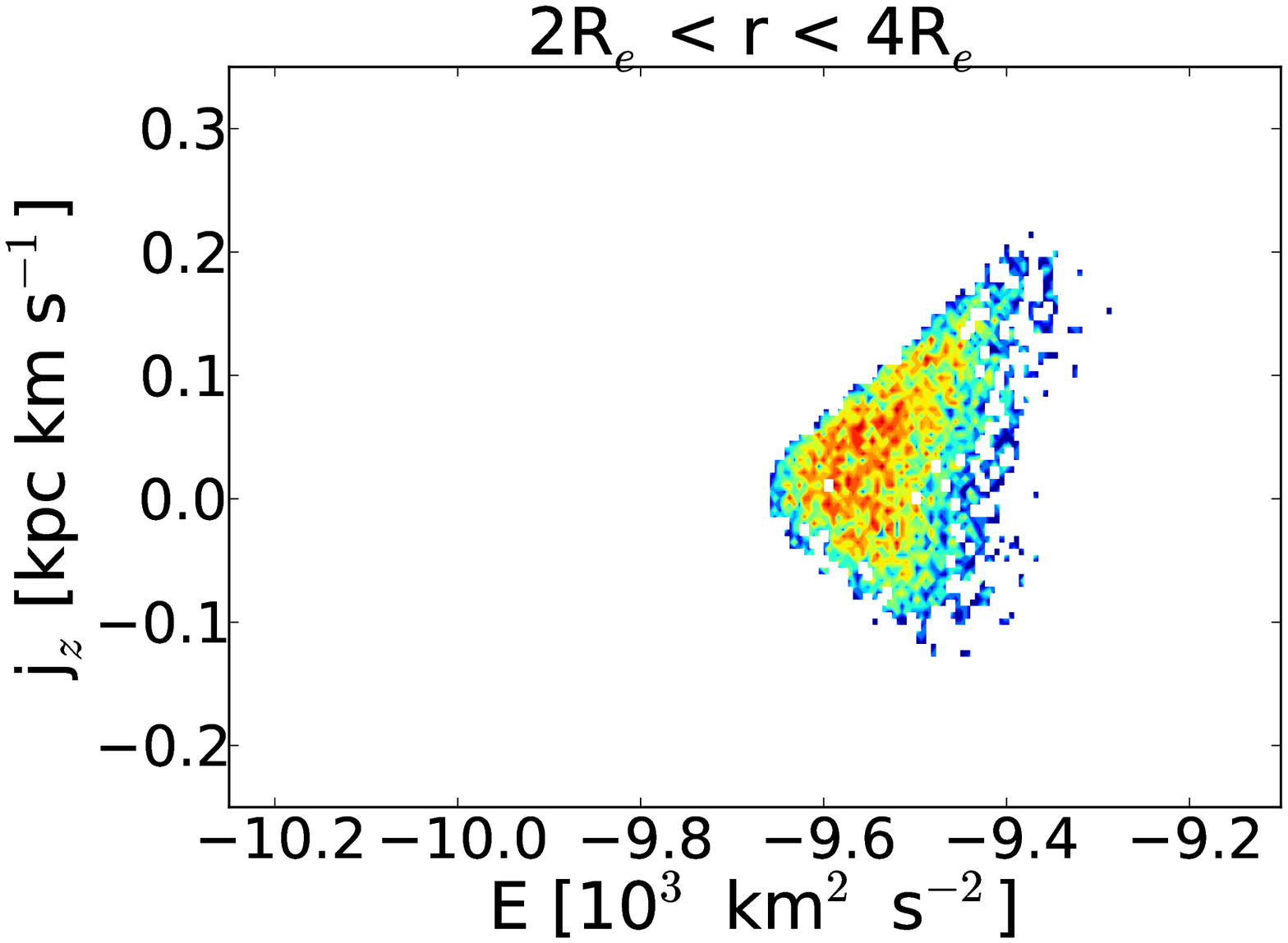}
\includegraphics[width=0.33\hsize,angle=0]{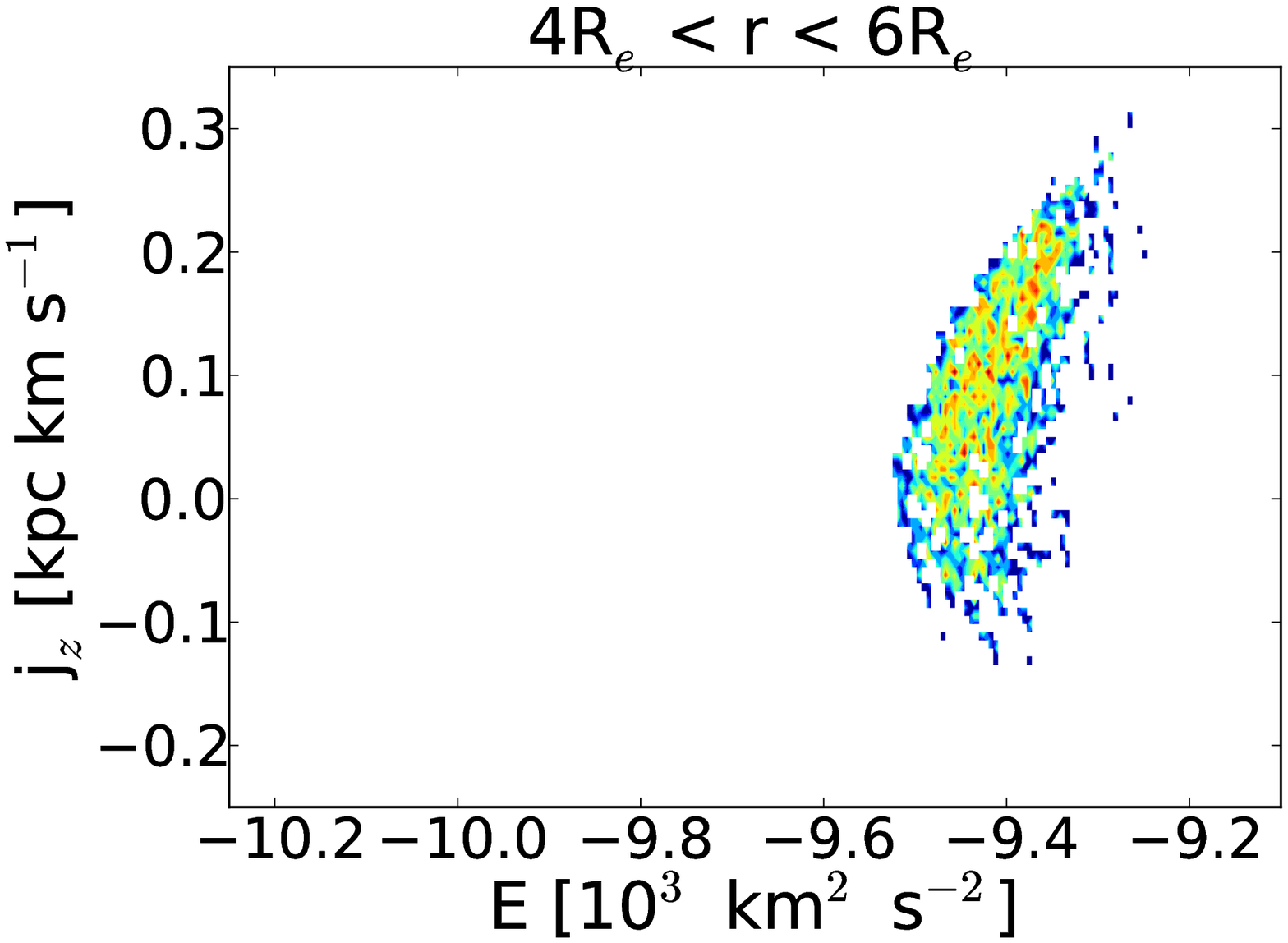} \\ 
\includegraphics[width=0.33\hsize,angle=0]{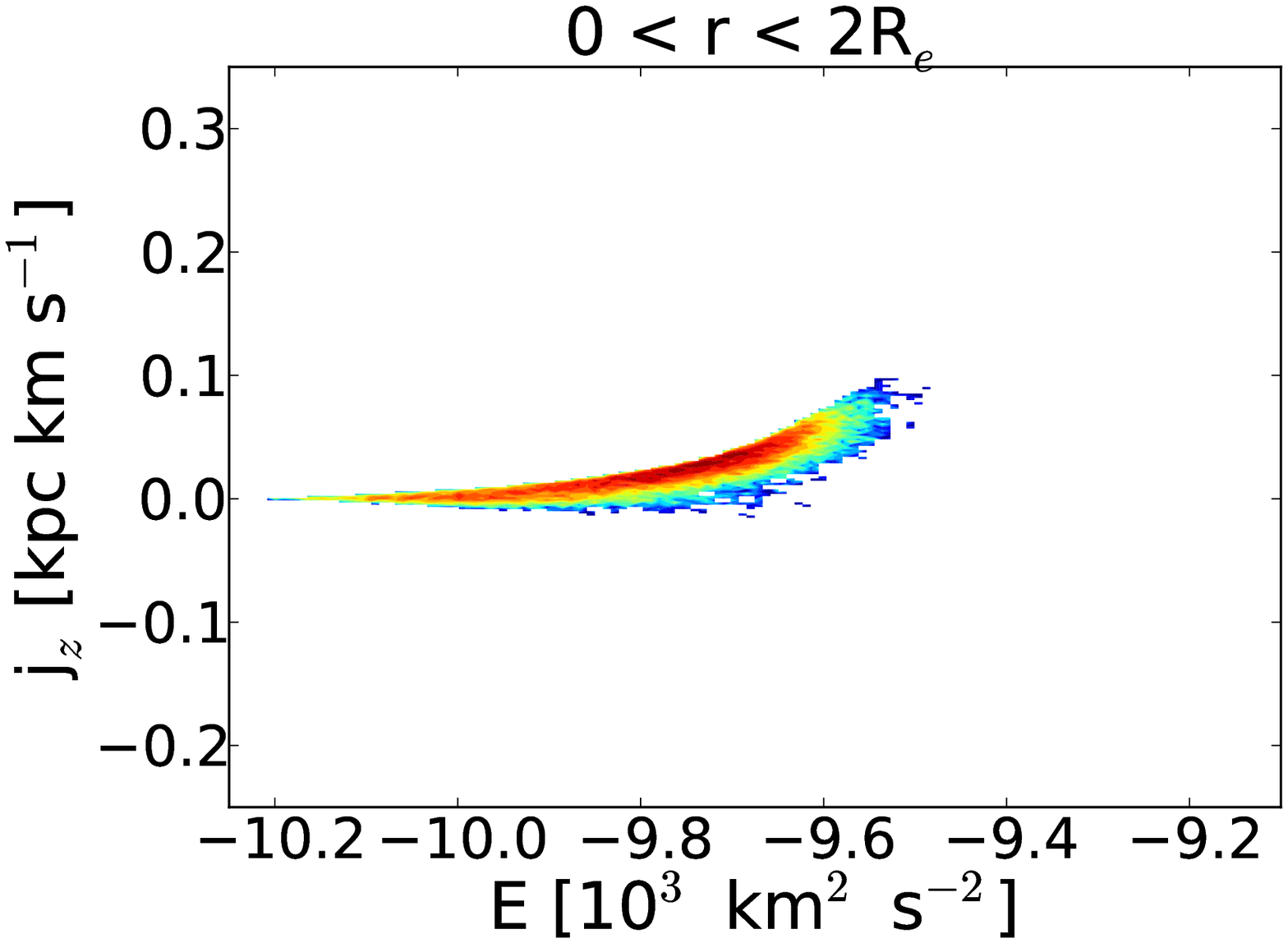} 
\includegraphics[width=0.33\hsize,angle=0]{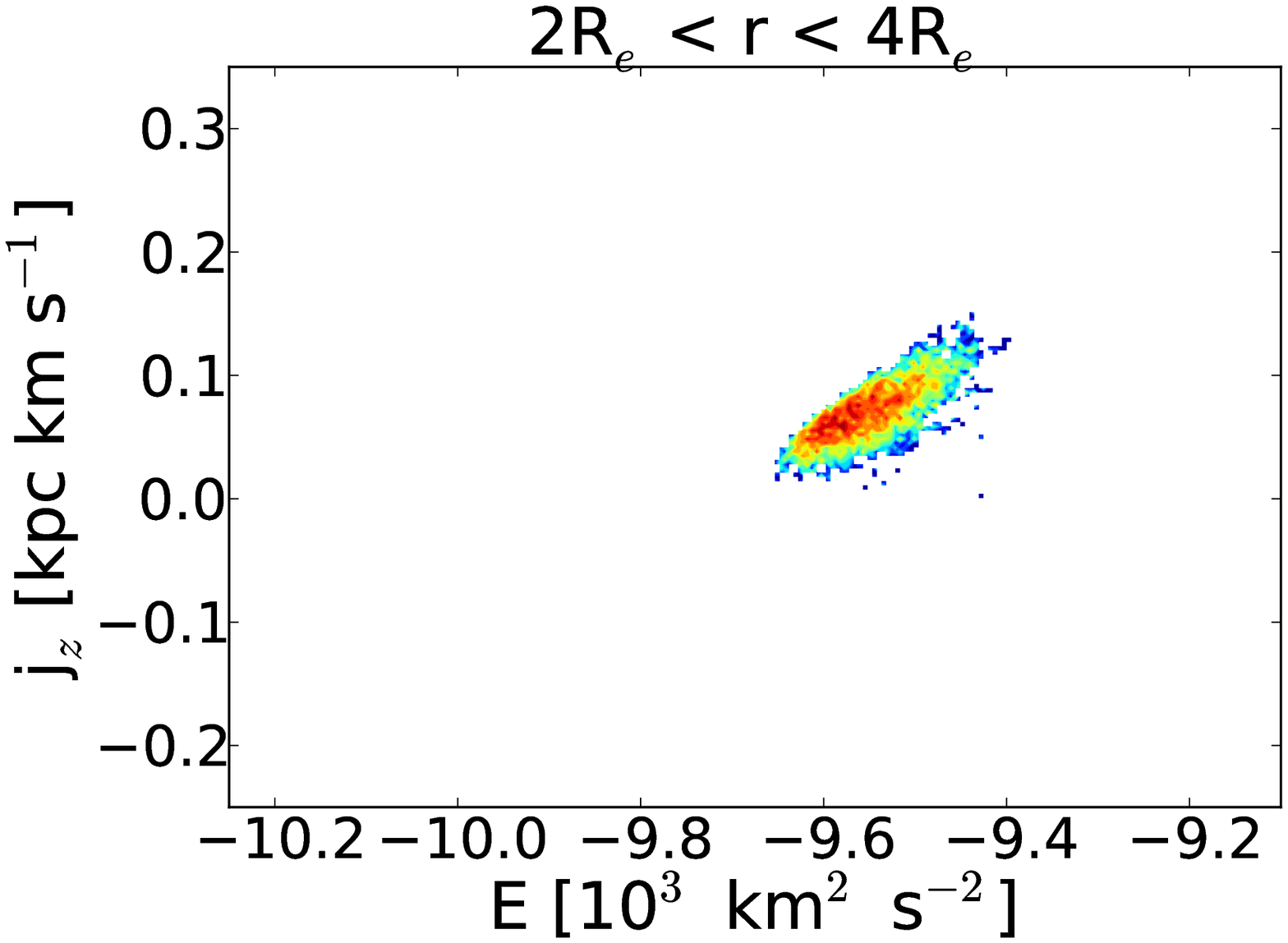}
\includegraphics[width=0.33\hsize,angle=0]{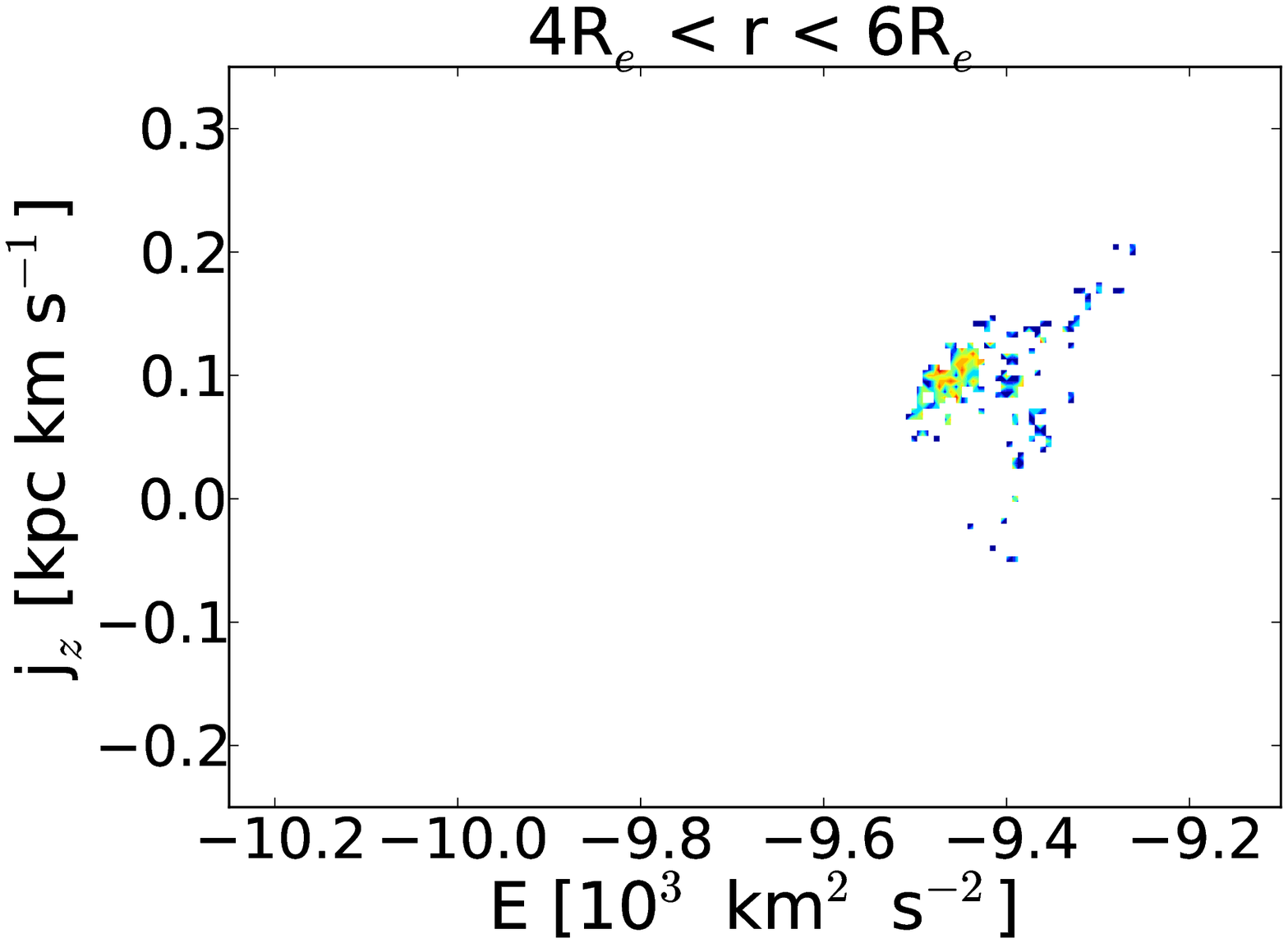}
\end{tabular}
\caption{ Plots of the vertical angular momentum j$_z$ versus energy
  for stars in three radial ranges at 1.56 Gyr for simulation 1. The
  left column shows stars between 0 and 2 $R_e$, the middle between 2
  and 4 $R_e$ and the right hand one shows stars between 4 and 6
  $R_e$. We show plots for all stars (top), for the stars originating
  in GC0 (middle) and GC5 (bottom). Stellar radii are measured in
  spherical coordinates. The number of stars in each bin in the plot
  is colour coded with the maximum shown in dark red and numbers
  decreasing moving through yellow to blue. }
\label{fig:jzE}
\end{figure*}

The top row in Figure \ref{fig:jzE} plots the angular momentum
perpendicular to the plane of overall rotation, $j_z$, versus energy
for stars in three spherical radial ranges at 1.56 Gyr. $j_z$ is
measured once the merger remnant has been centred and its angular
momentum vector aligned with the $z$ axis, resulting in any flattening
of the stars into a disc lying in the $x-y$ plane (c.f. Figure
\ref{fig:sdens}). Stars in the inner radial bin have lower $j_z$,
increasing outwards. There is no evidence of groupings of stars with
distinct angular momentum signatures in these plots which would be
indicative of separate populations. If we look at stars originating in
different star clusters the only one that shows a significant
difference in this plot is GC5, i.e. the last one to merge. The middle
and bottom rows of Figure \ref{fig:jzE} shows a comparison of $j_z$
versus energy for GC0 and GC5 in the same 3 radial bins. GC5 shows
significantly more stars with positive angular momentum in the inner
two radial bins however there is still significant overlap in stellar
distribution making this bias hard to observe.

\begin{figure}
\includegraphics[width=\hsize,angle=0]{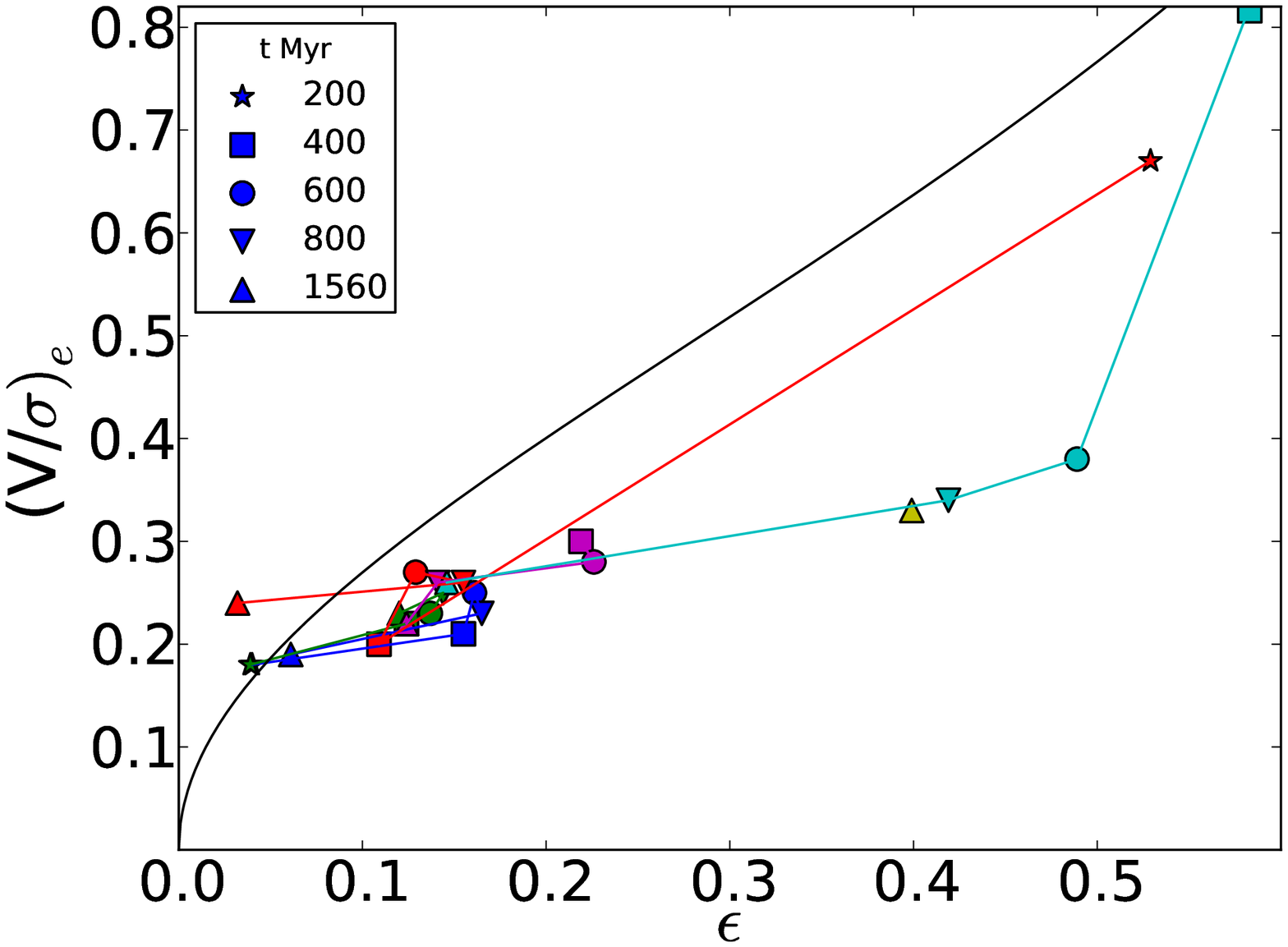}
\caption{Evolution of the merger remnant of model 1 on the
  ($V/\sigma$,$\epsilon$) diagram of \citet{Binney2005} measured as in
  \citet{Hartmann2011} within 2 R$_e$. The black line shows the
  location of edge-on oblate isotropic models. Each GC is indicated by
  a different colour. GC0 is blue, GC1 is green, GC2 is red, GC3 is magenta, GC4
  is cyan and GC5 is yellow. The different symbols show the values at
  different times. Note that GC3 and GC4 have not merged until 400 Myr
  and have no data for 200 Myr, and GC5 has just merged at the end of
  the simulation and shows no evolution. }
\label{fig:vosevo}
\end{figure}

Figure \ref{fig:vosevo} shows the evolution of the stars from each GC
on the ($V/\sigma$,$\epsilon$) diagram of \citet{Binney2005}. It has
been measured as described in \citet{Hartmann2011} within 2 R$_e$. The
radial profiles of the observables depicted in the diagram have been
calculated along a line of sight corresponding to a direction
perpendicular to the orientation of the angular momentum vector (i.e.,
edge-on). The region inside of 2 R$_e$ is divided into bins of equal
size and $V/\sigma$ is calculated as:

\begin{equation}
\left(\frac{V}{\sigma}\right)_e\equiv\frac{\langle{V}^2\rangle}{\langle\sigma^2\rangle}=\frac{\Sigma^N_{n=1}F_nV^2_n}{\Sigma^N_{n=1}F_n\sigma^2_n}
\end{equation}
and the ellipticity $\epsilon$ is found from:
\begin{equation}
(1-\epsilon)^2=q^2=\frac{\langle{y}^2\rangle}{\langle{x}^2\rangle}=\frac{\Sigma^N_{n=1}F_ny^2_n}{\Sigma^N_{n=1}F_nx^2_n}
\end{equation}
where $F_n$ is the mass in the $n$th bin and $V_n$ and $\sigma_n$ are
the corresponding mean velocity and velocity dispersion in that
bin. It can be seen that GCs when first merged can have large
differences in their location on the diagram, for instance GC2, GC4
and GC5. However as the remnant evolves, stars originating in
different GCs move closer together. Even though GC2 and GC4 are
initially located in very different parts of the diagram compared to
the other mass components, they subsequently evolve towards the same
region, corresponding to moderate flattening and mild rotation. In
particular, the evolution in the diagram of GC2 (from a condition of
high flattening and significant rotation) seems to be associated with
merger events of GC3 and GC4. The component GC4 evolves in a similar
way, but on a longer timescale, and it becomes comparable to the other
components only after the completion of all six merger events.

\subsubsection{Kolmogorov-Smirnov statistics}
\label{sec:KSsim1}

\begin{figure*}
\centering
\begin{tabular}{c}
\includegraphics[width=0.33\hsize,angle=0]{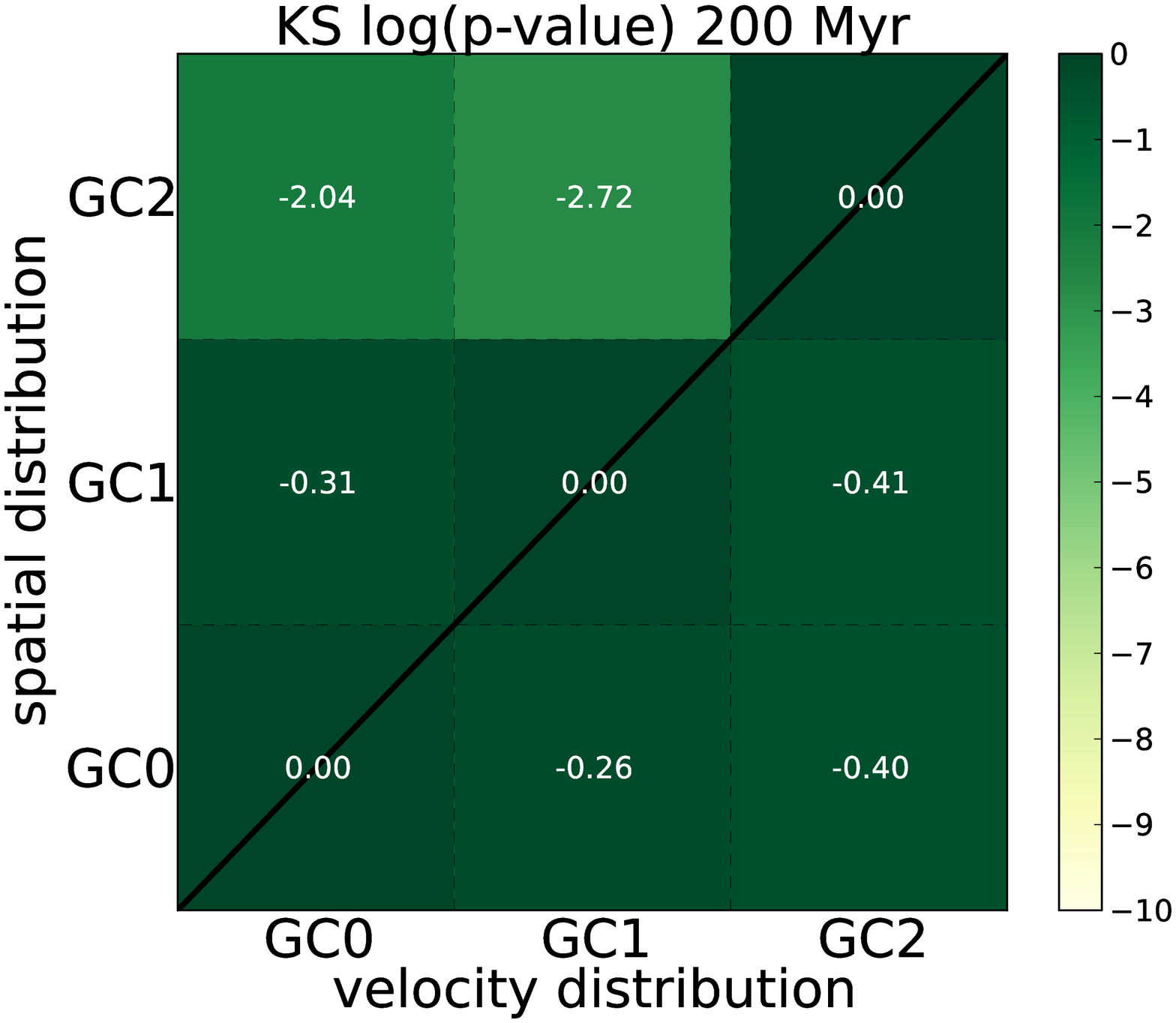}
\includegraphics[width=0.33\hsize,angle=0]{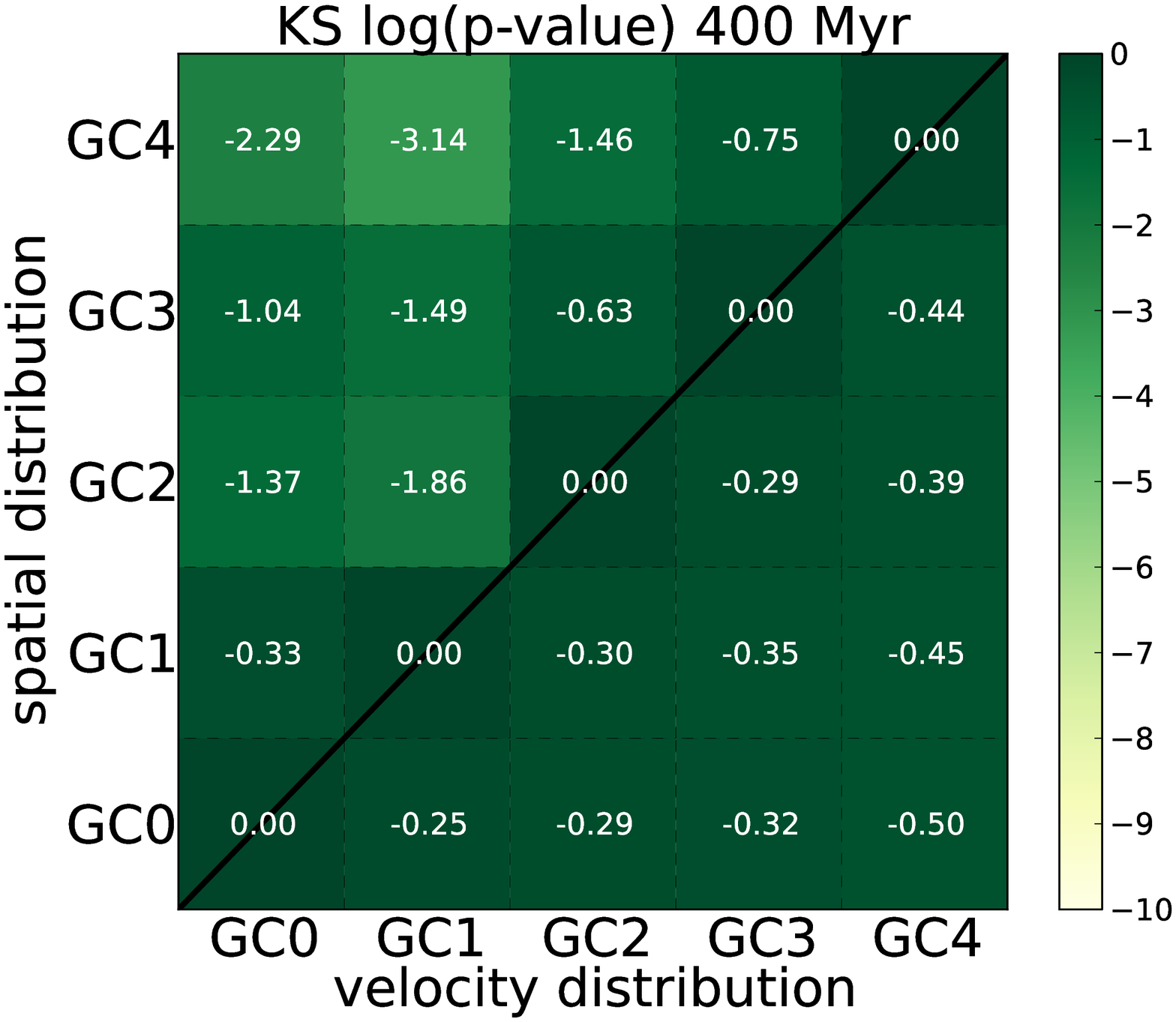}
\includegraphics[width=0.33\hsize,angle=0]{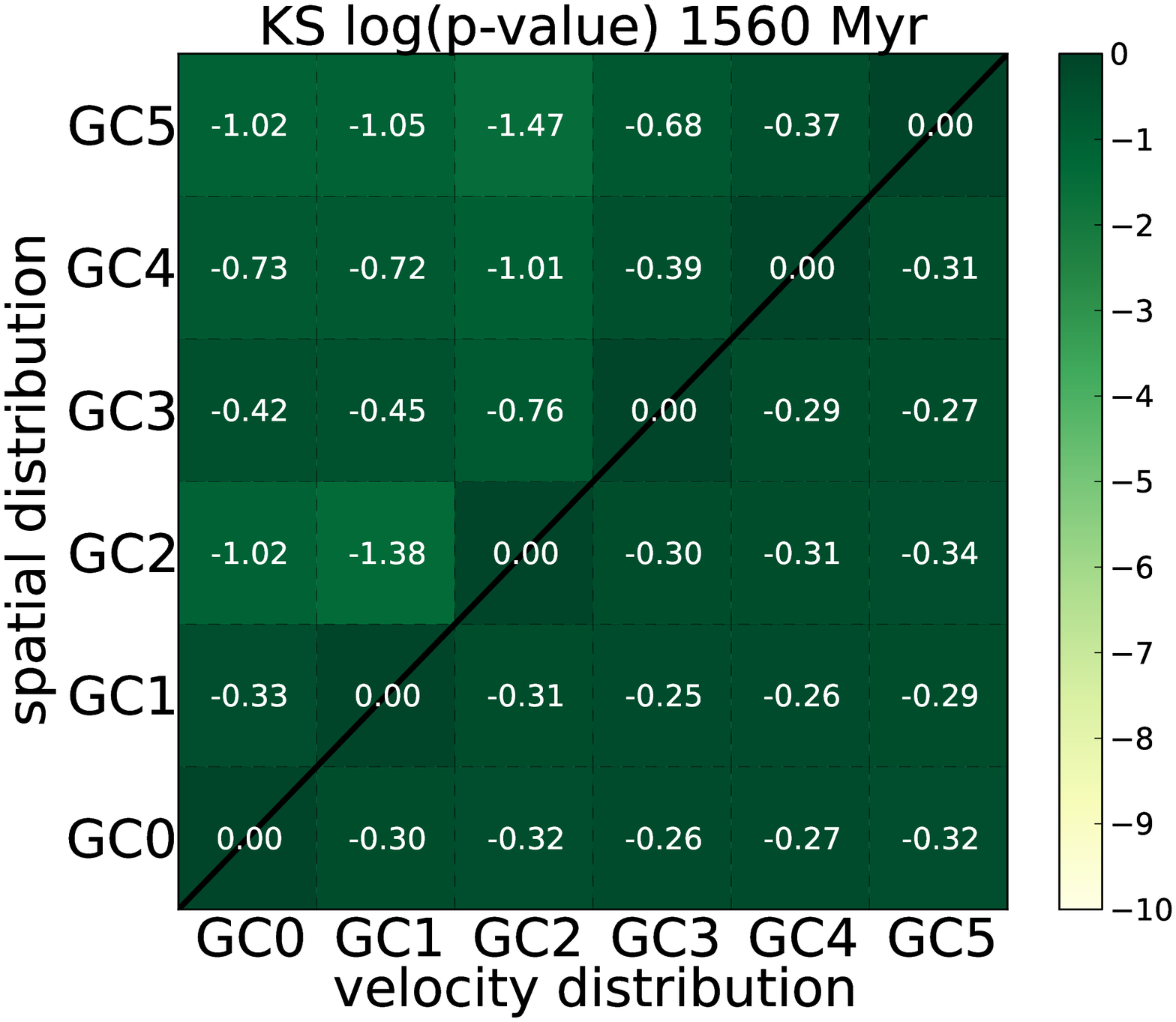}
\end{tabular}
\caption{ Model 1 log$_{10}$ of $p$-values for the cumulative spatial
  distribution of 200 stars within 4$R_e$ when the merger remnant
  contains 3 GCs (left), 5 GCs (middle) and 6 GCs (right) and
  cumulative absolute velocity distribution as seen by an observer
  viewing the system edge on to the average plane of the GCs' initial
  orbits. The $p$-value for any pair of GCs is found at the intersection
  of the appropriate row and column. The figure is colour coded so
  that high $p$-values are darker (green) and low $p$-values are
  lighter (yellow). }
\label{fig:KSme}
\end{figure*}

In this section we make some quantitative measurements of the
probability that stars originating in each of the GCs could be
distinguished from the overall distribution by using the
Kolmogorov-Smirnov (K-S) test on the fractional distribution of stars
within 4$R_e$. We shall assume that stars from each of the GCs can be
chemically identified and then find the probability that stars from
pairs of GCs have been drawn from the same population.

We first perform K-S tests on the spatial distribution of stars from
each GC at the point where 3, 5 and 6 GCs make up the merger
remnant. Figure \ref{fig:KSme} shows a graphical representation of the
probabilities, $p$ ($p$-value), in log$_{10}$ space, that stars
originating in pairs of merged GCs are drawn from the same
population. The $p$-values are calculated on the cumulative fraction of
a random selection of 200 stars taken 1000 times and averaged. Figure
\ref{fig:KSme} is composed of n$^2$ cells where n is the number of GCs
merged in the remnant at each time. The GCs are laid out along the x
and y-axes in order and the number displayed in the cell where the
appropriate row and column cross is the $p$-value for the probability
that the stars from those two GC populations are drawn from the same
population. The $p$-values in the upper left half of the figure (above
the diagonal) are for the spatial distribution of the GCs and those in
the lower right half (below the diagonal) are for the cumulative
fraction of stars versus radial velocity seen by an observer viewing
the system edge-on to the average plane of the GC's initial
orbits. The figure is colour coded so that high values of the $p$-value
are darker (green) and low values are lighter (yellow).

The $p$-values for the radial velocity show that there is a high
probability that any pair of GCs are drawn from the same
population. The lowest value has $p$-value of $\sim$0.32. The
$p$-values for the spatial distribution show a greater likelihood that
two GC populations could be distinguished. When the merger remnant
consists of 3 GCs GC0 and GC1 have a high probability that their stars
are drawn from the same population. However the most recently merged
GC2 has less than 1$\%$ probability that its stars are drawn from the
same population as GC0 or GC1 so it seems possible that it could be
distinguished by this method. We see a similar situation when 2 more
GCs have merged, when GC4 has less than 1$\%$ probability that its
stars are drawn from the same population as GC0 or GC1. However all
other $p$-values are $>$1$\%$ implying that it would be difficult to
distinguish separate populations. When the merger remnant contains 6
GCs even the most recently merged GC, GC5, has a greater than 2$\%$
probability that its stars are drawn from the same population as
GC2. Other $p$-values are $>$0.1 apart from GC2 which remains the most
distinguishable of the remaining GCs in its spatial distribution
having a 4$\%$ probability that its stars are drawn from the same
population as GC1.

%%%%%%%%%%%%%%%%%%%%%%%%%%%%%%%%%%%%%%%%%%%%%%%%%%%%%%%%%%%%%%%%%%%%%%
\subsection{Properties of the merger remnant in model 2}
\label{sec:sim2}

\begin{figure}
\centering
\begin{tabular}{l}
\includegraphics[width=0.5\hsize,angle=0]{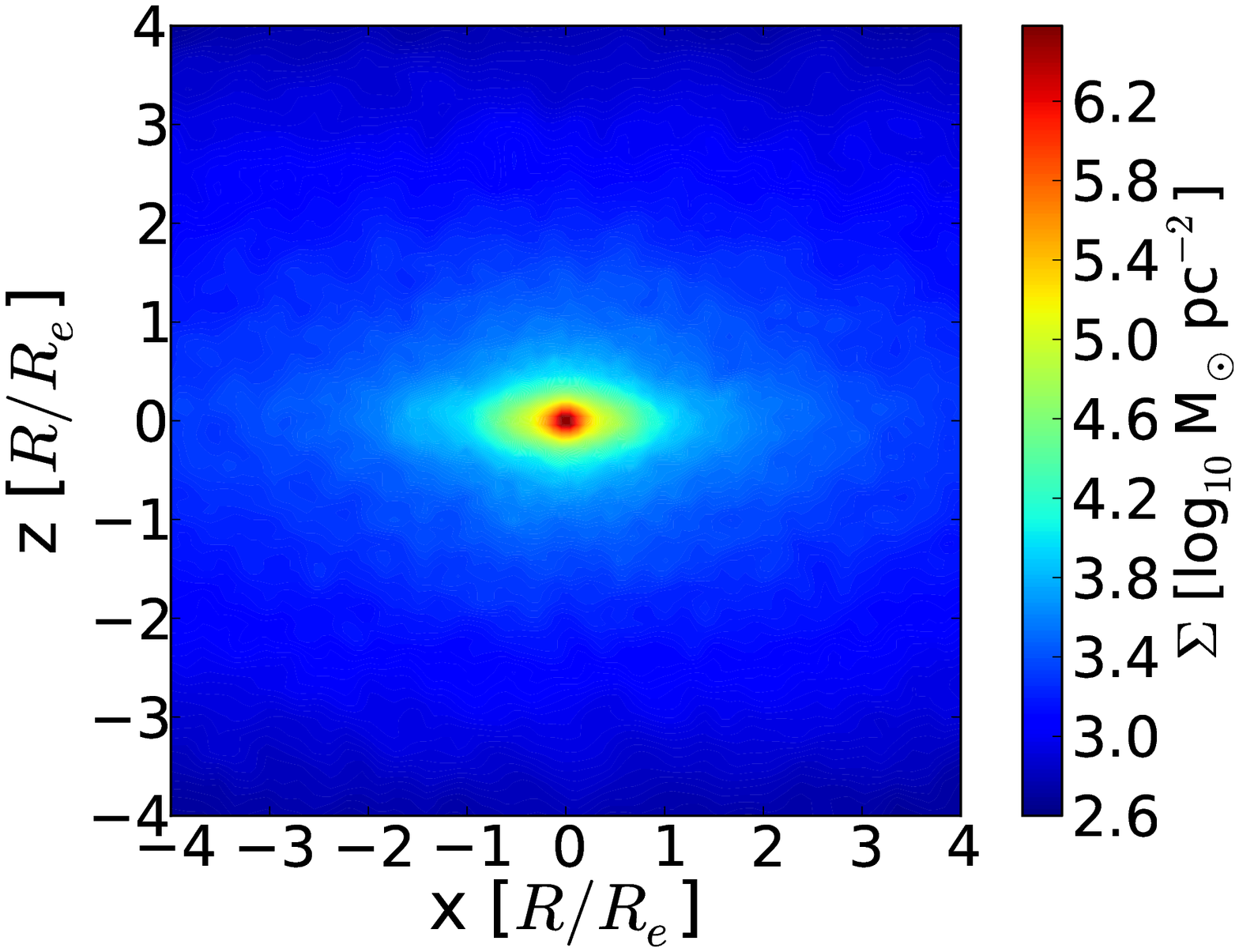} \\
\includegraphics[width=0.5\hsize,angle=0]{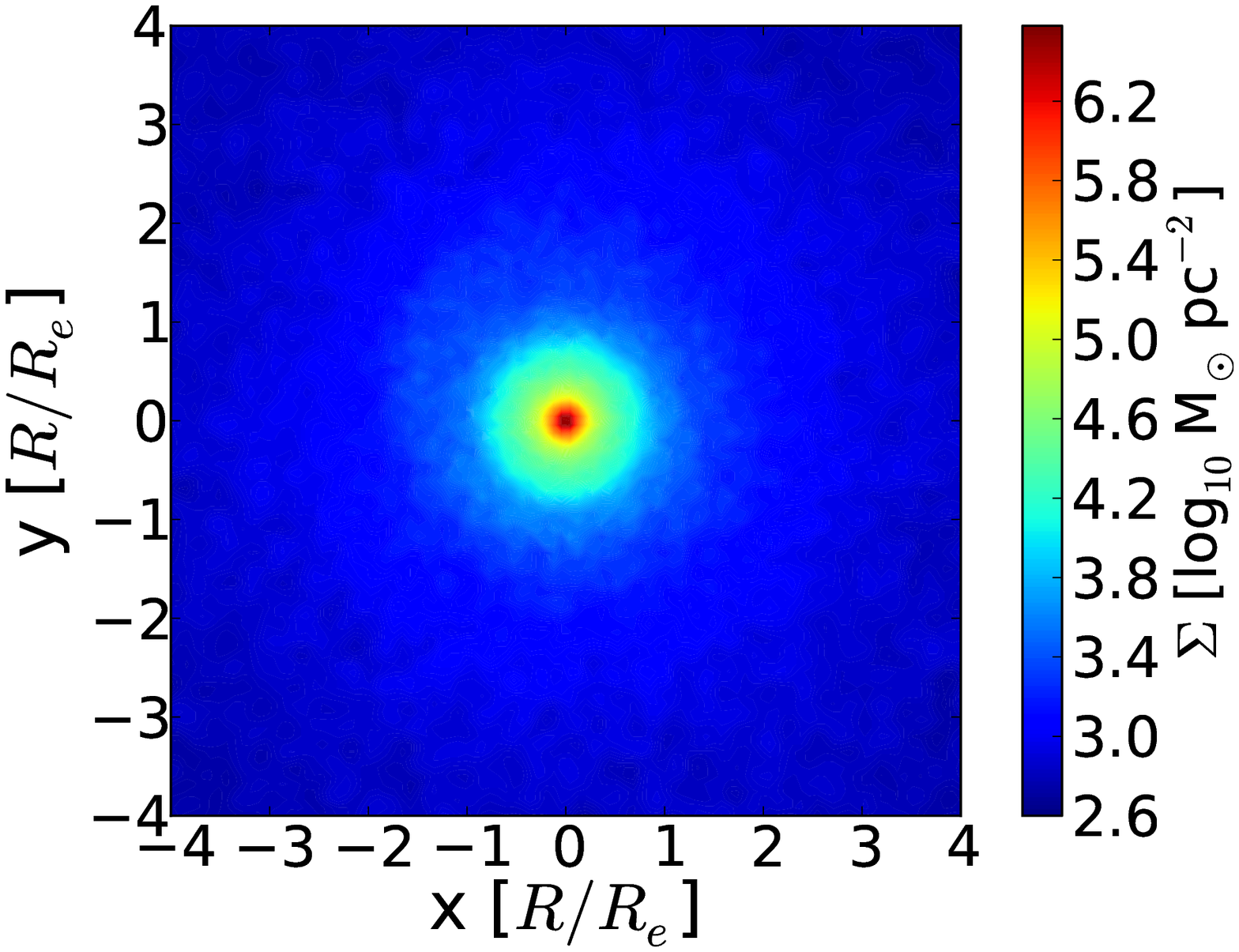} 
\includegraphics[width=0.5\hsize,angle=0]{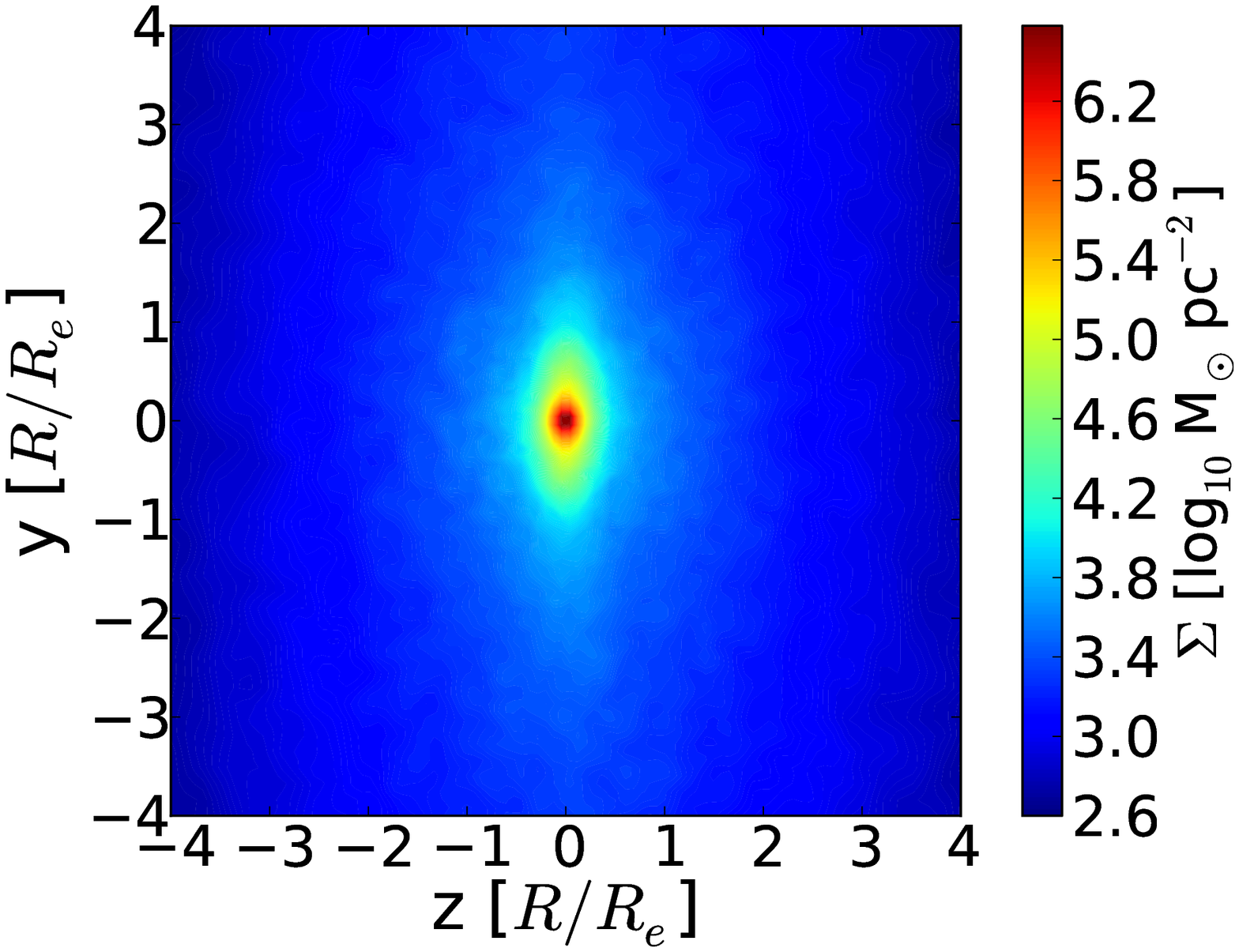} \\
\end{tabular}
\caption{Surface density map of the merger remnant from model 2
  (run A1 of \citet{Hartmann2011}) showing three orthogonal
  projections. The lower left hand plot is face-on to the plane of the
  orbits of the GCs and the other 2 are perpendicular to it. }
\label{fig:densmap2}
\end{figure}

We now examine the results for model 2 (run A1 of
\citet{Hartmann2011}). In this simulation a massive star cluster is
placed on an orbit close to the centre where it eventually settles. A
series of less massive star clusters are then placed on orbits at 32
pc from the centre, one at a time. Their orbits decay to the centre
where they merge with the central structure. Each one is allowed to
merge before the next one is added to the simulation. At the end of
the simulation 27 GCs have merged. Figure \ref{fig:densmap2} shows a
stellar density map for model 2 at the end of the simulation. The
merger remnant is much more flattened than that in simulation 1
because the GC orbits are all co-planar (compare with Figure
\ref{fig:profiles}). Table \ref{tab:evos2} shows the mass and half
mass radius for the merger remnant at 310 Myr, 600 Myr and 810 Myr
when 10, 20 and 27 clusters have merged. $R_e$ grows significantly
between 310 Myr and 600 Myr from 2.9 pc to 6.2 pc.

\begin{table}
  \begin{center}
    \begin{tabular}{lccc} 
      \hline
%      \hline#
      \multicolumn{1}{l}{Time} &
      \multicolumn{1}{c}{Mass} &
      \multicolumn{1}{c}{Half mass radius} &
      \multicolumn{1}{c}{$N_{GC}$} \\
      \multicolumn{1}{l}{Myr} &
      \multicolumn{1}{c}{$\times10^5$ M$_\odot$} &
      \multicolumn{1}{c}{pc} &
      \multicolumn{1}{c}{} \\[-0.5ex]
      \hline
\hline
310 & 16.22 & 2.9 & 10 \\
600 & 28.39 & 6.2 & 20 \\
810 & 36.89 & 6.95 & 27 \\
\hline
    \end{tabular}
  \end{center}
  \caption[]{\label{tab:evos2} Properties of the merger remnant in
    model 2 at three stages of its evolution. }
\end{table}

\begin{figure}
\centering
\begin{tabular}{c}
\includegraphics[width=\hsize,angle=0]{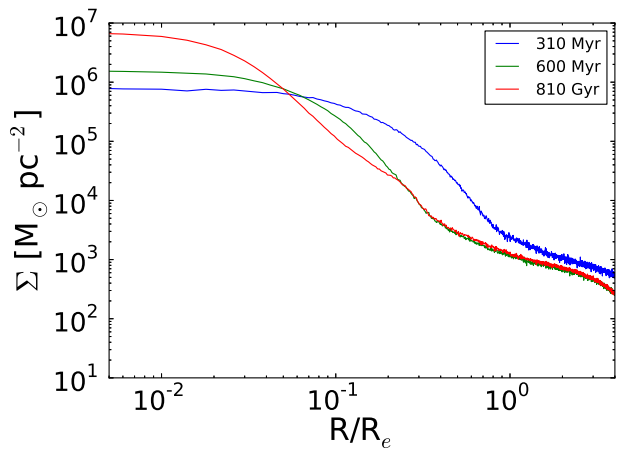} 
\end{tabular}
\caption{Evolution of the surface density for the merger remnant in
  model 2. }
\label{fig:densh2}
\end{figure}

\begin{figure*}
\centering
\begin{tabular}{c}
\includegraphics[width=0.33\hsize,angle=0]{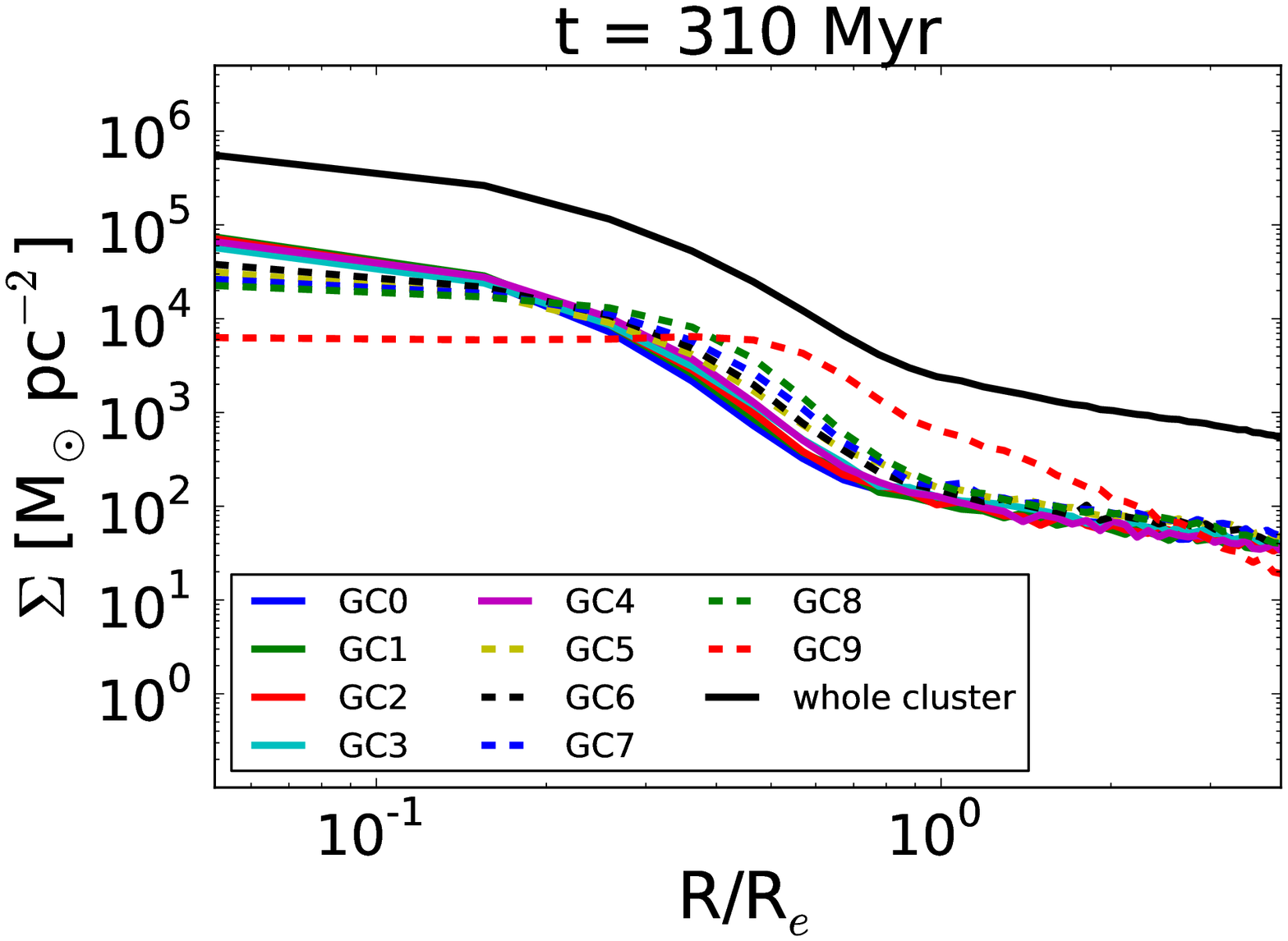} 
\includegraphics[width=0.33\hsize,angle=0]{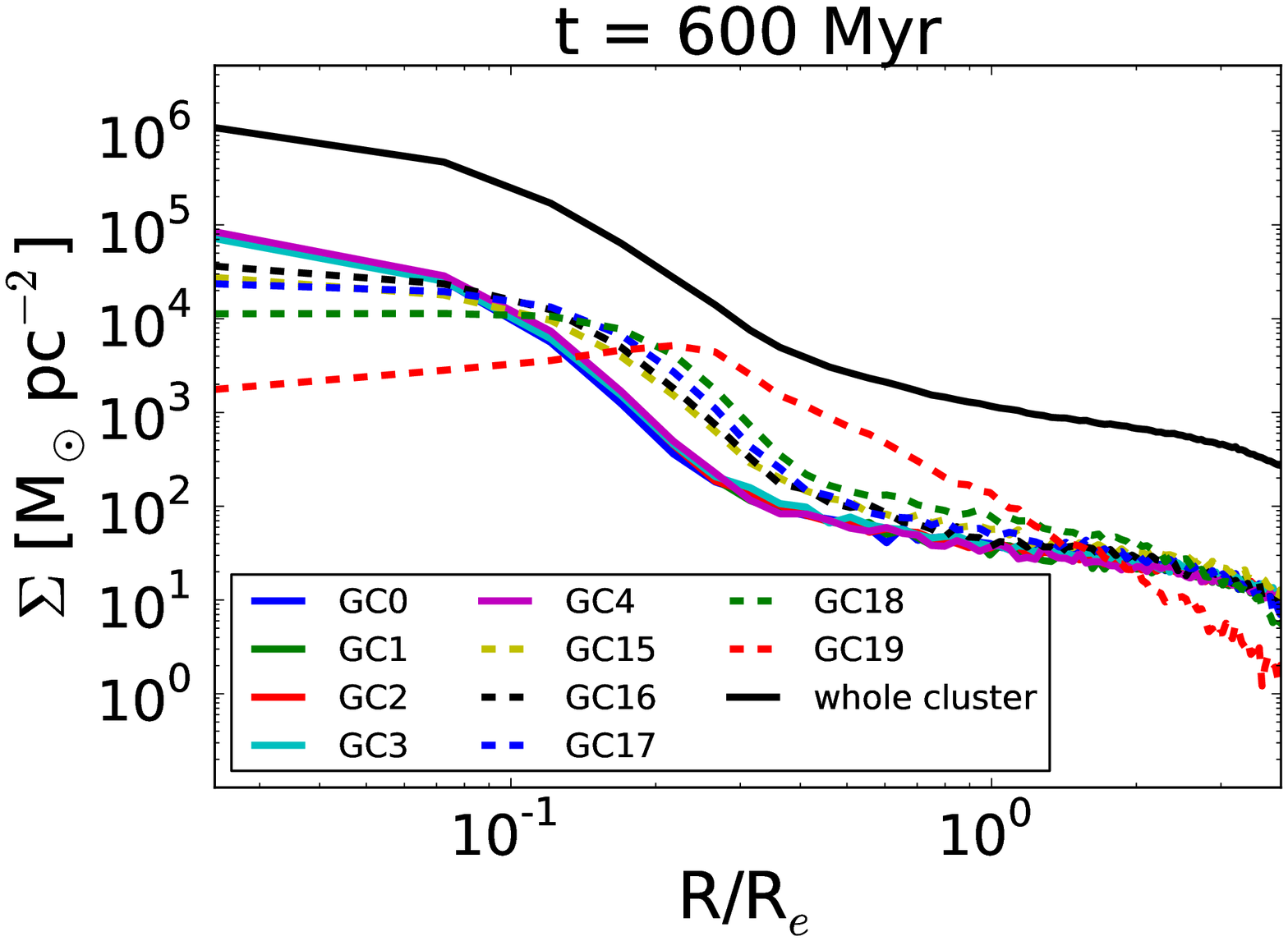} 
\includegraphics[width=0.33\hsize,angle=0]{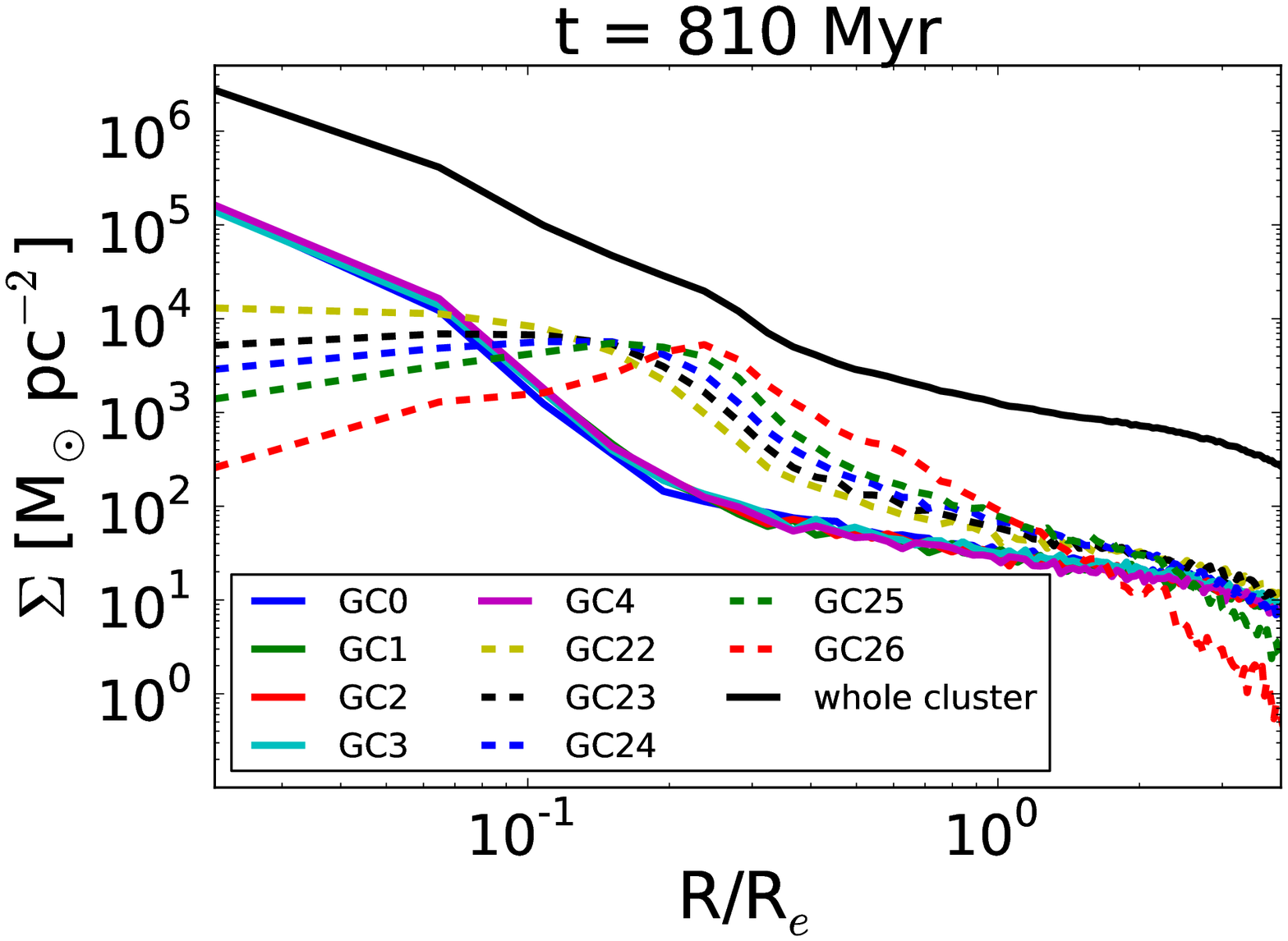} 
\end{tabular}
\caption{ The density profile at three different times for model
  2. The panels show when 10 GCs in the merger remnant on the left, 20
  GCs in the remnant in the centre and 27 GCs in the remnant on the
  right. }
\label{fig:denspr2}
\end{figure*}

Figure \ref{fig:densh2} shows the evolution of the surface density
profile for the merger remnant within 4$R_e$ showing increasing
density at the centre. Figure \ref{fig:denspr2} shows surface density
profiles for specific GCs at the same three times. The GCs which merge
first have very similar density profiles but we see larger differences
for the most recently merged GCs, especially at later times. The
density profile for these GCs is either flat in the middle or even
dropping towards the centre, showing that their density profiles will
evolve further.

\begin{figure*}
\centering
\begin{tabular}{c}
\includegraphics[width=0.33\hsize,angle=0]{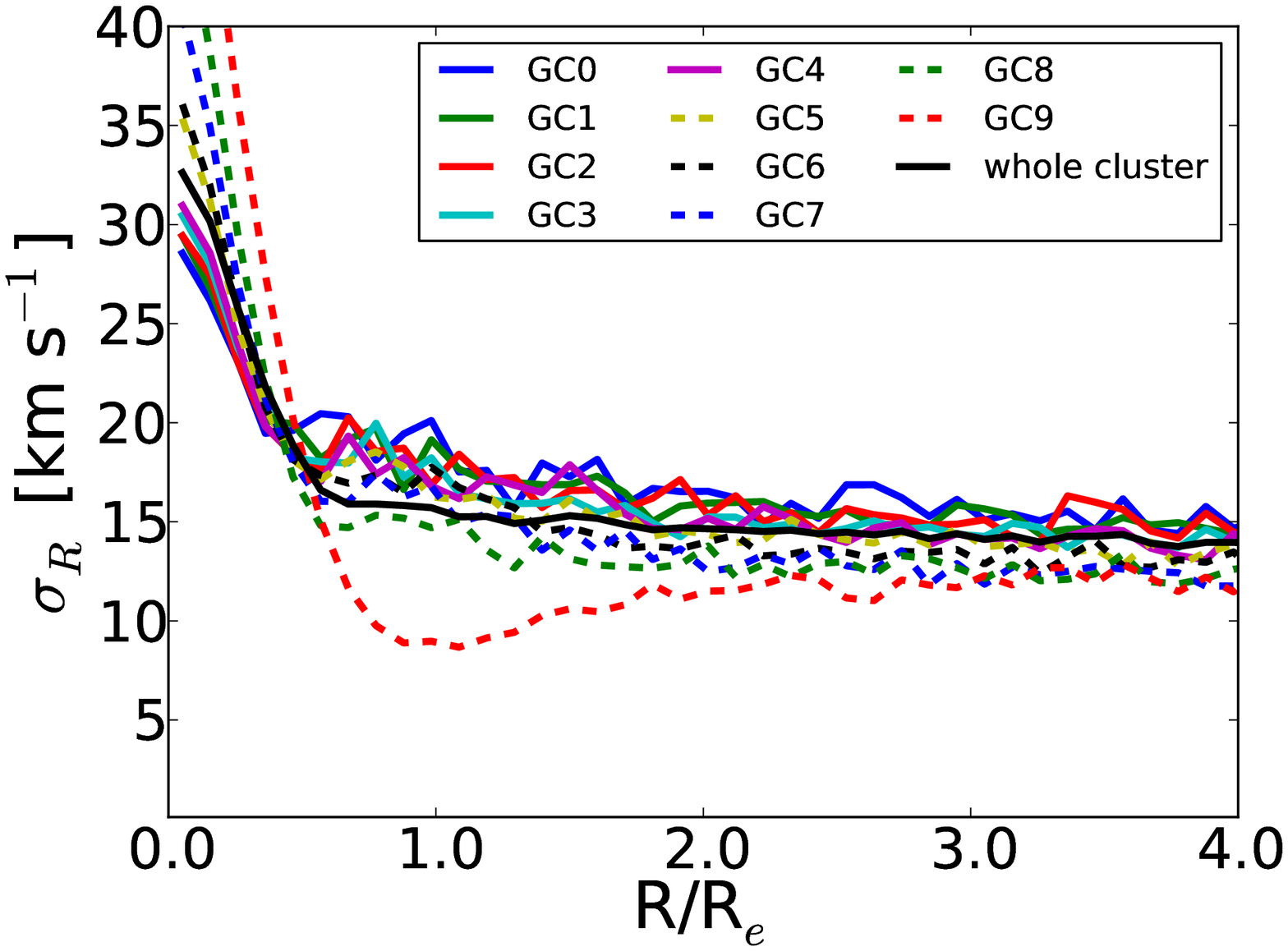} 
\includegraphics[width=0.33\hsize,angle=0]{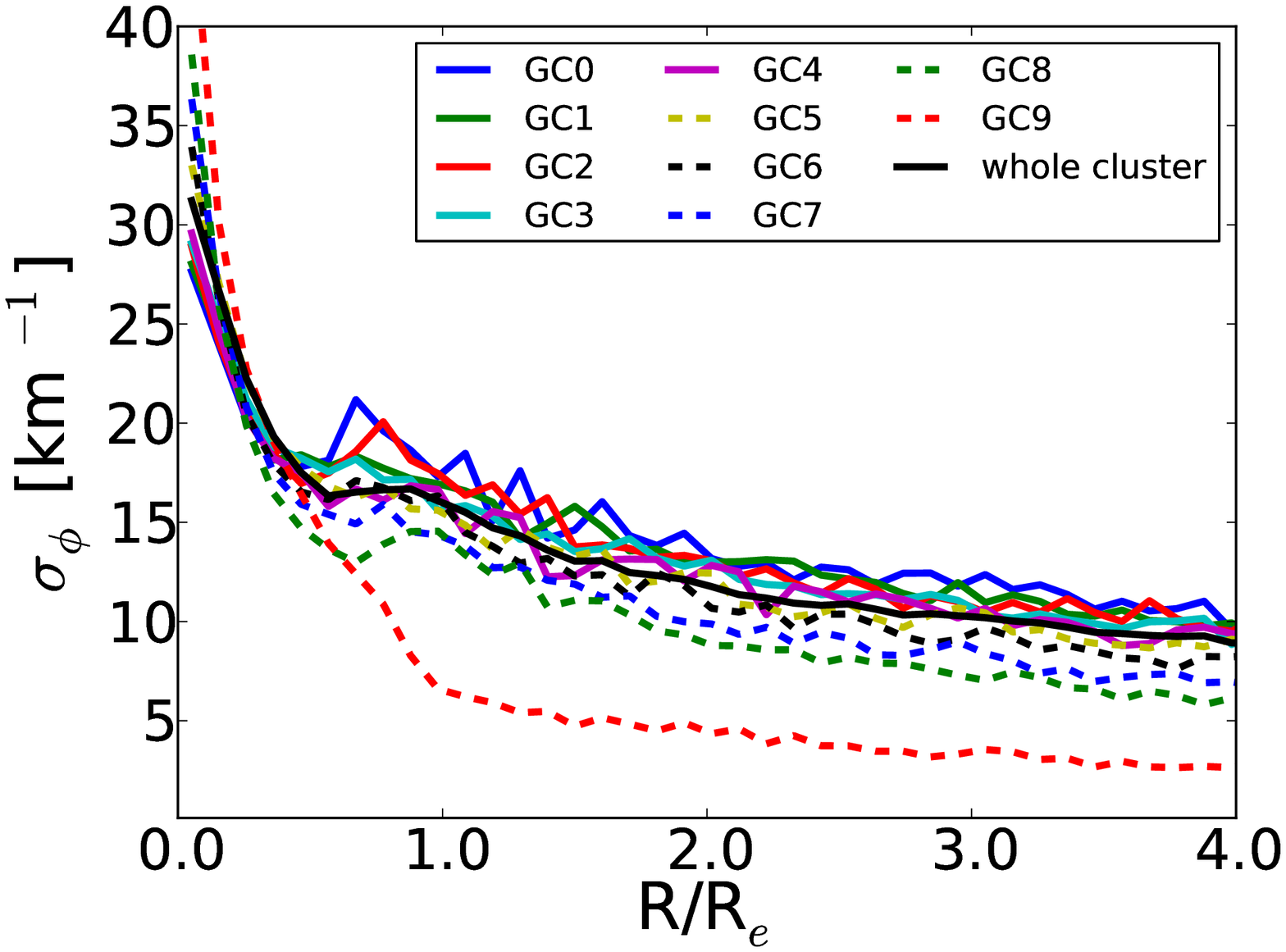} 
\includegraphics[width=0.33\hsize,angle=0]{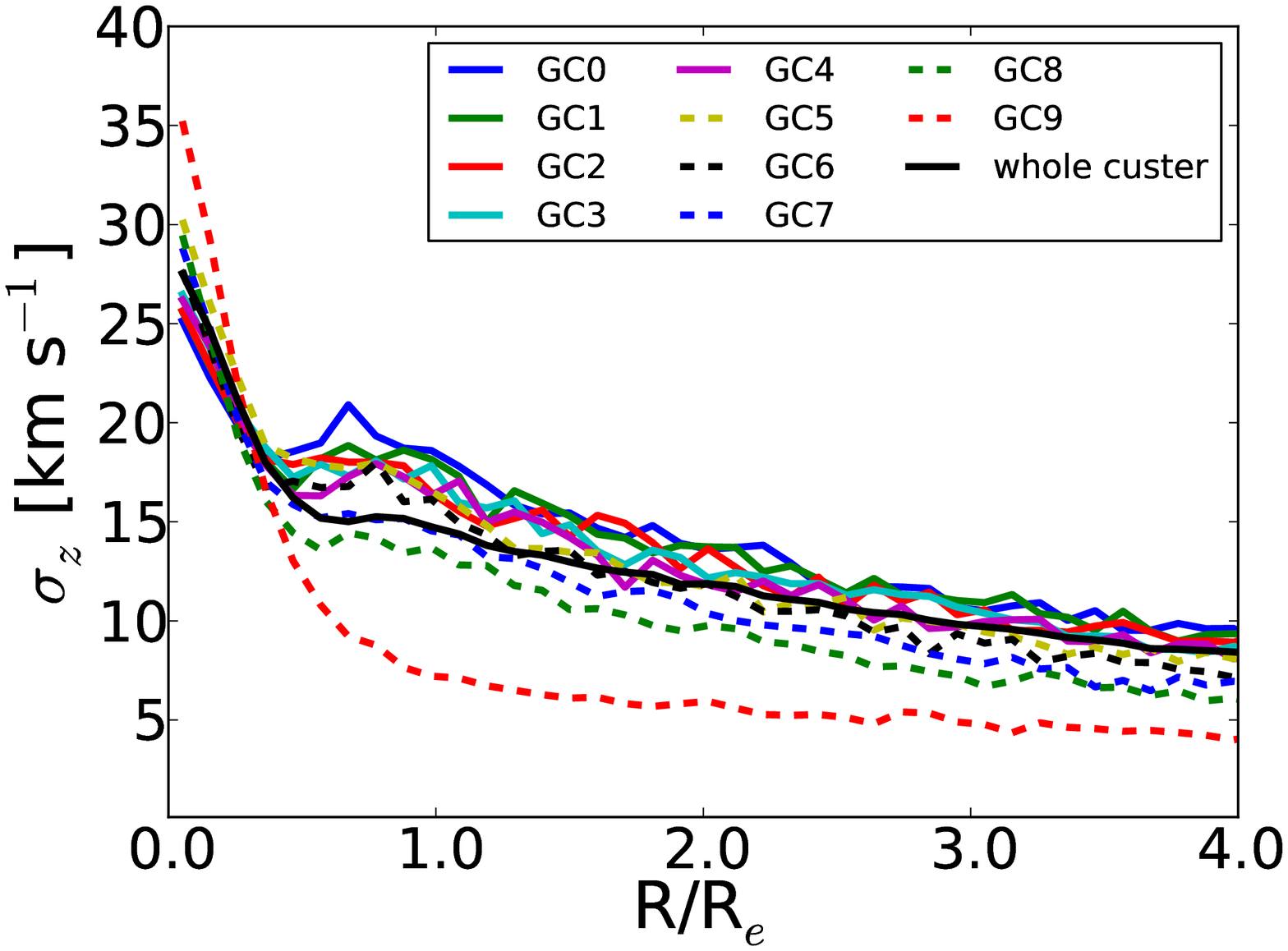} \\
\includegraphics[width=0.33\hsize,angle=0]{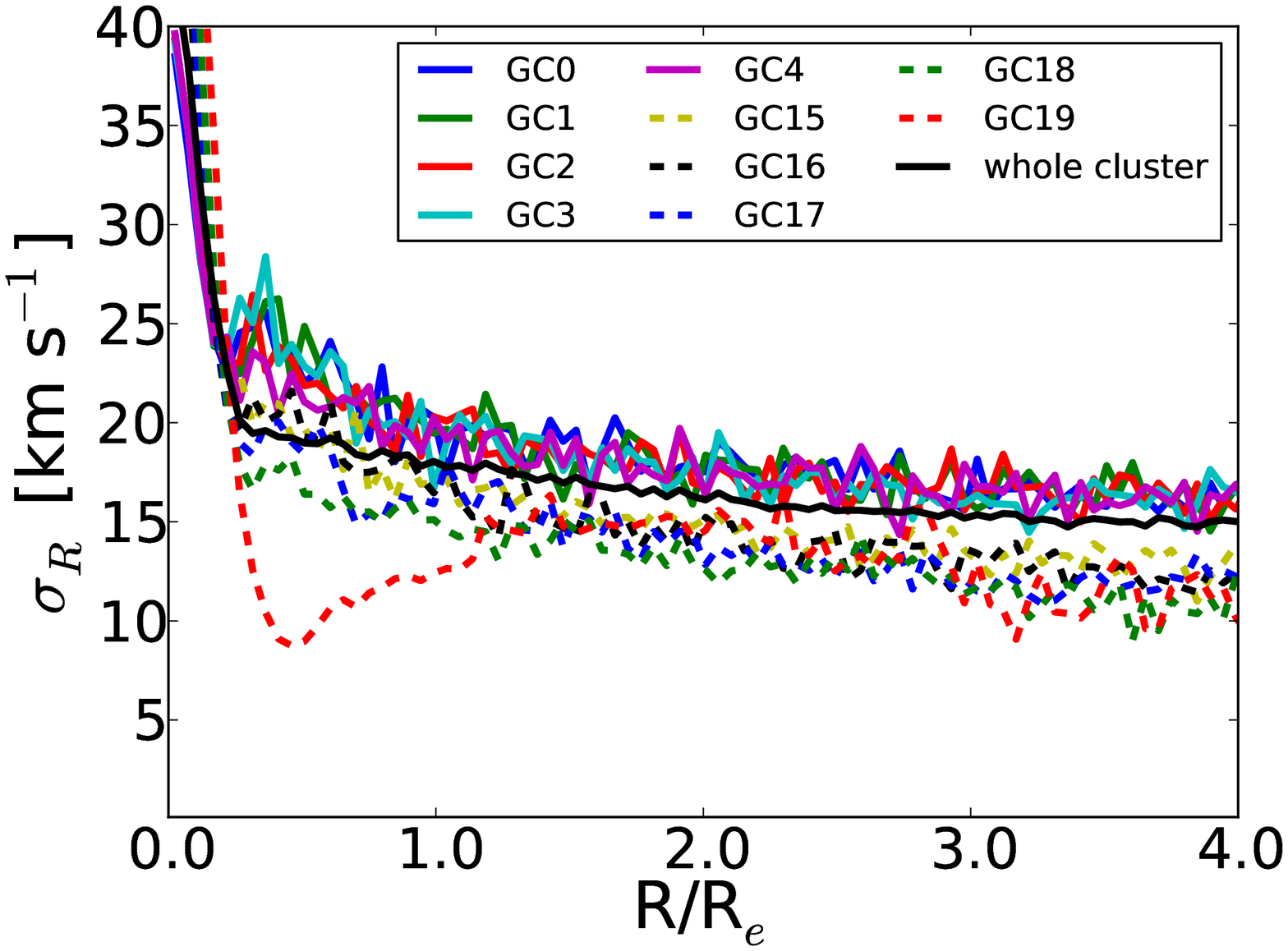} 
\includegraphics[width=0.33\hsize,angle=0]{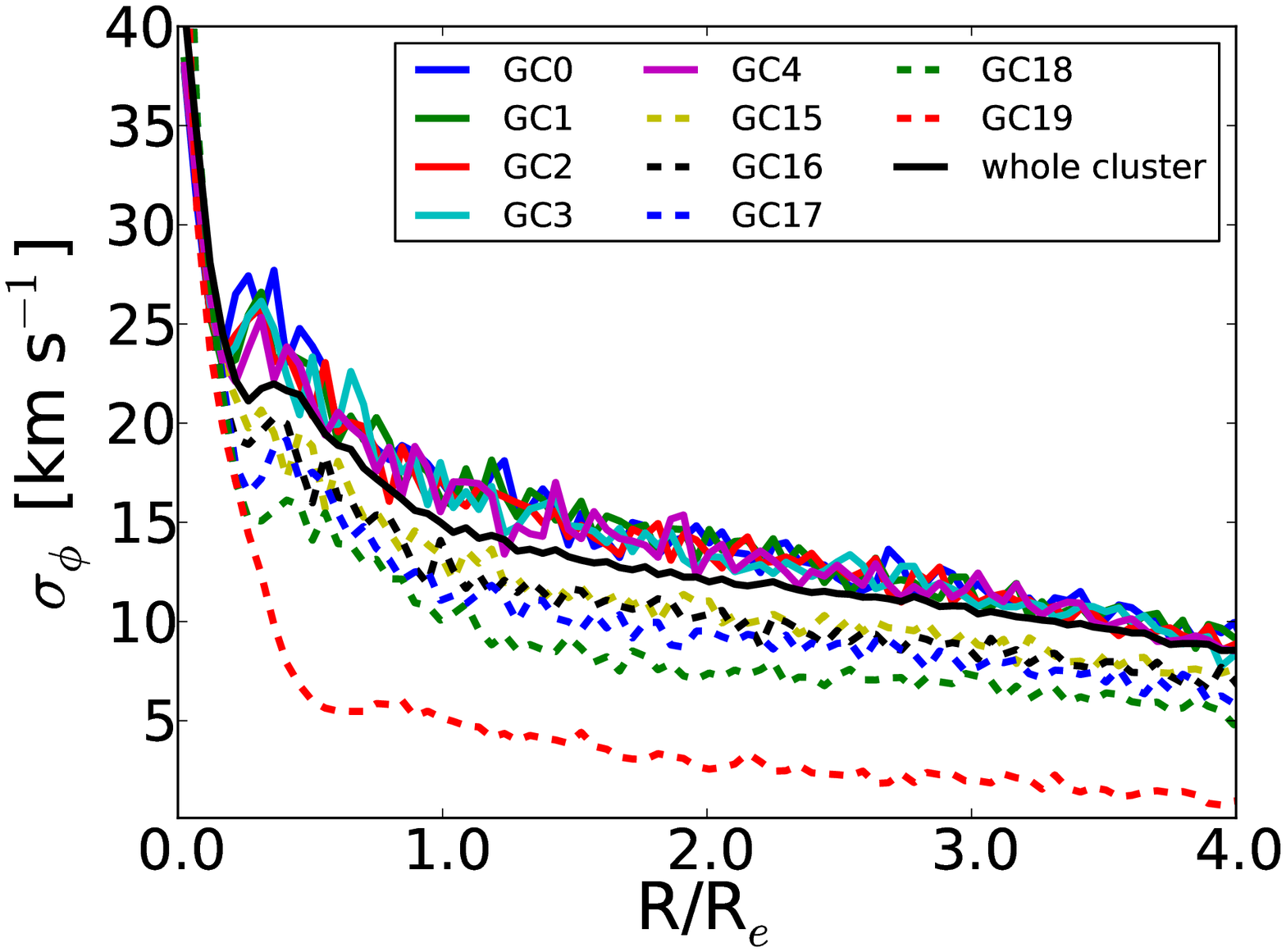} 
\includegraphics[width=0.33\hsize,angle=0]{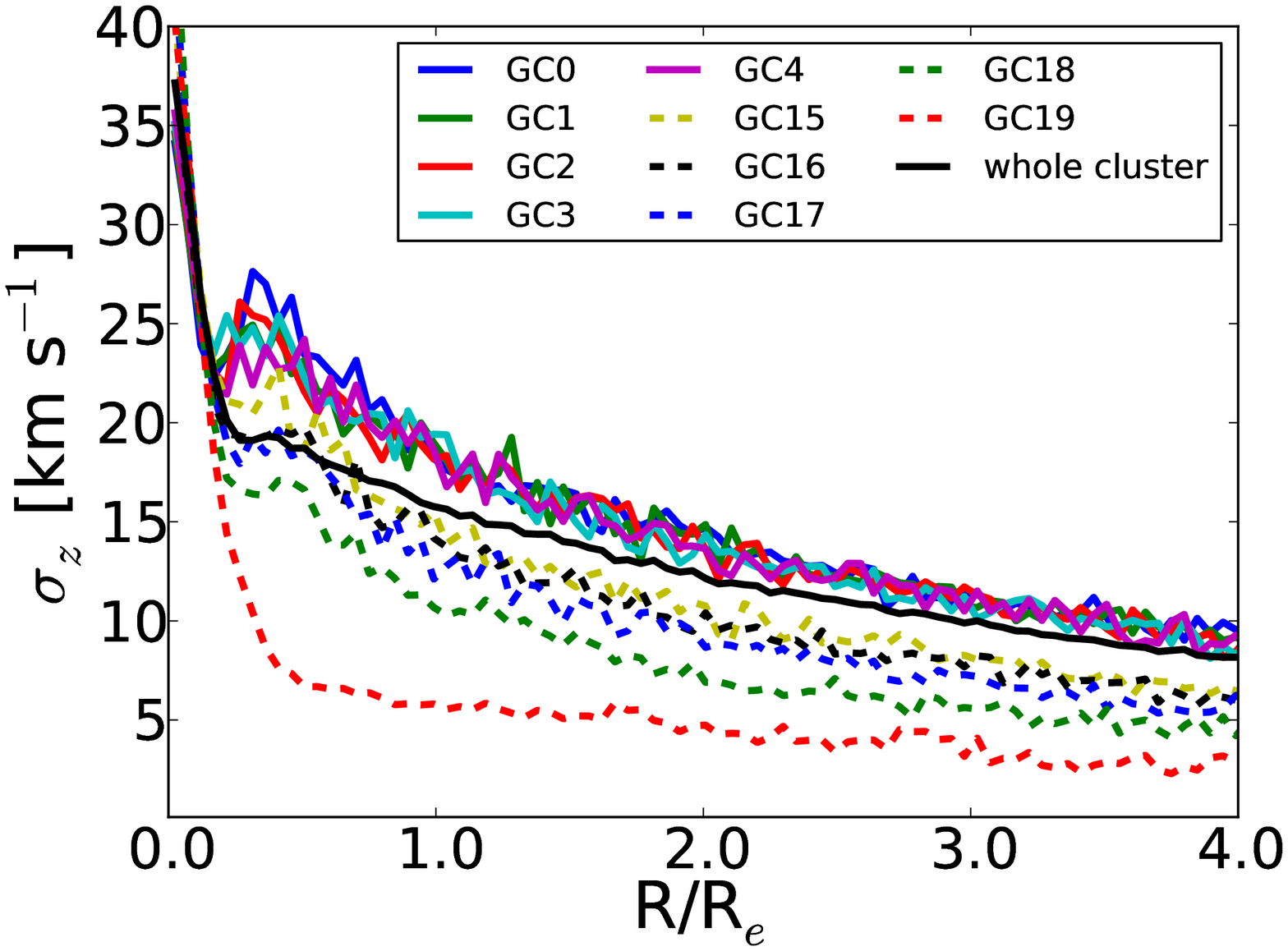} \\
\includegraphics[width=0.33\hsize,angle=0]{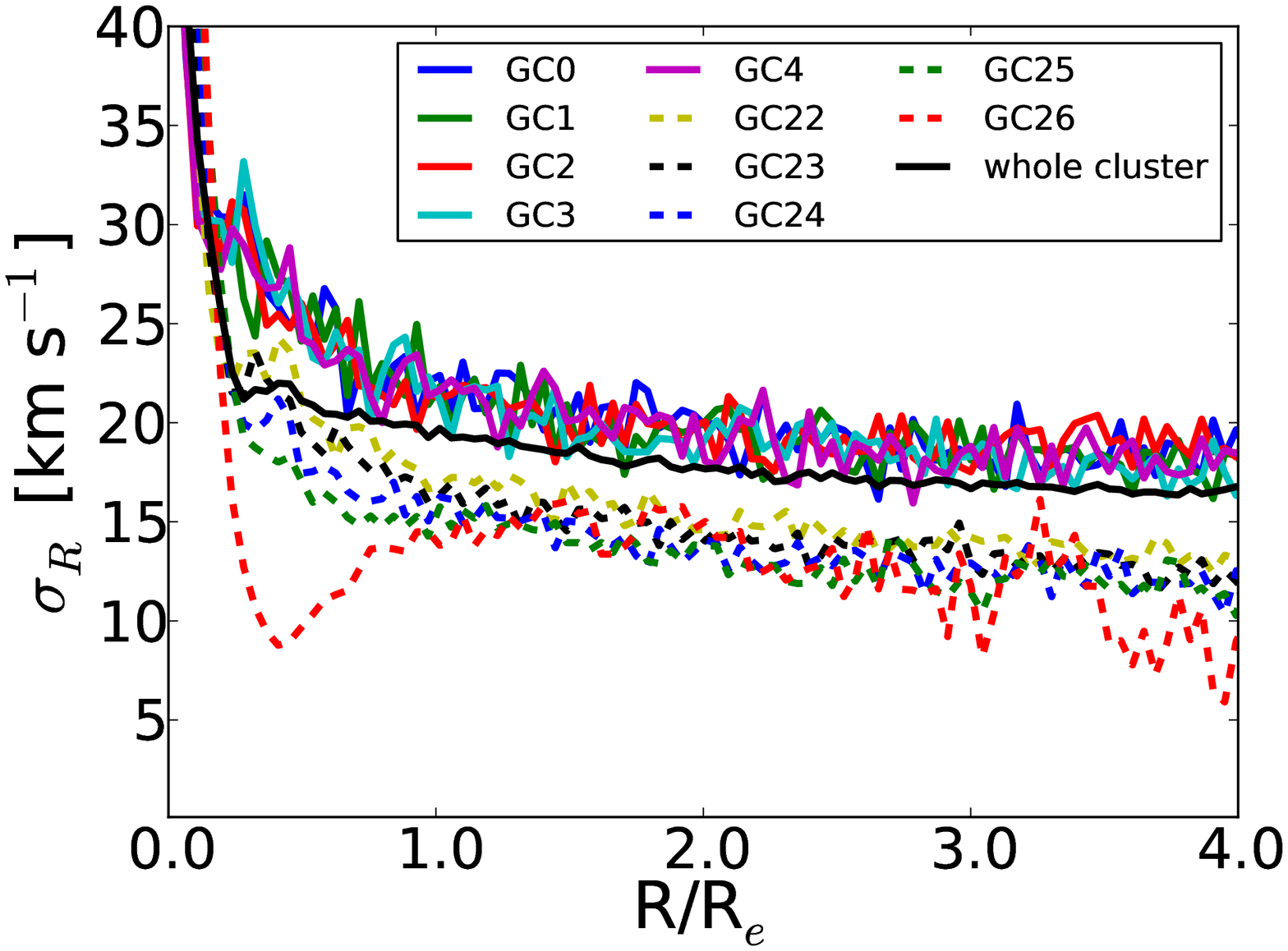} 
\includegraphics[width=0.33\hsize,angle=0]{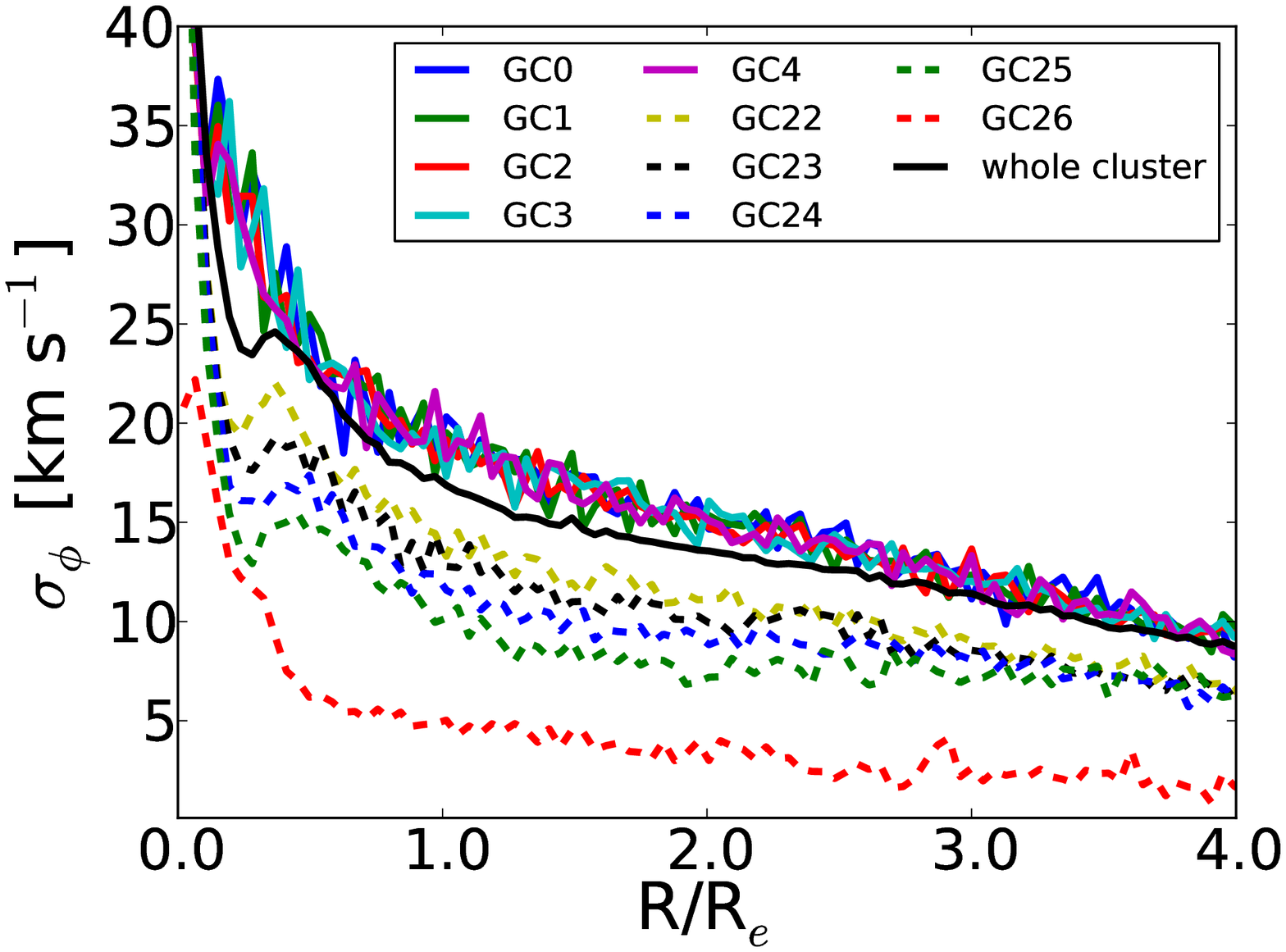} 
\includegraphics[width=0.33\hsize,angle=0]{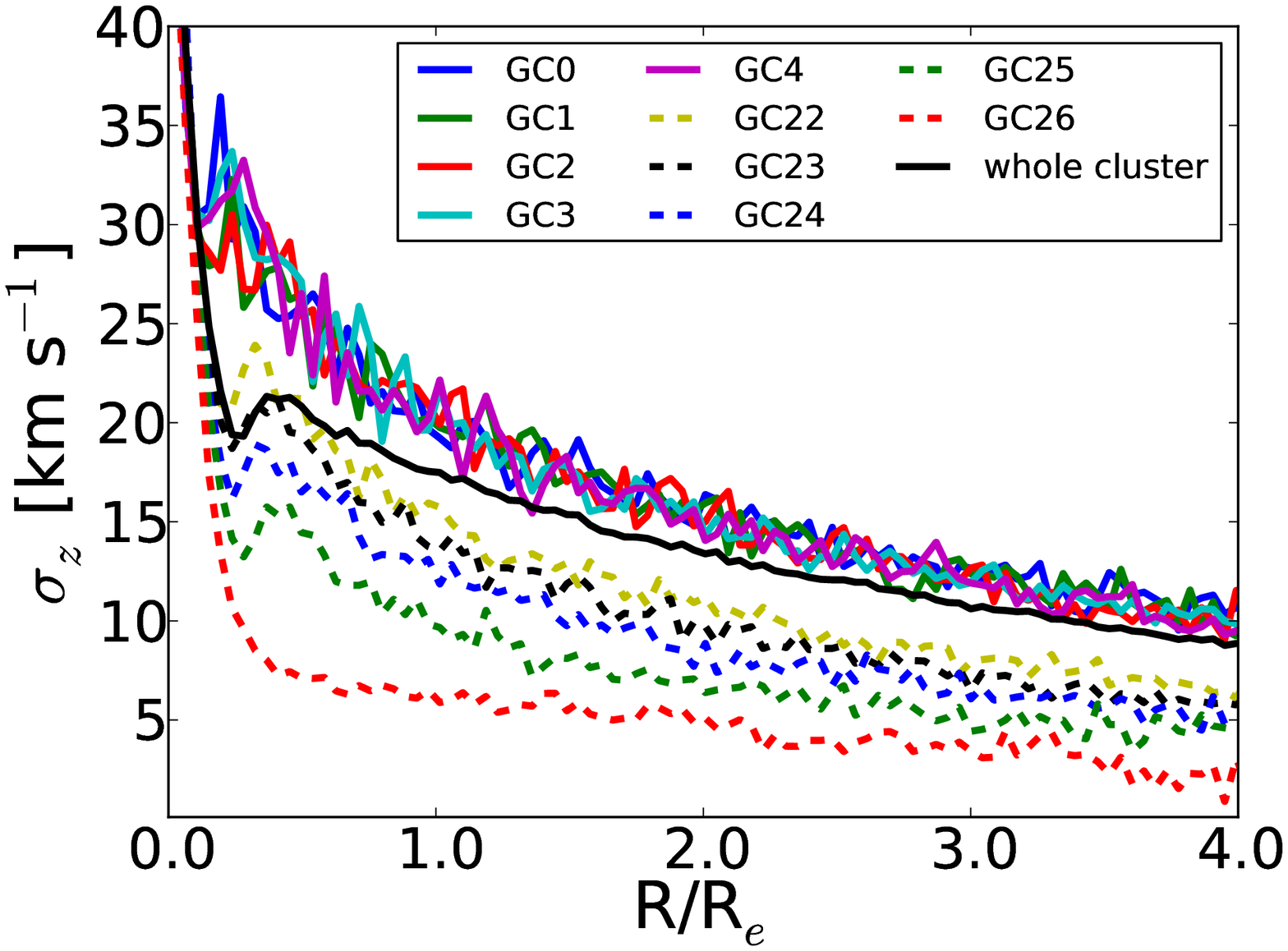} 
\end{tabular}
\caption{ Model 2 velocity dispersion profiles in cylindrical
  coordinates when the merger remnant contains 10 GCs (top), 20 GCs
  (middle) and 27 GCs (bottom) for model 2. Profiles for the first 5
  GCs to merge and the 5 most recently merged GCs are shown. Radial
  profiles are on the left, azimuthal profiles are in the centre and
  vertical profiles are on the right. 10 GCs have merged with the
  central cluster at this time. }
\label{fig:disph312}
\end{figure*}

Figure \ref{fig:disph312} depicts the velocity dispersion profiles for
the same GCs at the three times shown in Figs. 11 and 12 as well as
the velocity dispersion for the whole cluster. These profiles show a
similar pattern to the density and kinematic profiles. The first 5 GCs
to merge have very similar profiles and the most recent GC to have
merged always shows the biggest difference. There are also bigger
differences for the recently merged GCs at later times. The velocity
dispersions for the whole cluster increase significantly from when the
merger remnant contains 10 GCs to when it has 27 GCs.

At each time the 5 most recently merged GCs all show a greater
difference in density profile and kinematics from the average. From
inspection of the density profiles it appears that these components
are still experiencing significant evolution, and therefore it is not
surprising to notice some differences in their kinematics. After the
first ten merger events, only the last cluster component (GC9) is
still distinguishable from the global behaviour. Similarly, after
twenty merger events, the behaviour of the last five clusters still
retains some differences from the one of the previously merged
components (especially in the case of GC19). This applies also to the
subsequent components in the remnant, after it has experienced the
full series of 27 merger events. It appears that in model 2 there is a
greater likelihood that we would be able to distinguish individual
populations from specific recently merged GCs.

%%%%%%%%%%%%%%%%%%%%%%%%%%%%%%%%%%%%%%%%%%%%%%%%%%%%%%%%%%%%%%%%%%%%%%

\subsubsection{Kolmogorov-Smirnov statistics}
\label{sec:KSsim2}

\begin{figure*}
\centering
\begin{tabular}{c}
\includegraphics[width=0.33\hsize,angle=0]{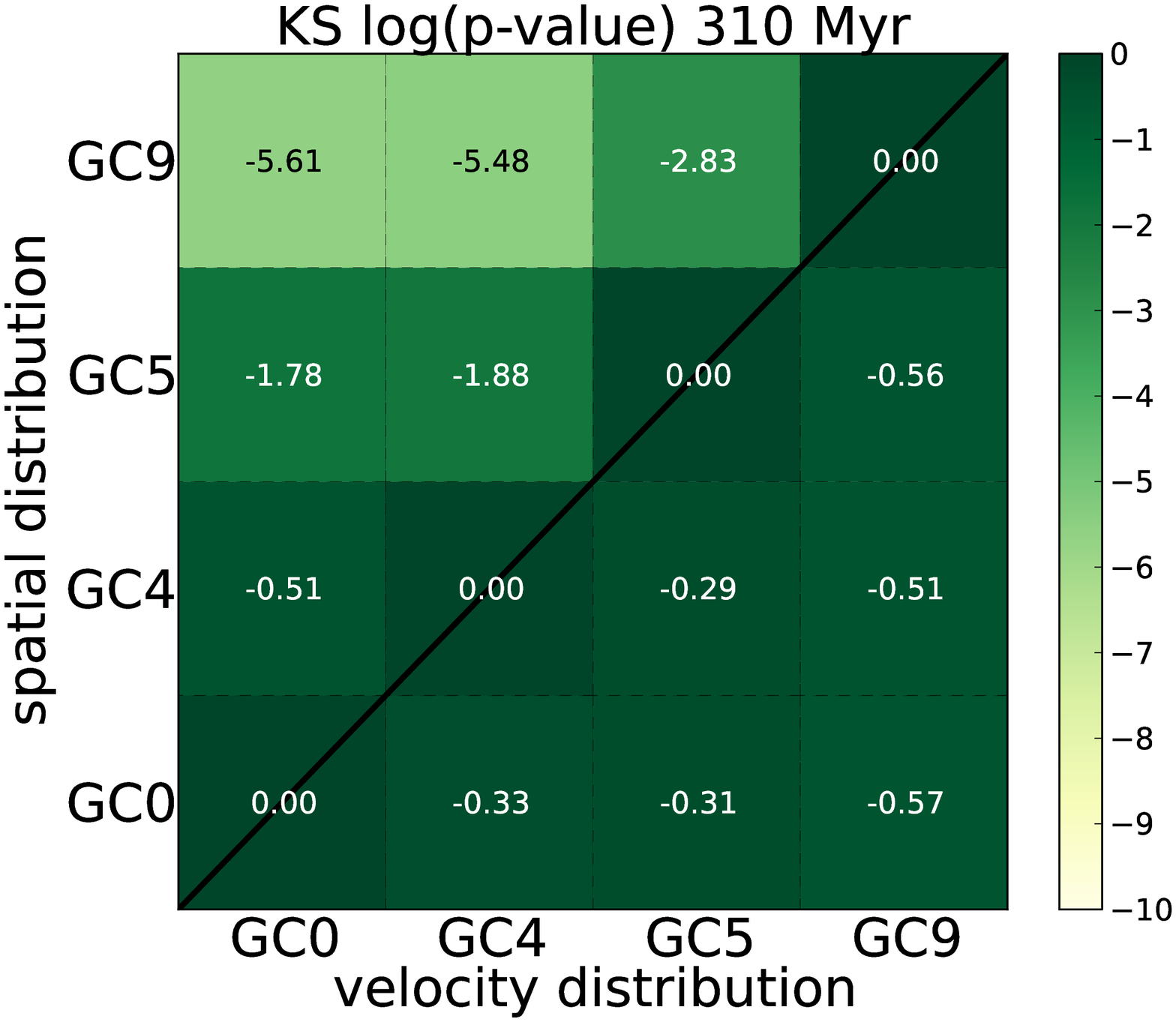}
\includegraphics[width=0.33\hsize,angle=0]{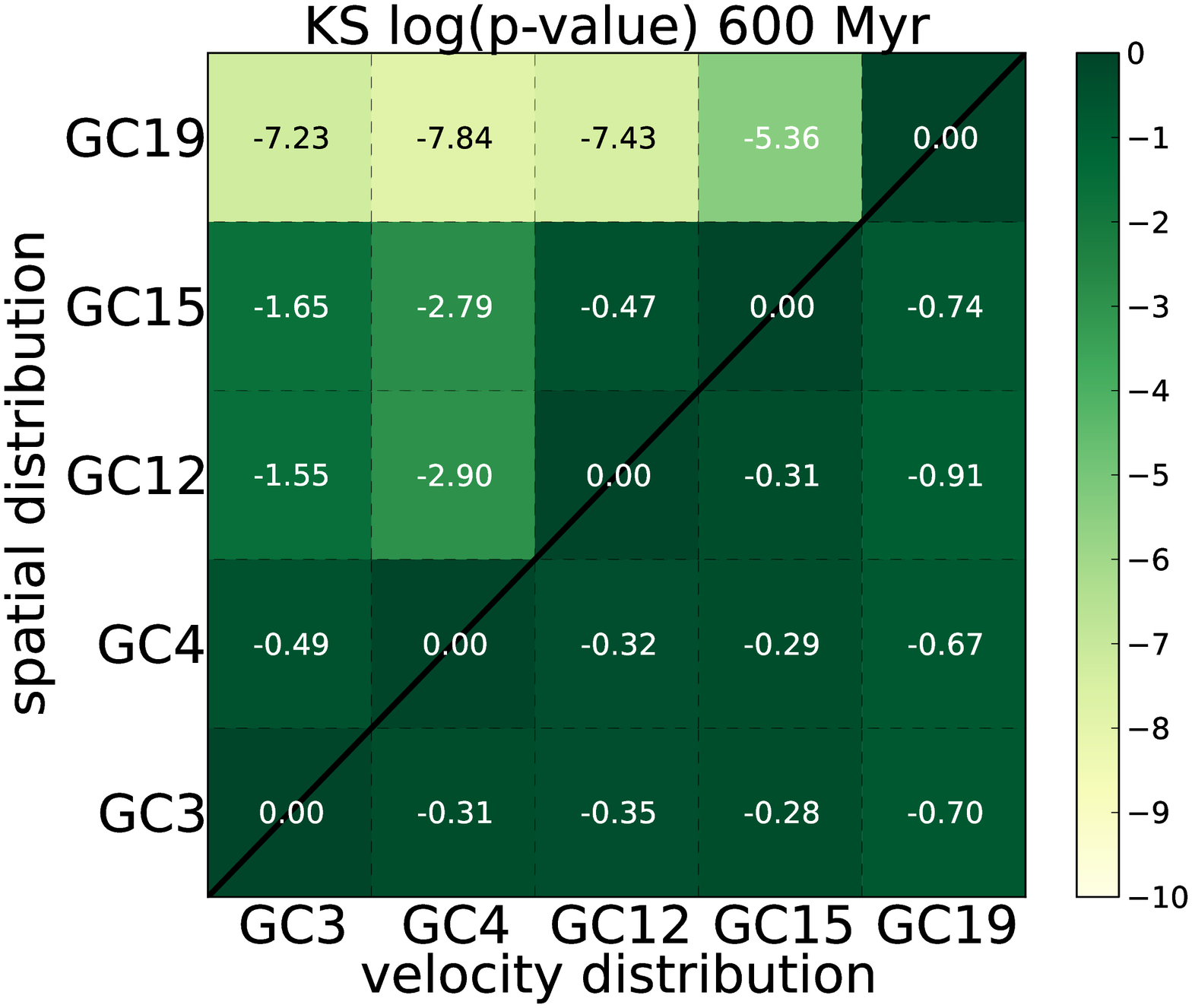}
\includegraphics[width=0.33\hsize,angle=0]{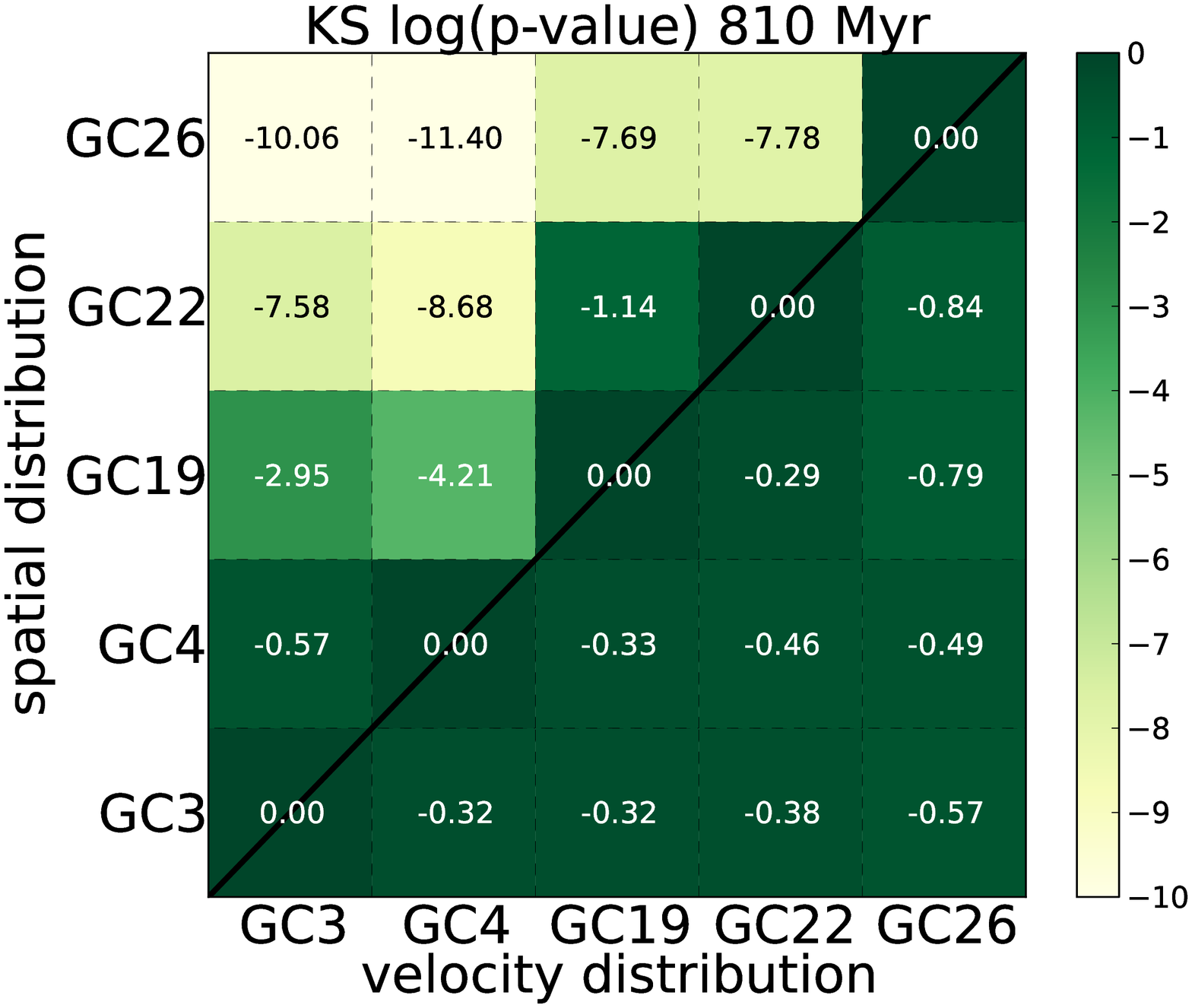}
\end{tabular}
\caption{ Model 2 log$_{10}$ of $p$-values for the cumulative spatial
  distribution of 200 stars within 4$R_e$ when the merger remnant
  contains 10 GCs (left), 20 GCs (middle) and 27 GCs (right) and
  cumulative absolute velocity distribution as seen by an observer
  viewing the system edge-on to the average plane of the GCs' initial
  orbits. The $p$-value for any pair of GCs is found at the intersection
  of the appropriate row and column. The figure is colour coded so
  that high $p$-values are green and low $p$-values are
  yellow. }
\label{fig:KSme2}
\end{figure*}

We performed K-S tests for model 2, looking at the spatial
distribution of stars originating in individual merged GCs after 10,
20 and 27 GCs have merged, drawing 200 stars in each sample. Figure
\ref{fig:KSme2} shows a representative sample of cumulative fractions
of stars originating in individual merged GCs at the same times
considered in Figure \ref{fig:KSme2}. The pairs are defined by
considering a selection of components which belong to either one of
the 5 earliest merged GCs or from one of the 5 most recently merged
clusters; such a prescription allowed us to perform a comparison
between the most similar and the most different distributions,
respectively. The pair from the 5 earliest merged GCs were chosen to
have the greatest apparent difference in distribution. After 20 GCs
have merged and 27 GCs have merged a further GC was chosen from an
intermediate merger event to sample a different stage of the
merger. The objective of this selection is to show the outliers of the
possible comparisons in spatial segregation.

Again the $p$-values in the upper left half of the figure (above the
diagonal) are for the spatial distribution of the GCs and those in the
lower right half (below the diagonal) are for the radial velocity seen
by an observer viewing the system edge-on to the average plane of the
GC's initial orbits. The $p$-values for the radial velocity tests
again show that there is a high probability that any pair of GCs are
indistinguishable using this K-S test. The lowest $p$-value $\sim$0.14
which still demonstrates a high likelihood that the 2 GCs are drawn
from the same population. When the merger remnant contains 10 GCs the
K-S tests based on the spatial distribution show that the earliest
merged GCs, which are GC0, GC4 and GC5, have $p$-values $>$1$\%$
whereas GC9 has a low probability that it is drawn from the same
population as the other three. Similarly when 20 GCs have merged GC3,
GC4, GC12 and GC15, which have been merged for longest, have
$p$-values $>0.001$ in tests between each other whereas GC19 has a
$p$-value $<$5$\times$10$^{-6}$ that its stars are drawn from the same
population as any of the others. When 27 GCs have merged GC3 and GC4
have a similar spatial distribution and GC19, GC22 and GC26 have low
likelihoods of being drawn from the same population as GC3 and GC4.

\begin{figure}
\centering
\begin{tabular}{c}
\includegraphics[width=1.\hsize,angle=0]{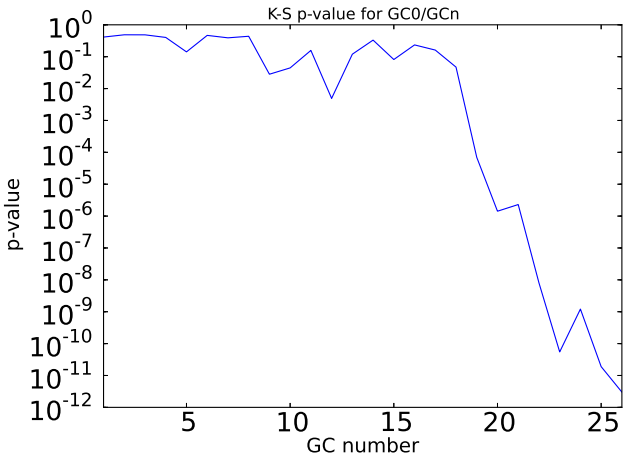}
\end{tabular}
\caption{ The probability when the merger remnant is made of 27 GCs
  that the earliest merged GC0 in model 2 is drawn from the same
  population as all the others based on spatial distribution. }
\label{fig:KStrend}
\end{figure}

At each time we find that the 6 to 8 most recently merged GCs have a
low probability that they are drawn from the same population as
earlier merged GCs based on their spatial distribution. When GCs have
merged prior to this in the merger sequence their stars have a spatial
distribution which is similar to that of the overall merger
remnant. An illustration of this is shown in Figure
\ref{fig:KStrend}. This shows the probability that the earliest merged
GC, GC0, is drawn from the same population as each of the others when
the merger remnant contains stars from 27 GCs. This $p$-value is
generally $\sim$0.2 to 0.5 but always $>$1$\%$ for all GCs up to GC18
and then falls sharply from GC19 (probability, p$<$10$^{-5}$) to GC26
(just merged). This implies that stars from a GC which has merged in
the most recent 8 mergers could be distinguished by their spatial
distribution but stars from a GC which underwent a prior merger could
not.

\begin{figure*}
\centering
\begin{tabular}{c}
\includegraphics[width=1.0\hsize,angle=0]{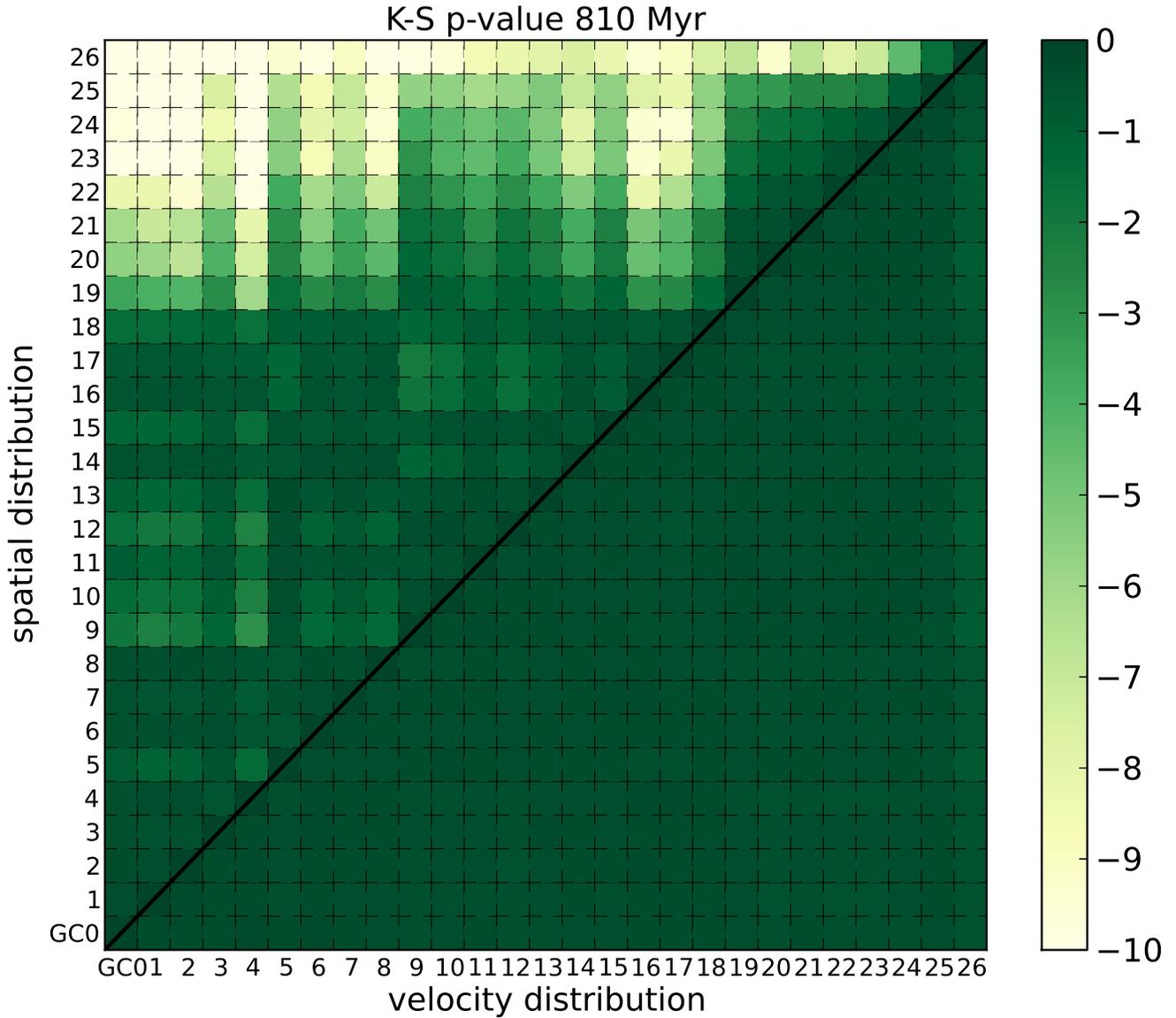}
\end{tabular}
\caption{ $p$-values for the cumulative spatial distribution of 200
  stars within 4$R_e$ at the end of the simulation for model 2, for
  all pairs of GCs. The $p$-value for any pair of GCs is found at the
  intersection of the appropriate row and column. The figure is colour
  coded so that high $p$-values are green and low $p$-values are yellow. }
\label{fig:KSallGCs}
\end{figure*}

Figure \ref{fig:KSallGCs} shows the $p$-values laid out as before for
all pairs of GCs at the end of the simulation in model 2. The lower
right half of the figure shows the $p$-values for velocity
distribution. These $p$-values are all $>$10$\%$ showing we cannot use
their velocity distribution to distinguish stars from different
GCs. The upper left half of the figure shows $p$-values for spatial
distribution. The $p$-values of the last 6 to 8 GCs to merge when tested
with one of the first 18 GCs to merge are low. $p$-values for GC0 to
GC18 taken in pairs show higher likelihood that these GCs are drawn
from the same population.

If we look for groups of GCs which all have $p$-values of $>$10$\%$ when
tested with each other then we find two large groups made up of 11 and
7 GCs and three small groups of 3, 3 and 2 GCs. The two large groups
are made up from the first 18 GCs to merge. If we set the $p$-value
threshold at 1$\%$ we still find 2 large groupings made up of 15 and 6
GCs (again made up from the earlier GCs to merge) and two smaller made
up of 3 and 2 GCs. This implies that if observations find kinematic
substructure then it is not because of mergers.

\subsection{Dependence of K-S results on number of observable data points}
\label{sec:KSdisc}

We have chosen, somewhat arbitrarily, 200 stars to compare our
distributions. Given that we have simulations with $\sim$40000 star
particles representing each GC we could have used up to thousands of
data points to perform our K-S tests. Though the profiles we have
examined would not have changed by using more points it is true that
using more points would result in lower $p$ values for the same value
of D. Observational uses of the K-S test are limited by the number of
stars observed and so for our purposes we should ensure that we are
performing our tests with values of $N_p$ which are comparable to
observations. \citet{Kucinskas2014} perform their K-S tests on 101
main sequence turn-off stars in 47 Tuc meaning they are comparing
subsamples with tens of stars. \citet{Lardo2011} used K-S tests to
distinguish different populations of stars in the $u,\,g,\,r$ SDSS bands.
They compared samples containing from 10s to several hundred stars.
In order to assess the effect of using larger numbers of stars we
repeated some of our K-S tests again with 1000 stars.

\begin{figure*}
\centering
\begin{tabular}{c}
\includegraphics[width=0.33\hsize,angle=0]{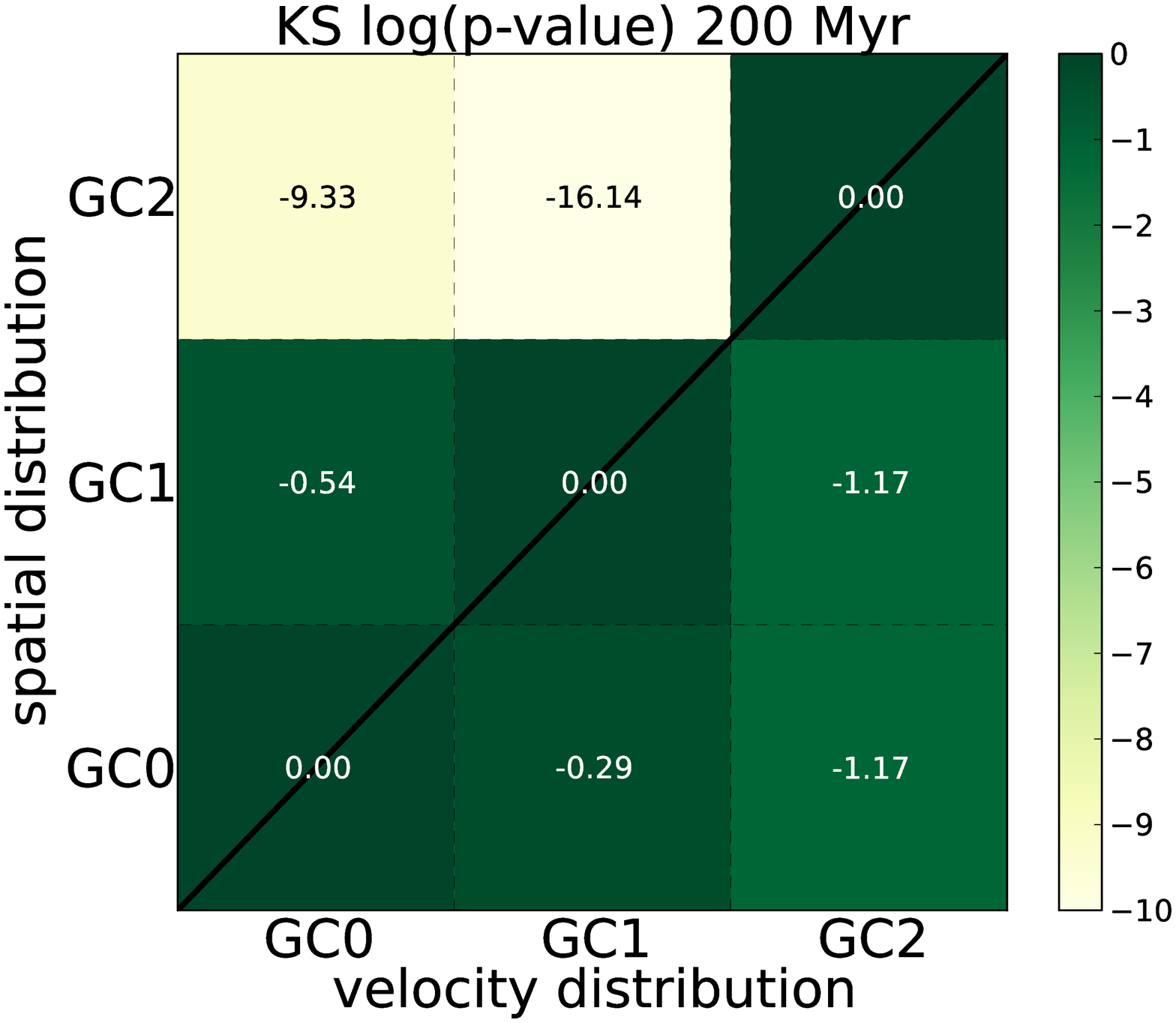}
\includegraphics[width=0.33\hsize,angle=0]{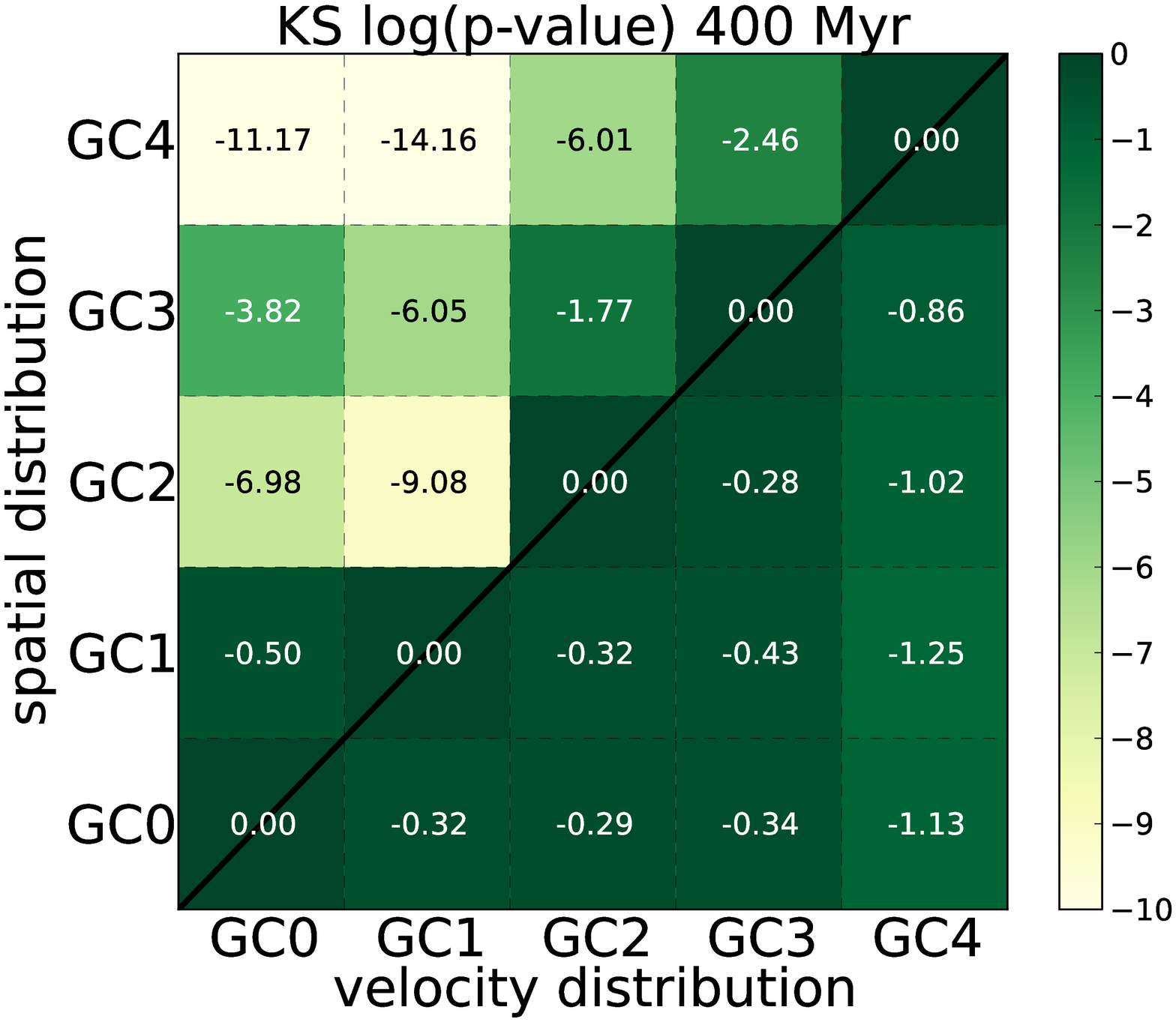}
\includegraphics[width=0.33\hsize,angle=0]{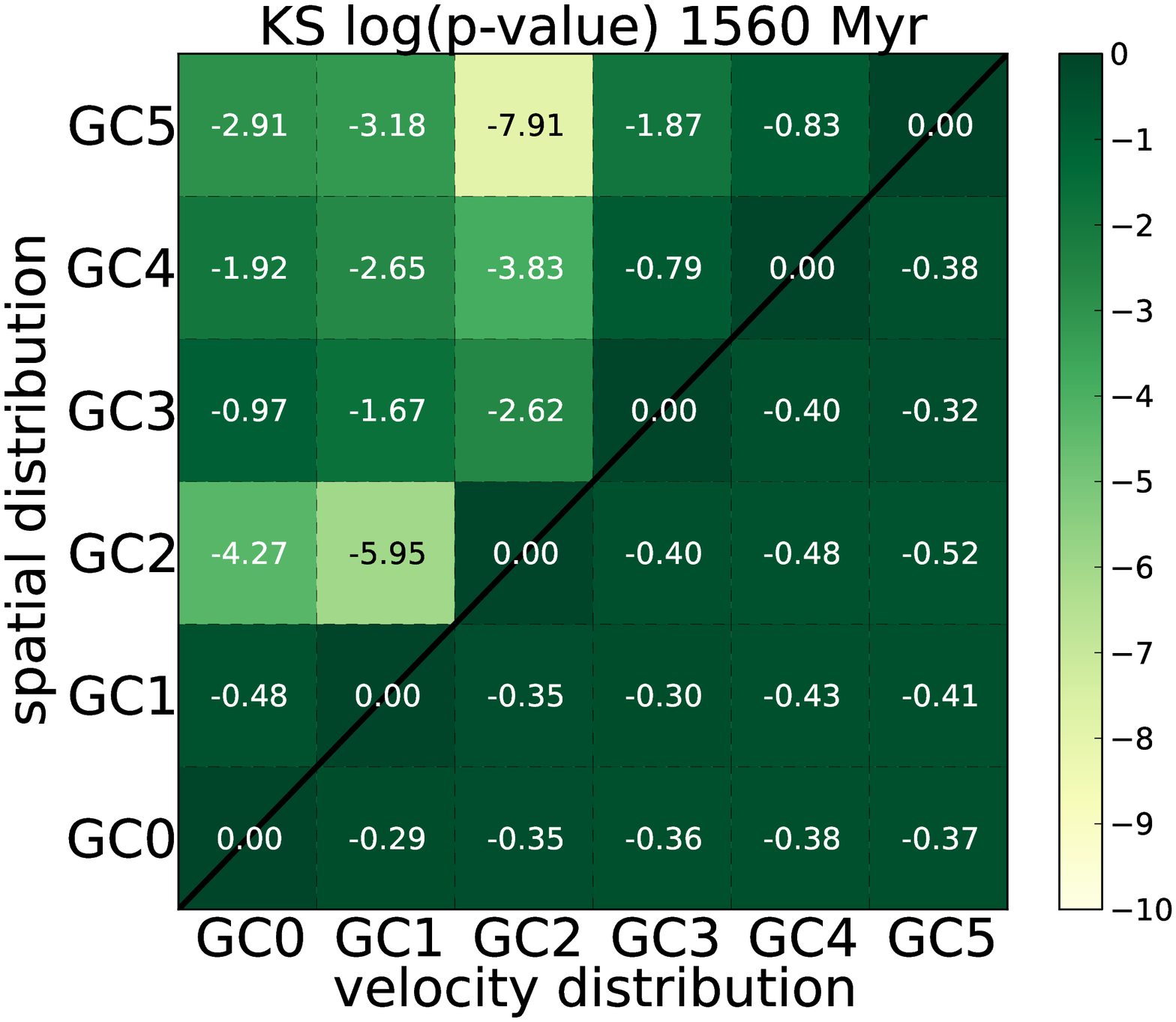}
\end{tabular}
\caption{ Model 1 log$_{10}$ of $p$-values for the cumulative spatial
  distribution of 1000 stars within 4$R_e$ when the merger remnant
  contains 3 GCs (left), 5 GCs (middle) and 6 GCs (right) and
  cumulative absolute velocity distribution as seen by an observer
  viewing the system edge on to the average plane of the GCs' initial
  orbits as in Figure \ref{fig:KSme}. The $p$-value for any pair of GCs
  is found at the intersection of the appropriate row and column. The
  figure is colour coded so that high $p$-values are green and low
  $p$-values are yellow. }
\label{fig:KS1000}
\end{figure*}

Figure \ref{fig:KS1000} shows the $p$-values as described in Section
\ref{sec:KSsim1} but now with 1000 stars in each distribution. At 200
Myr the results are similar to the results with 200 stars. As
previously GC2 is distinguishable by its spatial distribution. At 400
Myr however most GC pairs have a spatial $p$-value $<$10$^{-6}$ and are
now distinguishable by their spatial distribution. Only GC0/GC1 and
GC2/GC3 have a $p$-value for their spatial distribution $>$1.5$\%$. We
see a similar effect at 1560 Myr though not quite as marked. However
at this time GC2, GC4 and GC5 are distinguishable by their spatial
distribution with GC3, GC4 and GC5 having relatively high $p$-values that
they are drawn from the same population. Increasing the number of
stars has had a significant effect on our ability to identify
different populations by their spatial distribution.

\section{Discussion}
\label{sec:discuss}

We have shown that the stars originating in individual GCs which
merge can be difficult to identify from their spatial or velocity
distributions with currently observable sample sizes. It is often the
most recently merged GCs which are distinguishable by
observations. Stars from the most recently merged few GCs have a low
probability, based on their spatial distribution that they are drawn
from the same population as earlier merged GCs. However this
probability increases quickly as more GCs merge and soon become
spatially distributed and kinematically similar to the rest of the
cluster.

Our simulations are examples of violent relaxation
\citep{LyndenBell1967} where the final distribution of particles is
the result of the star particles being scattered by the rapidly
changing gravitational potential produced by a merger. Studies of
violent relaxation in galaxy mergers have found that radial abundance
and colour gradients can survive the mixing of stellar populations but
that they are reduced
\citep{White1980,Barnes1988,Mihos1994,Barnes1996}.

Our simulation remnants are collisionless, non-spherical
systems. \citet{Merritt1996} found that collisionless mixing in
triaxial potentials representative of elliptical galaxies occurs with
characteristic times of $10-30$ dynamical times. For our systems the
dynamical time at R$_e\sim{5}\times10^5$ yr giving a mixing time of 5
to 15 Myr. \citet{Valluri2007} studied the mechanisms responsible for
mixing in collisionless mergers (in their case dark matter
halos). They found that the mixing in phase space is driven by the
exchange of energy and angular momentum at pericentric passage due to
tidal shocks and dynamical friction. They find that in the merger
remnant most particles retain a memory of their original kinetic
energy and angular momentum but there are changes due to the tidal
shocks. Importantly they do not find more large scale mixing in radius
compared to an isolated halo and conclude that radial gradients in
stellar properties such as metallicity can survive such mergers.  This
supposes that such gradients exist prior to a merger, which is very
likely for galaxies. In the case of NSCs (and potentially GCs too) we
would like to know if the merging of stellar systems composed of a
single population can produce multiple populations distinguishable by
their spatial and velocity profiles. If mono-abundance GCs merge to
form NSCs they will retain some of their kinetic energy and angular
momentum profiles. Stars from different merging GCs will have similar
spatial and velocity distributions prior to the merger and retain
these afterwards. Our results indicate that creating a merger remnant
results in the stars from different GCs having a similar spatial and
velocity distributions except for recent mergers, implying that if our
clusters were made up of distinct stellar populations they would be
difficult to detect by their spatial and velocity distributions.

\citet{Kobayashi2004} studied the chemodynamic evolution of elliptical
galaxies following mergers and showed that metallicity gradients have
the largest change when the galaxies are of comparable mass. They find
that when the mass ratio of the two galaxies is more than 20$\%$ then
the metallicity gradient change is $\gtrsim$0.5
dex. \citet{DiMatteo2009} investigated dry mergers of early-type
galaxies with a variety of properties using $N$-body simulations. They
found that such mergers do flatten the metallicity gradient of the
merger remnant but that ellipticals can retain their pre-merger
metallicity gradient if one of the merging galaxies has a steep
pre-merger slope. Should a small metallicity gradient exist in our
merger remnant the repeated merging in our simulations would be likely
to continually reduce any stellar population gradients making any
remnant gradient hard to observe. Building a NSC from GCs would
require early mergers to have mass ratios more than 20$\%$ maximising
the reduction in the existing gradient.

From the globular clusters perspective, the characterisation of the
process of mixing of different stellar populations plays a crucial
role for the interpretation of the spatial and kinematical properties
of present-day Galactic star clusters. The key physical driver of the
mixing is represented by two-body collisional relaxation processes,
which, during the course of the long-term dynamical evolution of the
systems, may gradually erase any intrinsic difference in the spatial
and kinematical distribution of different stellar populations. Within
the formation scenario in which the asymptotic giant branch stars are
the "polluters" contributing to enrich the gas from which the second
generation is formed, \citet{Vesperini2013} have explored, by means of
direct $N$-body simulations, the time-scales and the dynamics of the
spatial mixing of two different populations, and their dependence on
the initial concentration of the ``second generation'' stars.  They
found that the time-scale for complete mixing indeed depends on the
initial concentration of the second generation, but that, in general,
complete mixing is expected only for clusters in the late stages of
their evolution, after they have lost a significant fraction of their
initial mass due to relaxation-driven processes. Such a theoretical
investigation therefore supports the observational evidence that, in
several present-day star clusters, different populations are
characterised by distinguishable spatial distributions (with the
helium-enriched population being the more centrally concentrated
one). In particular, \citet{Kucinskas2014} in a study of 47 Tuc found
that a K-S test of the fractional distribution of the different
generations of stars plotted against radius from the centre of the
cluster gives a probability p = 6.0 $\times10^{-7}$ that the
primordial and chemically enriched distributions are drawn from the
same population (for a total of 101 stars). K-S tests of the absolute
radial velocities of the different stellar generations also give low
probabilities that they are drawn from the same population (p =
7.0$\times10^{-7}$). We emphasise that K-S tests with a greater number
of stars from mass components associated with specific GCs in our
simulations find much higher likelihoods that they are drawn from the
same population both for spatial and velocity distributions.

As for the globular clusters kinematical properties, it has been
shown, again with the support of direct $N$-body simulations, that
different populations may be characterised by different kinematical
properties, as established by the effects of two-body relaxation
process \citep[see Section 4 in][]{Bellini2015}. In particular, it
results that the diffusion from the innermost regions to the outer
parts of the clusters of the most centrally concentrated population is
associated with the growth of radial anisotropy in such a population,
in agreement with recent observational studies of selected Galactic
globular clusters (47 Tuc, \citet{Richer2013}; NGC~2808,
\citet{Bellini2015}). One additional question is related to the
kinematic imprints (and their survival) of different formation
scenarios for multiple stellar populations in
GCs. \citet{HenaultBrunet2015} have addressed such a question by using
direct N-body simulations, and they found that different formation
mechanisms show distinct kinematical signatures that can persist for a
Hubble time. In summary, in the context of the formation and dynamical
evolution of globular clusters, there is convincing evidence, mostly
based on $N$-body models, that spatial and kinematical differences,
either intrinsically associated with the formation scenarios or
induced by collisional relaxation processes, may persist for several
half-mass relaxation times.

Bearing in mind the fundamental differences between the formation
scenarios (and the intrinsic nature) of globular clusters and nuclear
star clusters, we have performed an investigation of the structural
and kinematical properties of the mass components associated with
different proto-clusters, progressively merged to form a single
central stellar systems. Motivated by specific cases of peculiar star
clusters which, in light of their chemical and dynamical complexity,
have been suggested to be stripped nuclei of dwarf galaxies (e.g., M
54, $\omega$ Cen ), we wished to assess the existence and persistence of
any spatial or dynamical signature associated with the merger
histories considered in our two simulations. Our analysis shows that
such a differentiation is difficult with currently available numbers
of observations if NSCs formed by merging alone, except for recent
mergers.

%     A C K N O W L E D G E M E N T

\section{Acknowledgements}

Simulations in this paper were carried out on the COSMOS Shared Memory
system at DAMTP, University of Cambridge operated on behalf of the
STFC DiRAC HPC Facility. This equipment is funded by BIS National
E-infrastructure capital grant ST/J005673/1 and STFC grants
ST/H008586/1, ST/K00333X/1. DRC and VPD were supported by STFC
Consolidated grant \#ST/J001341/1. We made use of pynbody
(https://github.com/pynbody/pynbody) in our analysis for this
paper. ALV acknowledges support from the Royal Commission for the
Exhibition of 1851. We would like to acknowledge the Lorentz Centre in
Leiden which hosted the meeting ``Nuclear Star Clusters in Galaxies,
and the Role of the Environment'', 30 June to 4 July 2014, where
discussions stimulated our ideas set forth in this paper.

%     R E F E R E N C E S                                                                                                                                                                                   

\bibliographystyle{mn2e1}
\bibliography{omc}{}

%\bibliography{allrefs}                                                                                                                                                                                     

\end{document}